\documentclass[12pt]{amsart}
\usepackage{amsmath}
\usepackage{amsxtra}
\usepackage{amscd}
\usepackage{amsthm}
\usepackage{amsfonts}
\usepackage{amssymb}
\usepackage{eucal}
\def\({\left(}
\def\){\right)}
\def\[{\left[}
\def\]{\right]}
\newcommand{\ket}[1]{{| #1 \rangle}}      

\newcommand{\ao}{\mathbf{a}}
\newcommand{\half}{\textstyle{\frac 1 2 }}

\newcommand{\nn}{\nonumber}
\newcommand{\bea}{\begin{eqnarray}}
\newcommand{\ena}{\end{eqnarray}}
\def\bel{\begin{eqnarray}}
\def\enl{\end{eqnarray}}
\newcommand{\be}{\begin{eqnarray*}}
\newcommand{\en}{\end{eqnarray*}}
\newcommand{\ba}{\begin{array}}
\newcommand{\ea}{\end{array}}

\newcommand{\R}{{\mathbb R}}
\newcommand{\bb}{\mathbf{b}}
\newcommand{\mb}{\mathbf{m}}
\newcommand{\kb}{\mathbf{k}}
\newcommand{\fb}{\mathbf{f}}
\newcommand{\ab}{\mathbf{a}}
\newcommand{\bc}{\mathbf{c}}
\newcommand{\cb}{\mathbf{c}}

\newcommand{\bt}{\mathbf{t}}
\newcommand{\tb}{\mathbf{t}}
\newcommand{\qb}{\mathbf{q}}
\newcommand{\gb}{\mathbf{g}}

\newcommand{\ub}{\mathbf{u}}
\newcommand{\C}{{\mathbb C}}
\newcommand{\Z}{{\mathbb Z}}

\newcommand{\G}{\mathbb{G}}

\newcommand{\cR}{\mathcal{R}}

\newcommand{\bbS}{\mathbb{S}}

\newcommand{\bT}{\mathbb{T}}

\newcommand{\Tb}{\mathbb{T}}

\newcommand{\alb}{\mbox{\boldmath$\al$}}

\newcommand{\scalb}{\mbox{\boldmath$\scriptstyle\al$}}

\def\alb{\mbox{\boldmath$\alpha $}}
\newcommand{\Ob}{\mbox{\boldmath$\Omega $}}

\newcommand{\slt}{\mathfrak{sl}_2}
\newcommand{\slth}{\widehat{\mathfrak{sl}}_2}
\newcommand{\res}{{\rm res}}
\newcommand{\id}{{\rm id}}

\newcommand{\tr}{{\rm tr}}

\newcommand{\Tr}{{\rm Tr}}

\newcommand{\bS}{\mathbb{S}}
\newcommand{\End}{\mathop{\rm End}}

\newenvironment{tenumerate}{
  \begin{enumerate}
  
  }{\end{enumerate}}
\newcommand{\bi}{\begin{tenumerate}}
\newcommand{\ei}{\end{tenumerate}}
\newcommand{\isoto}[1][]%
{{\mathop{\buildrel{\sim}\over\longrightarrow}\limits_{#1}}}

\newcommand{\la}{\lambda}

\newcommand{\al}{\alpha}

\newcommand{\s}{\sigma}
\newcommand{\z}{\zeta}

\renewcommand{\mod}{\text{mod}\ (\z^2-1)^{l-m}}
\numberwithin{equation}{section}
\newtheorem{thm}{Theorem}[section]

\newtheorem{lem}[thm]{Lemma}
\newtheorem{rem}[thm]{Remark}
\newtheorem{cor}[thm]{Corollary}

\newcommand{\UAB}{U_{AB}}
\newcommand{\YAB}{Y_{AB}}
\newcommand{\UBA}{U_{BA}}
\newcommand{\YBA}{Y_{BA}}

\newcommand{\taub}{\mbox{\boldmath$\tau$}}
\newcommand{\bL}{\mathbb{L}}

\renewcommand{\sb}{\mathbf{s}}
\newcommand{\bR}{\mathbb{R}}
\newcommand{\J}{\mathbb{J}}
\newcommand{\xb}{\mathbf{x}}
\newcommand{\ctb}{\bar{\mathbf{c}}}
\newcommand{\W}{\mathcal{W}}
\newcommand{\bP}{\mathbb{P}}

\newcommand{\rb}{\mathbf{r}}
\newcommand{\D}{{\mathbb D}}
\newcommand{\A}{{\mathbb A}}
\newcommand{\Wal}{\W^{(\al)}}
\newcommand{\quasi}{{\rm quasi}}

\begin{document} 

\begin{title}[Grassmann Structure in XXZ Model]
{Hidden Grassmann Structure in the XXZ Model II: 
Creation Operators}
\end{title}
\date{\today}
\author{H.~Boos, M.~Jimbo, T.~Miwa, F.~Smirnov and Y.~Takeyama}
\address{HB: Physics Department, University of Wuppertal, D-42097,
Wuppertal, Germany\footnote{
on leave of absence from 
Skobeltsyn Institute of Nuclear Physics, 
MSU, 119992, Moscow, Russia
}}\email{boos@physik.uni-wuppertal.de}
\address{MJ: Graduate School of Mathematical Sciences, The
University of Tokyo, Tokyo 153-8914, Japan;
Institute for the Physics and Mathematics of the Universe, 
Kashiwa, Chiba 277-8582, Japan}\email{jimbomic@ms.u-tokyo.ac.jp}
\address{TM: Department of Mathematics, Graduate School of Science,
Kyoto University, Kyoto 606-8502, 
Japan}\email{tetsuji@math.kyoto-u.ac.jp}
\address{FS\footnote{Membre du CNRS}: Laboratoire de Physique Th{\'e}orique et
Hautes Energies, Universit{\'e} Pierre et Marie Curie,
Tour 16 1$^{\rm er}$ {\'e}tage, 4 Place Jussieu
75252 Paris Cedex 05, France}\email{smirnov@lpthe.jussieu.fr}
\address{YT:  Department of Mathematics, 
Graduate School of Pure and Applied Sciences, Tsukuba University, 
Tsukuba, Ibaraki 305-8571, Japan}
\email{takeyama@math.tsukuba.ac.jp}
\dedicatory{Dedicated to the Memory of Alexei Zamolodchikov}

\begin{abstract}
In this article we unveil a new structure in 
the space of operators of the XXZ chain. 
For each $\al$ we consider the space $\W_\al$
of all quasi-local operators,  
which are products of the disorder field
$q^{\al\sum_{j=-\infty}^0\sigma ^3_j}$ 
with arbitrary local operators.
In analogy with CFT the disorder operator itself is considered
as primary field.
In our previous paper, we have introduced
the annhilation operators $\bb(\z)$, $\cb(\z)$ which 
mutually anti-commute and kill the ``primary field''. 
Here we construct the creation counterpart
$\bb^*(\z)$, $\cb ^*(\z)$ 
and prove the canonical anti-commutation
relations with the annihilation operators. 
We conjecture that the creation operators
mutually anti-commute, thereby upgrading the Grassmann structure
to the fermionic structure. 
The bosonic operator $\tb^*(\z)$ is the generating function of  
the adjoint action by local integrals of motion,  
and commutes entirely with the fermionic creation 
and annihilation operators. Operators $\bb^*(\z)$, $\cb ^*(\z)$, $\tb^*(\z)$ create quasi-local operators starting from the primary field.
We 
show that the ground state averages of quasi-local operators 
created in this way are given by determinants.


\end{abstract}

\maketitle

\bigskip 
\section{Introduction}\label{sec:1}

The present paper is a continuation of our previous article
\cite{HGSI}. 
We consider the infinite XXZ spin chain  with the Hamiltonian
\begin{eqnarray}
H_{\rm XXZ}=\textstyle{\frac{1}{2}}\sum\limits_{k=-\infty}^{\infty}
\left( 
\sigma_{k}^1\sigma_{k+1}^1+
\sigma_{k}^2\sigma_{k+1}^2+
\Delta\sigma_{k}^3\sigma_{k+1}^3
\right), \quad \Delta =\half(q+q^{-1})\,,
\label{Ham}
\end{eqnarray}
where $\sigma^a \, (a=1,2,3)$ are 
the Pauli matrices. 
In this paper we shall mostly consider the critical 
regime where $q=e^{\pi i \nu}$, $\nu\in\mathbb{R}$. 

Let us recall briefly the main definitions and results of the
paper \cite{HGSI}.
Consider the vacuum expectation values (VEVs) 
\begin{align}
\frac{\langle\text{vac}|q^{2\al S(0)}
\mathcal{O}|\text{vac}\rangle}{\langle\text{vac}|q^{2\al S(0)}
|\text{vac}\rangle}\,,
\label{exp}
\end{align}
where $S(k)=\textstyle{\frac 1 2} \sum_{j=-\infty}^k\sigma ^3_j$,
and $\mathcal{O}$ is a local operator (an operator 
localized on a finite portion of the chain). 
We call $X=q^{2\al S(0)} \mathcal{O}$ 
quasi-local operator with tail $\al$.
In other words, an operator $X$ is quasi-local if 
there exist $k\le l$ such that 
$X$ stabilizes as $q^{\al\sigma ^3_j}$ for $j < k$ and as 
the identity $I_j$ for $j>l$. 
The length of $X$ is defined to be the minimum of $l-k+1$. 
The spin of $X$ is the eigenvalue of $\mathbb{S}=[S,\cdot]$ 
where $S=S(\infty)$ is the total spin. 
It will be helpful to think of the operator $q^{2\al S(0)}$ 
as a lattice analog of the primary field in CFT.

Denote by $\mathcal{W}_{\al}$ 
the space of all quasi-local operators with tail $\al$, 
and by $\mathcal{W}_{\al,s}$ the subspace of those with spin $s$. 
Let us introduce the following formal object
\begin{align}
\mathcal{W}=\bigoplus\limits _{\al\in\mathbb{C}}
\mathcal{W}_{\al}\,. 
\label{defW}
\end{align}
We introduce also the operator $\alb$ on $\mathcal{W}$ 
which acts as $\al$ times the identity 
on each summand $\mathcal{W}_{\al}$.
%
%

In \cite{HGSI} we have defined 
anti-commuting one-parameter families of 
operators $\bb (\z)$, $\cb (\z)$ acting on $\mathcal{W}$. 
For reasons which will be clear later, 
we shall call them annihilation operators. 
The annihilation operators have the following block structure
$$
\bb (\z):\ \mathcal{W}_{\al-1,s+1}\to
\mathcal{W}_{\al,s}, 
\quad \bc (\z ):\ \mathcal{W}_{\al+1,s-1}\to 
\mathcal{W}_{\al,s}
\ .
$$
Clearly they commute with $\alb +\mathbb{S}$. 
All other operators considered in this paper have this property. 
Hence, in the actual working, we shall restrict ourselves 
to each eigenspace of $\alb +\mathbb{S}$ with 
 fixed eigenvalue $\alpha\in\C$,  
$$
\mathcal{W}^{(\al)}
=\bigoplus\limits _{s=-\infty}^{\infty}\mathcal{W}_{\al-s,s}
\,.
$$

As we have said, $\bb (\z)$, $\cb(\z)$ are two completely anti-commuting
families of operators:
\begin{align}
\[\bb (\z_1),\bb (\z_2)\]_+=\[\bb (\z_1),\bc (\z_2)\]_+=\[\bc (\z_1),\bc (\z_2)\]_+=0\ .\nn
\end{align}
The annihilation operators have the following structure as functions of
the spectral parameter:
\begin{align}
&\bb (\z)
=\z^{-\scalb-\bbS}
\bigl(\bb _0+\sum\limits _{p=1}^{\infty}(\z ^2-1)^{-p}\bb _p\bigr),\quad
\cb (\z)
=\z^{\scalb+\bbS}
\bigl(\cb _0+\sum\limits _{p=1}^{\infty}(\z ^2-1)^{-p}\cb _p\bigr)\,.\nn
\end{align}
Here the operators $\bb _0$ and $\cb_0$ are written separately  
because they are not independent 
of $\bb _p$ and $\cb_p$ with $p>0$, 
and do not enter the final formulae. 
Besides the anti-commutativity,  
the most important property of the annihilation 
operators $\bb _p$, $\cb _p$ ($p>0$) is:
\begin{align}
\bb _p\bigl(X\bigr)=0,\  \bc _p\bigl(X\bigr)=0\ \text{for}\ \ 
p>\text{length}\bigl(X\bigr)\,.
\label{annihil}
\end{align}
In particular they vanish on the `primary fields' $q^{2\al S(0)}$
(and their translations)  whose
length equals zero. 
The property (\ref{annihil}) explains the nam
`annihilation operators': every monomial of 
$\bb _p$, $\cb_p$ of degree larger than 
$2\,\text{length}(X)$ vanishes on $X$. 

The main result of \cite{HGSI} is the following formula.
Introduce the  linear functional on $\mathcal{W}_{\al}$ 
$$
\mathbf{tr}^{\al}(X)=\cdots\tr _1^{\al}\ \tr _2^{\al}\ 
\tr _3^{\al}\cdots (X)\,,
$$
where we set for $x\in \End(\mathbb{C}^2)$
\begin{align}
\text{tr}^{\alpha}(x)=
\text{tr} \bigl(q^{-\frac 1 2 \al \sigma ^3}x\bigr)/
\text{tr} \bigl(q^{-\frac 1 2 \al \sigma ^3}\bigr)\,.
\nn
\end{align}
Then the VEV is expressed as 
\begin{align}
\frac{\langle\text{vac}|q^{2\al S(0)}
\mathcal{O}|\text{vac}\rangle}{\langle\text{vac}|q^{2\al S(0)}
|\text{vac}\rangle}\ =
\mathbf{tr}^{\al}\(e^{\mbox{\scriptsize\boldmath{$\Omega$}}}\(q^{2\al S(0)}\mathcal{O}\)\)\,,
\label{main}
\end{align}
where $\Ob$ is an operator acting on $\mathcal{W}^{(\al)}$, 
\footnote{We change slightly the 
normalization of $\bb$, $\bc$ from \cite{HGSI}, 
but $\boldsymbol\Omega$ remains unchanged.}
\begin{align}
\Ob=
 \res _{\z_1^2=1}\res _{\z_2^2=1}
\(\omega \(\z_1/\z_2,\al\)\bb (\z _1)\bc (\z _2)
\frac{d\z _1^2}{\z _1^2}\frac{d\z _2^2}{\z _2^2}\)\,,\nn
\end{align}
and 
\bea
&&
\omega(\z,\alpha)=\omega_{\rm trans}(\z,\alpha)-\frac{4q^{\alpha}}{\(1-q^{\alpha}\)^2}\,\omega_0(\z,\alpha)\,  \nn
\ena 
is a scalar function. 
For future convenience $\omega(\z,\alpha)$ 
is split into two pieces,  
the transcendental part and the elementary part.
The trancendental part is given by
\be
&&\omega_{\rm trans}(\z,\alpha)
=P\!\!\int\limits_{-i\infty}^{i\infty}\z ^{u+\alpha}
\frac {\sin \frac {\pi} 2\(u-\nu(u+\alpha)\)}{\sin \frac {\pi} 2 u
\cos \frac {\pi\nu} 2\(u+\alpha\)}du\,. 
\nn
\en
The elementary part is defined by
\bea
\omega_{0}(\z,\al)=-
\left(\frac{1-q^{\alpha}}{1+q^{\alpha}}\right)^{2}
\Delta_{\z}\left(
\psi (\z,\al)\right)\,,\label{omega0}
\ena
where we introduced two important  notations:
$$\Delta _\z (f(\z))=f(\z q)-f(\z q^{-1}), \quad
\psi (\z ,\al)=\z ^{\al}\ \frac {\z ^2+1}{2(\z ^2-1)}\,.$$

In the present paper we complete the construction of \cite{HGSI}
introducing the creation operators. 
Along with the homogeneous chain
described by the Hamiltonian (\ref{Ham}), 
we consider also the inhomogeneous one. 
The latter case is often very useful, 
but in this Introduction we shall deal only with
the homogeneous case which has a clearer physical meaning. 

The creation operators must generate the entire space
$\mathcal{W}^{(\al)}$ from the primary field $q^{2\al S(0)}$, 
and must have nice commutation relations 
with the annihilation operators.
Obvious examples of this sort were discussed in \cite{HGSI}.
First, there are the operators 
$\taub$  ($\taub ^{-1} $) of right (left)
shift along the chain which change neither the length 
of quasi-local operators nor their VEVs. 
Second, there is the adjoint action of local 
integrals of motion on $\mathcal{W}^{(\al)}$. 
By this adjoint action, we do  
create operators with larger length from 
a given quasi-local operator. 
However, their VEVs 
vanish for a clear reason. 
These facts are consistent with the right hand side of  
(\ref{main}), for, as has been explained in \cite{HGSI},  
$\taub ^{\pm1}$ and the adjoint action of local integrals 
of motion commute with $\bb $, $\cb$ as well as $q^{-\al S}$ 
which enters the definition of ${\bf tr}^\al$ .  
The first creation operator which we shall describe
in this paper is $\tb ^*(\z )$, 
the adjoint action of the usual transfer matrix. 
In other words, $\log(\taub^{-1}\tb^*(\z ))$ is the generating function 
for the adjoint action of local integrals of motion.
Obviously  $\tb ^*(\z )$ is block diagonal:
$$
\tb ^*(\z ):\ \ 
\mathcal{W}_{\al ,s}\rightarrow \mathcal{W}_{\al ,s}.
$$
This operator satisfies the commutation relations
$$
[ \tb ^*(\z _1 ),\tb^* (\z _2)]=0,
\quad[ \tb ^*(\z _1 ),\bb (\z _2)]
=[ \tb ^*(\z _1 ),\bc (\z _2)]=0\,,
$$
and has the expansion in $\z ^2-1$, 
$$
\tb ^*(\z)=\sum\limits _{p=1}^{\infty} (\z ^2-1)^{p-1}\tb ^* _p
\,,
$$
where $\tb ^*_1=2\taub$. 
The operators $\tb^*_p$ satisfy
the main property of our creation operators:
they increase the length of operators, but do this in a 
controllable way, namely 
$$
\text{length}\(\tb^*_p(X)\)\le 
\text{length}\(X\)+p\,.
$$

Now we come to the description of the main part of 
our construction.
We define the operators $\bb ^*(\z)$, $\cb ^*(\z )$
acting on $\mathcal{W}$
with the following block structure 
$$
\bb ^*(\z):\ \mathcal{W}_{\al+1,s-1}\to
\mathcal{W}_{\al,s}\,, 
\quad \bc ^*(\z ):\ \mathcal{W}_{\al-1,s+1}\to 
\mathcal{W}_{\al,s}
\ ,
$$
and the dependence on $\zeta$:
\begin{align}
&\bb ^*(\z)
=\z^{\scalb+\bbS+2}
\ \sum\limits _{p=1}^{\infty}(\z ^2-1)^{p-1}\bb ^*_p,\quad
\cb ^*(\z)
=\z^{-\scalb-\bbS-2 }\ 
\sum\limits _{p=1}^{\infty}(\z ^2-1)^{p-1}\cb _p^*\,.\nn
\end{align}

The definition of the annihilation operators is 
a result of a long chain of transformations 
from the multiple integral formulae for VEV \cite{JMbk}. 
In contrast, the way we define the creation operators 
cannot be explained absolutely logically.  
Their definition is a result of many experiments, 
mistakes, dead ends, {\it etc.}. 
Even after the correct operators have been found, 
the proof of their properties took some time. 
The first property explains that the operators  
$\bb ^*(\z)$, $\cb ^*(\z )$
are creation operators acting on $\mathcal{W}$:
\begin{align}
&\text{length}\bigl(\bb _p^*\bigl(X\bigr)\bigr)\le\text{length}\bigl(X\bigr)+p
\label{create}\,\\
&\text{length}\bigl(\cb _p^*\bigl(X\bigr)\bigr)
\le\text{length}\bigl(X\bigr)+p\,.
\nn
\end{align}
The second property is the commutation relations with the
annihilation operators:
\begin{align}
&[\bb (\z _1),\cb ^*(\z _2)]_+=[\cb (\z _1),\bb ^*(\z _2)]_+=0
\,,\label{comancr}\\
&[\bb (\z _1 ),\bb ^*(\z _2)]_+
=-\psi (\z_2/\z_1,\alb+\mathbb{S})\,,\nn\\
&[\cb (\z _1 ),\cb ^*(\z _2)]_+=
\psi (\z_1/\z_2,\alb+\mathbb{S})\,.
\nn
\end{align}
The third property consists in the following fact:
\begin{align}
\mathbf{tr}^{\al}\bigl(e^{\mbox{\scriptsize\boldmath{$\Omega$}}_0}
\bb ^*(\z )
(q^{2(\al +1)S(0)}\mathcal{O}_1)\bigr)=0,\ \ 
\mathbf{tr}^{\al}\bigl(e^{\mbox{\scriptsize\boldmath{$\Omega$}}_0}
\cb ^*(\z )
(q^{2(\al -1)S(0)}\mathcal{O}_2)\bigr)=0\,,\label{altr}
\end{align}
where $\mathcal{O}_1$, $\mathcal{O}_2$ have respectively spins $-1$
and $1$,
\begin{align}
\Ob  _0=-
 \res _{\z_1^2=1}\res _{\z_2^2=1}
\(\omega _0 \(\z_1/\z_2,\al\)\bb (\z _1)\bc (\z _2)
\frac{d\z _1^2}{\z _1^2}\frac{d\z _2^2}{\z _2^2}\)\,,\nn
\end{align}
and $\omega _0$ is the simple function defined in (\ref{omega0}).
Let us restrict our considerations to $\mathcal{W}^{(\al)}$.
The primary field $q^{2\al S(0)}$ 
is in the common kernel of  the annihilation operators and 
plays the role of the `right vacuum'. 
On the other hand, the linear functional
$\mathrm{v}^{\al}$ on $\mathcal{W}^{(\al)}$ given by
$$
\mathrm{v}^{\al}(\ \cdot\ )
=\mathbf{tr}^{\al}\bigl(e^{\mbox{\scriptsize\boldmath{$\Omega$}}_0}(\ \cdot\ )\bigr)
$$
plays the role of the `left vacuum': 
it vanishes on the image of creation operators.

Starting from the primary field $q^{2\al S(0)}$,  
let us define inductively quasi-local operators  
$$
X^{\epsilon_1\cdots \epsilon _k}(\z_1,\cdots ,\z _k;\al)
=\begin{cases}
\bb ^*(\z _k)X^{\epsilon_{1}\cdots \epsilon _{k-1}}
(\z_1,\cdots ,\z _{k-1};\al) & (\epsilon _k=+), \\
\cb ^*(\z _k)(-1)^{\mathbb{S}}
X^{\epsilon_{1}\cdots \epsilon _{k-1}}
(\z_1,\cdots ,\z _{k-1};\al) & (\epsilon _k=-),\\
\half\tb ^*(\z _k)
X^{\epsilon_{1}\cdots \epsilon _{k-1}}
(\z_1,\cdots ,\z _{k-1};\al)& (\epsilon _k=0). \\
\end{cases}
$$
Actually,
$X^{\epsilon_1\cdots \epsilon _n}(\z_1,\cdots ,\z _n,\al)$  is
rather a generating function of quasi-local operators:
\begin{align}
X^{\epsilon_1\cdots \epsilon _n}(\z_1,\cdots ,\z _n;\al)=\prod
\z _j^{\epsilon _j\al}
\sum _{p_1,\cdots ,p_{n}}(\z_1 ^2-1)^{p_1-1}\cdots 
(\z_n ^2-1)^{p_n-1}
X^{\epsilon_1\cdots \epsilon _n}_{p_1,\cdots ,p_n}
(\al)\,,
\nn
\end{align}
where the coefficients
$X^{\epsilon_1,\cdots,\epsilon_n}_{p_1,\cdots,p_n}(\al)$ are 
quasi-local operators from $\W_{\al-s,s}$, with $s=\#(j:\epsilon _j=+)-\#(j:\epsilon _j=-)$.
Rewriting the formula (\ref{main}) as 
\begin{align}
\frac{\langle\text{vac}|q^{2\al S(0)}
\mathcal{O}|\text{vac}\rangle}{\langle\text{vac}|q^{2\al S(0)}
|\text{vac}\rangle}\ =
\mathrm{v}^{\al}\(e^{\mbox{\scriptsize\boldmath{$\Omega$}}
-\mbox{\scriptsize\boldmath{$\Omega$}}_0}\(q^{2\al S(0)}\mathcal{O}\)\)\,,
\label{main2}
\end{align}
we get immediately 
\begin{align}
\frac{\langle\text{vac}|
X^{\epsilon_1\cdots \epsilon _n}(\z_1,\cdots ,\z _n,\al)|\text{vac}\rangle}{\langle\text{vac}|q^{2\al S(0)}
|\text{vac}\rangle}= \det \((\omega
-\omega _ 0)(\z _{j^+_p}/\z _{j^-_q},\al)\)_{p,q=1.\cdots ,l}\nn
\end{align}
where 
$j_1^+<\cdots <j_l^+$ are the indices with  
$\epsilon _{j_p^+}=+$ 
and $j_1^-<\cdots <j_l^-$ are those with $\epsilon _{j_p^-}=-$.

At the moment we do not have a proof of 
the completeness of 
$X^{\epsilon_1\cdots \epsilon _n}(\z_1,\cdots ,\z _n,\al)$ 
in $\mathcal{W}^{(\al)}$, but we conjecture that it holds.
This conjecture is supported by the consideration of the
inhomogeneous case, for which completeness 
is easy to prove.

Let us comment on the commutation relations of the creation operators. We prove that
$$[\tb ^*(\z_1),\bb ^*(\z _2)]=[\tb ^*(\z_1),\cb ^*(\z _2)]=0\,. $$
These commutation relations are crucial for our construction:
they show that the fermionic operators commute completely with
the adjoint action of the local integrals of motion.

We do not prove, but only conjecture
the remaining commutation relations:
\begin{align}
&[\bb ^*(\z_1),\bb ^*(\z _2)]_+=[\bb ^*(\z_1),\cb ^*(\z _2)]_+=[\cb ^*(\z_1),\cb ^*(\z _2)]_+=0\,.
\label{comcreation}
\end{align}
We already know from our construction that 
these commutation relations hold in a weak sense, i.e., 
when we consider pairings with elements of 
the subspace of $\mathcal{W}^{(\al)*}$
created from $\mathrm{v}^{\al}$ by right action 
of the annihilation operators. 
This is enough for our goals. 
So, we decided to leave the direct proof of (\ref{comcreation})
for future work. 
We think that the reader will forgive this after passing 
through extremely complicated calculation of  
Section \ref{sec:4} devoted to the commutation relations.
Nevertheless, 
computer experiments suggest that (\ref{comcreation}) hold
generally. 

In summary, we have fermions 
which are (conjecturally) completely canonical, and  
the (adjoint) integrals of motion commuting with them.  
It would be very interesting to find the conjugate 
operators for the latter.  When all this has been done, 
we would have a novel description of the space 
of quasi-local operators: 
it is simply the tensor product of Fock spaces 
of fermions and  bosons.
For the descendant operators created 
by the latter, the VEV's can be computed as in free theory. 
Hence it is important to know how to express a given quasi-local 
operator 
in terms of these descendants. This remains a major open problem. 

Finally let us comment 
on the paper by Bazhanov-Lukyanov-Zamolodchikov 
\cite{BLZ}. It contains besides deep analytic conjectures  
a remarkable algebraic construction. 
Namely these authors relate Baxter's 
$Q$-operators to transfer matrices 
constructed via the $q$-oscillator 
representation of the Borel subalgebra 
of the quantum affine algebra $U_q(\slth )$. 
The BLZ treatment of the $q$-oscillator representation 
is a cornerstone of our algebraic construction. 
Unlike the usual considerations, however, 
we introduce operators not on the space 
of states but rather on the space of quasi-local operators. 
So our philosophy is closer to that of CFT than to the 
usual approach of QFT. 
In order to define such operators,  
we use transfer matrices in the adjoint representation. 
Our main message is that a correct understanding of 
the $q$-oscillator transfer matrices allows one 
to define fermionic operators 
in addition to the usual commutative families of $Q$-operators.  
In a recent work \cite{Wuppertal}, it is conjectured and checked
on examples that at least in the limit $\al \to 0$ 
the same creation-annihilation operators describe 
the thermal averages, 
only the function $\omega (\z,\al)$ becomes dependent
on the temperature. 
This suggests the universal character of
our algebraic construction. 

The plan of the paper is as follows. 

In Section 2 we introduce our notation. 
Working with a fixed interval $[k,l]$,  
we define various transfer matrices 
acting on the space of local operators
and state their basic properties. 
These operators typically have poles at $\z^2=\xi_j^2$ and/or  
$\z^2=q^{\pm2}\xi_j^2$, where $\xi_j$'s are the inhomogeneity parameters. 
On the basis of this pole structure, 
we then define the annihilation and creation operators on $[k,l]$. 

When the interval $[k,l]$ is extended to the left as  
$[k',l]$ ($k'<k$), operators on the larger interval 
are simply related to those on the smaller. 
On the other hand, extension to the right $[k,l']$ ($l'>k$)
is non-trivial. 
We call these (the left and right) reduction relations 
and study them in Section 3. 
Using the reduction relations, 
we extend the operators in Section 2 to those 
on the space of operators on the whole infinite chain. 
While the annihilation operators 
$\cb$, $\bar\cb$, $\bb$, $\bar\bb$ 
are defined in 
the same way both for homogeneous and inhomogeneous chains, 
the creation operators need to be treated separately.
We explain the difference of the 
construction first in the simpler case of $\tb^*$,   
and then proceed to $\bb^*$, $\cb^*$. 

In Section 4, we study the commutation relations. 
We shall mainly discuss homogeneous chains. 
We show that $\bt^*$ commutes with creation and annihilation operators,  
and that the annihilation operators
 $\cb$, $\bar\cb$, $\bb$, $\bar\bb$ mutually anti-commute. 
Proof of the anti-commutation relations between creation and annihilation 
operators is technically quite involved, and occupies 
a substantial part of the section.  
The main results are Theorems \ref{thcb*}, \ref{thbb*}. 
The commutation relations between creation operators 
remain as conjecture. We prove the simplest case between 
$\tb^*$ and $\bb^*,\bc^*$ in Theorem \ref{tht*b*}. 
Results about the inhomogeneous case are stated
as Theorem \ref{thm:comm-inhom} at the end of the section. 

We use these results to construct a fermionic basis in Section 5,  
and evaluate their VEV's. 
The determinant formula for VEV's is given as Theorem \ref{determinant},
\ref{det-homog}.

The text is followed by 4 appendices. 
In Appendix A we collect some necessary facts concerning the 
quantum affine algebra $U_q(\slth)$ and R matrices. 
In Appendix B we give a proof of a technical Lemma in Section 3. 
When we deal with the $q$ oscillator representations, 
one of the technical complications is that the $R$ matrix does not exist 
for the tensor product $W^+$ and $W^-$ (see Appendix A for the notation). 
We explain in Appendix C 
that the original BLZ construction offers a way around,  
and deduce exchange relations of monodromy matrices under the trace
which are used in the main text.  
The definition of the annihilation operators adopted in this paper
slightly differs from the one in the previous work \cite{HGSI}, \cite{FB}. 
In Appendix D, we give the precise connection between the two. 

\medskip

This paper is dedicated to the memory of Alexei Zamolodchikov.
His premature decease was a great shock for all of us.

Aliosha, thinking about you, generosity is the word which comes to our mind.
You had a great talent, which you shared with the scientific community,
and at the same time you were a kind and open man.
This is how we shall always remember you.

\section{Creation and annihilation operators in finite intervals}\label{sec:2}

In this section we introduce various operators which act on the space of 
linear operators on a finite tensor product of $\C^2$.
In subsequent sections, we will study their basic properties, 
such as the reduction and commutation relations,
and compute the expectation values of their products.
\subsection{Twisted transfer matrices}
First, let us explain the basic construction in a general setting
using the representation theory of $U_q\slth$.
We fix $q\in\C$, which is not a root of unity. We leave some details on the
representation theory to Appendix \ref{sec:appA}.

We use the universal $R$ matrix $\mathcal{R}$ for the quantum affine algebra
$U_q\slth$. Set
$\mathcal R':=\mathcal R\cdot q^{c\otimes d+d\otimes c}
\in U_q\mathfrak{b}^+\otimes U_q\mathfrak{b}^-$.
The Borel subalgebra $U_q\mathfrak{b}^+$ is generated by $e_0,e_1,t_0,t_1$,
and $U_q\mathfrak{b}^-$ by $f_0,f_1,t_0,t_1$.
In this paper, we consider the level zero case where $c=0$.
Take two representations,
$\pi_{\rm aux}:U_q\mathfrak{b}^+\rightarrow\End(V_{\rm aux})$ and
$\pi_{\rm qua}:U_q\mathfrak{b}^-\rightarrow\End(V_{\rm qua})$.
The former is called the `auxiliary' space and the latter the `quantum' space.
Set
\be
L_{V_{\rm aux}\otimes V_{\rm qua}}
:=(\pi_{\rm aux}\otimes\pi_{\rm qua})(\mathcal R').
\en
This is called the $L$ operator. The most basic property of the $L$ operator
is the commutativity
\bea
[q^{\pi_{\rm aux}(h_1)}\otimes q^{\pi_{\rm qua}(h_1)},
L_{V_{\rm aux}\otimes V_{\rm qua}}]=0.\label{H}
\ena
For $X\in\End(V_{\rm qua})$, set
\bea
&&\tb_{V_{\rm qua}}(\al)(X):={\rm trace}_{V_{\rm aux}}
\left\{L_{V_{\rm aux}\otimes V_{\rm qua}}\cdot
\left(q^{\al\pi_{\rm aux}(h_1)}\otimes X\right)\cdot
\left(L_{V_{\rm aux}\otimes V_{\rm qua}}\right)^{-1}\right\}.\label{TTM}
\ena
We thus obtain an operator acting on the space of operators on
the quantum space. We call this operator the twisted transfer matrix.
This is different from the usual setting where the transfer matrix is acting
on the quantum space itself.

Now, we specify our auxiliary and quantum spaces.
Let $V$ be a two-dimensional vector space over $\C(q^\al)$,
where $q^\al$ is a formal variable. The reason for introducing $q^\al$ is
clear from \eqref{TTM}.

Fixing a basis of $V$,
we identify $M=\End(V)$ with the algebra of $2\times 2$ matrices.
With each $j\in\Z$ we associate $V_j\simeq V$, $M_j\simeq M$
and $\xi_j\in\C^\times$.
The tensor product $\otimes_{j\in\Z}V_j$ with the `inhomogeneity'
parameters $\xi_j$ is called the inhomogeneous chain. The parameter
$\xi_j$ is used to specify the action of $U_q'\slth$ on $V_j$. This point
will be explained shortly.

For each finite interval $[k,l]\subset \Z$, we denote by 
\bea
M_{[k,l]}:=M_k\otimes\cdots\otimes M_l\,\label{Vkl}
\ena
the space of operators in the interval $[k,l]$.
This is the space on which the twisted transfer matrices act when
$V_{\rm qua}=V_k\otimes\cdots\otimes V_l$ is chosen.

We define an action of $U_q\mathfrak{b}^-$ on $V$:
\be
\pi^{(1)}_\z=\pi^{(1)}\circ ev_\z:U_q\mathfrak{b}^-\rightarrow M,
\en
where the notation is explained in Appendix \ref{sec:appA}.
We use $\z=\xi_j$ for $V_j$.

Let us fix the notation for $L$ operators.
We shall consider two kinds of auxiliary spaces:
representations of $U_q'\slt$
and representations of the $q$-oscillator algebra $Osc$.
Let $M_a$ be a copy of $M$. We set
\be
L_{a,j}(\z/\xi):=(\pi^{(1)}_\z\otimes\pi^{(1)}_\xi)
\mathcal R'\quad\in M_a\otimes M_j.
\en
We have
\be
L_{a,j}(\z)=\rho(\z)L^\circ_{a,j}(\z)
\en
where $L^\circ_{a,j}(\z)$ is the standard trigonometric $R$ matrix
\bea
&&
L^\circ_{a,j}(\z):=
\begin{pmatrix}  1&0&0&0\\
0&\beta(\z) &\gamma(\z)&0\\
  0&\gamma(\z)&\beta(\z) &0\\
  0&0&0&1
\end{pmatrix},
\label{eq:Rmat}\\
&&\beta(\z):=\frac {\z -\z ^{-1}}{q \z - q ^{-1}\z^{-1}}\,,
\quad
\gamma(\z):=\frac {q-q ^{-1}}{ q \z-q ^{-1}\z^{-1} }\,.\nn
\ena
The inverse of $L^\circ_{a,j}(\z)$ is given by
$L^\circ_{a,j}(\z)^{-1}=L^\circ_{a,j}(\z^{-1})$.

See Appendix \ref{sec:appA}, \eqref{SIGMA} 
for the normalization factor 
$\rho(\z)$. We also use the notation $R_{j_1,j_2}(\z):=L^\circ_{j_1,j_2}(\z)$ especially
when both of the tensor components are from the inhomogeneous chain.
This occurs when we specialize the spectral parameter
$\z$ of the auxiliary space to $\xi_j$.

The $q$-oscillator algebra $Osc$ is also defined over $\C(q^\al)$.
It has the generators $\ao,\ao^*,q^{\pm D}$ and the relations
\bea
&&q^D\ao\ q^{-D}=q^{-1}\ao,\quad 
q^D\ao^* q^{-D}=q\ \ao^*\,,
\label{Osc}
\quad\ao\ \ao^*=1-q^{2D+2},\quad
\ao^* \ao=1-q^{2D}\,.
\nn
\ena

We have a homomorphism $o_\z:U_q\mathfrak{b}^+\to Osc$ given by
\bea
&&o_\z(e_0)=\frac{\z}{q-q^{-1}}\ao,\
o_\z(e_1)=\frac{\z}{q-q^{-1}}\ao^*,\
o_\z(t_0)=q^{-2D},\ o_\z(t_1)=q^{2D}\,.\label{osc}
\ena
Let $Osc_A$ be a copy of $Osc$. We set
\bea
L_{A,j}(\z/\xi):=
\left(o_{\z}\otimes\pi_{\xi}\right)\mathcal R'
\quad\in Osc_A\otimes M_j.\label{LOP}
\ena
We have
\bea
L_{A,j}(\z)=\sigma(\z)L^\circ_{A,j}(\z)\label{Lcirc}
\ena
where
\bea
&&L^\circ_{A,j}(\z):=
\begin{pmatrix}
1-\z^2q^{2D_A+2}&-\z\ab_A\\-\z\ab^*_A&1
\end{pmatrix}_j
\begin{pmatrix}q^{-D_A}&0\\0&q^{D_A}
\end{pmatrix}_j,\label{eq:Lplus}\\
&&L^\circ_{A,j}(\z)^{-1}:=
\frac1{1-\z^2}
\begin{pmatrix}q^{D_A}&0\\0&q^{-D_A}\end{pmatrix}_j
\begin{pmatrix}
1&\z\ab_A\\\z\ab^*_A&1-\z^2q^{2D_A}
\end{pmatrix}_j.
\ena
We consider two representations $W^\pm$ of $Osc$, but we do not use them
before Section \ref{sec:4}. See Appendix \ref{sec:appA} for
their definitions.

In what follows, we shall use indices $a,b,\cdots$ as labels
for $M$ or its representation $V$, and
$A,B,\cdots$ for $Osc$ or its representations $W^\pm$. They are the auxiliary
space indices.
We use the indices $j,k,\cdots$ for the quantum spaces,
the components of the inhomogeneous chain.

Here we make some notational principles on suffixes.
We denote by $X_{[k,l]},Y_{[k,l]},\cdots$ operators which belong to
$M_{[k,l]}$. We denote by $x_a,y_a,\cdots,$ $2\times2$ matrices
which belong to $V_a$. We use also $L_{a,j}$, $T_{A,[k,l]}$, {\it etc.}.
They are some special operators which belong to $M_a\otimes M_j$,
$Osc_A\otimes M_{[k,l]}$, {\it etc.}. We do not drop suffixes in these cases.

In Section 3 we introduce the spaces of operators $\W_\al$, $\W^{(\al)}$, {\it etc.},
for which $k=-\infty$ and $l=\infty$. We denote by $X,Y,\cdots$ operators
which belong to these spaces, without putting suffixes.

We denote by boldface letters $\bb,\cb,\cdots$ or 
`blackboard boldface' letters $\bT,\bS,\cdots$
the operators acting on the spaces $M_a$, $Osc_A$, $M_{[k,l]}$, {\it etc.}.
We also put suffixes $a$, $A$, $[k,l]$ indicating the spaces on which they act.
However, if they are written with operands in quantum spaces,
say, $X_{[k,l]}$, we may drop the suffix $[k,l]$ from these operators.
There are two exceptions for this rule: if the interval for the operand
is larger than that of the operator (this happens when we divide
the latter in two parts), we do not drop the suffix in the latter;
if an operator $\xb_{[k,l]}$ acts on $X_{[k,m]}\id_{[m+1,l]}$,
where $\id_{[m+1,l]}\in M_{[m+1,l]}$ is the identity operator, we write
$\xb_{[k,l]}(X_{[k,m]})$ to mean $\xb(X_{[k,m]}\id_{[m+1,l]})$.
We do not drop the suffixes for auxiliary spaces.

We stop talking just about notations.
Now we define the twisted transfer matrices on $M_{[k,l]}$.
When we choose $V_a$ as the auxiliary space the twisted transfer matrix
\eqref{TTM} is written as
\be
&&
\tb^*(\z,\al)(X_{[k,l]}):=\Tr_a\left[\mathbb{T}_a(\z,\al)(X_{[k,l]})\right]\,,
\label{tstar}\\
&&\bT_a(\z,\al)(X_{[k,l]}):=T_{a,{[k,l]}}(\z)q^{\al\s^3_a}X_{[k,l]}
T_{a,{[k,l]}}(\z)^{-1}\,,\label{Ta}\\
&&T_{a,{[k,l]}}(\z):=L_{a,l}(\z/\xi_l)\cdots L_{a,k}(\z/\xi_k),
\en
where $X_{[k,l]}\in M_{[k,l]}$.
Here $\Tr_a:M_a\rightarrow\C(q^\al)$ is defined by the usual trace on
the two dimensional space. Later we use the notation
\bea
\bL_{a,j}(\z/\xi_j)(X_{[k,l]}):=
L_{a,j}(\z/\xi_j)\cdot X_{[k,l]}\cdot L_{a,j}(\z/\xi_j)^{-1}.\label{Lba}
\ena
Here we take $X_{[k,l]}\in M_a\otimes M_{[k,l]}$. In other words, the operator
$\bL_{a,j}$ belongs to $\End(M_a)\otimes \End(M_{[k,l]})$.
It is not considered as an element of $M_a\otimes \End(M_{[k,l]})$.
Therefore, we have
\be
\bT_a(\z,\al)(X_{[k,l]})=\bL_{a,l}(\z/\xi_l)\cdots\bL_{a,k}(\z/\xi_k)
(q^{\al\s^3_a}X_{[k,l]})
\en
for $X_{[k,l]}\in M_{[k,l]}$. We set $\bT_{a,[k,l]}(\z):=\bT_{a,[k,l]}(\z,0)$.
Notice that $\bL_{a,j}(\z)$ has poles at $\z^2=q^{\pm2}$.

Products of $L$ operators such as $T_{a,{[k,l]}}(\z)$ are called monodromy matrices.
The twisted transfer matrix $\tb^*_{[k,l]}(\z,\al)$
is essentially the trace of the adjoint action
of the monodromy matix $T_{a,{[k,l]}}(\z)$. We note that the choice of
the normalization factor $\rho(\z)$ is irrelevant for the adjoint action,
though in some calculations the properties of the universal $R$ matrix
help us.

Define the total spin operator $\bS_{[k,l]}\in\End(M_{[k,l]})$ by
\be
\bS(X_{[k,l]}):=[S_{[k,l]},X_{[k,l]}],\quad 
S_{[k,l]}:=\half\sum_{j\in[k,l]}\sigma^3_j. 
\en
We say an operator $X_{[k,l]}$ is of spin $s$ if and only if
\bea
\bS(X_{[k,l]})=sX_{[k,l]}.\label{SPINX}
\ena

When a representation of $Osc_A$ is used for the auxiliary space,
we modify \eqref{TTM} by insertion of $\z^{\al-\bS_{[k,l]}}$ and
the left multiplication by $q^{-2S_{[k,l]}}$:
\bea
&&
\qb(\z,\al)(X_{[k,l]}):=\Tr_A\left[\bT_A(\z,\al)
\z^{\al-\bS}(q^{-2S_{[k,l]}}X_{[k,l]})\right]\,,\label{qstar}\\
&&\bT_A(\z,\al)(X_{[k,l]}):=T_{A,{[k,l]}}(\z)
\left(q^{2\al D_A}X_{[k,l]}\right)T_{A,{[k,l]}}(\z)^{-1}\,,\label{TA}\\
&&T_{A,{[k,l]}}(\z):=L_{A,l}(\z/\xi_l)\cdots L_{A,k}(\z/\xi_k)\,,
\ena
where $X_{[k,l]}\in M_{[k,l]}$. Here the trace
$\Tr_A:q^{2\al D_A}Osc_A\rightarrow\C(q^\al)$ is defined
in Appendix \ref{sec:appA}.
We define
the  operator $\bL_{A,j}(\z)$ like in (\ref{Lba}), it has a pole at $\z^2=1$.

The reason for putting $\z^{\al-\bS_{[k,l]}}$
in the definition of $\qb_{[k,l]}(\z,\al)$
is that with this insertion the Baxter equation looks nicer:
\bea
{\bf q}_{[k,l]}(q\z ,\al)+{\bf q}_{[k,l]}(q ^{-1}\z,\al)-
\tb^*_{[k,l]}(\z,\al){\bf q}_{[k,l]}(\z,\al)=0\,.\label{TQ}
\ena
The reason for the insertion of $q^{-2S_{[k,l]}}$ can be understood only
when we discuss the reduction relation in Section \ref{sec:3}.
\subsection{$R$ matrix symmetry and spin selection rule}\label{symmetry}
By construction, it is obvious that 
$\tb^*_{[k,l]}$ enjoys the $R$ matrix symmetry
\bea
\sb_i\,\tb^*_{[k,l]}(\z,\al)=\tb^*_{[k,l]}(\z,\al)\,\sb_i.\label{RSYM}
\ena
Here
\be
&&\sb_i:=K_{i,i+1}\check \bR_{i,i+1}(\xi_i/\xi_{i+1}),\\
&&\check\bR_{i,i+1}(\xi_i/\xi_{i+1})(X):=\check R_{i,i+1}(\xi_i/\xi_{i+1})X
\check R_{i,i+1}(\xi_i/\xi_{i+1})^{-1},\\
&&\check R_{i,i+1}(\z):=P_{i,i+1}R_{i,i+1}(\z),
\en
where
$K_{i,j}$ stands for the transposition of arguments $\xi_i$ and $\xi_j$, and 
$P_{i,j}\in\End(V_i\otimes V_j)$ for that of vectors.

A similar remark applies to $\qb_{[k,l]}$
and other operators which will appear
in Subsection \ref{k*}, so we will not repeat it.

Another general remark is on the spin selection rules.
Our operators $\tb^*_{[k,l]}$, $\qb_{[k,l]}$
as well as those which will be introduced
in later subsections satisfy spin selection rules in the following sense.

We say an operator $\xb_{[k,l]}\in\End(M_{[k,l]})$ has spin $s$ if
\be
[\bS_{[k,l]},\xb_{[k,l]}]=s\xb_{[k,l]}.
\en
If $\xb_{[k,l]}$ has spin $s$, we denote $s(\xb)=s$.
If an operator $X_{[k,l]}\in M_{[k,l]}$ has spin $s$,
the operator $\xb(X_{[k,l]})\in M_{[k,l]}$ has spin $s+s(\xb)$.
We have
\be
s(\tb^*)=0,\quad s(\qb)=0.
\en
For convenience sake we list $s(\xb)$ for those $\xb$
which will be introduced in the following sections.
\begin{align}
s(\xb)=\begin{cases}
1&\hbox{ if $\xb=\kb,\fb,\cb,\bar\cb,\bb^*$};\\
-1&\hbox{ if $\xb=\bb,\bar\bb,\cb^*$}.
\end{cases}
\label{spins}
\end{align}
\subsection{Spin reversal transformation}
The operator $\tb^*_{[k,l]}(\z,\al)$ is invariant under the spin reversal
coupled to the change of $\al$ to $-\al$. However, the other operators
$\qb_{[k,l]}(\z,\al)$, $\kb_{[k,l]}(\z,\al)$ are not.
We can introduce new operators
by such a transformation. Let us define the transformation.

For $X_{[k,l]}\in M_{[k,l]}$ we define
\be
\J(X_{[k,l]}):=
\prod_{j\in[k,l]}\sigma^1_j\cdot\,X_{[k,l]}\cdot\prod_{j\in[k,l]}\sigma^1_j.
\en
Then we have
\be
\J_{[k,l]}\circ\tb^*_{[k,l]}(\z,-\al)\circ\J_{[k,l]}=\tb^*_{[k,l]}(\z,\al).
\en
Set
\bea
N(x):=q^{-x}-q^{x}\,.\label{Nal}
\ena
For an operator $\xb_{[k,l]}(\z,\al)\in\End(M_{[k,l]})$
we define the transformation by
\be
&&\phi_\al(\xb_{[k,l]}(\z,\al)):=q^{-1}N(\al-\bS_{[k,l]}-1)\circ
\J_{[k,l]}\circ\xb_{[k,l]}(\z,-\al)\circ\J_{[k,l]}.
\en
We use $\phi(\xb)$ for the operator symbol of the operator
$\phi_\al(\xb_{[k,l]}(\z,\al))$:
\be
\phi(\xb)_{[k,l]}(\z,\al):=\phi_\al(\xb_{[k,l]}(\z,\al)).
\en
The second solution to the Baxter equation
\be
\xb_{[k,l]}(q\z,\al)+\xb_{[k,l]}(q^{-1}\z,\al)
-\tb^*_{[k,l]}(\z,\al)\xb_{[k,l]}(\z,\al)=0
\en
is given by $\phi(\qb)_{[k,l]}(\z,\al)$.
The convenience of the normalization factor $q^{-1}N(\al-\bS_{[k,l]}-1)$
will be understood when we discuss the commutation relations of our operators.
\subsection{Fusion relation and off-diagonal transfer matrix}\label{k*}
The product of $L_{a,j}(\z)$ and $L_{A,j}(\z)$
can be brought into a triangular matrix in $M_a$.
\bea
L_{\{a,A\},j}(\z)
&:=&
(F_{a,A})^{-1}L_{a,j}(\z)L_{A,j}(\z)F_{a,A}\label{eq:Lfusp}\\
&=&
\begin{pmatrix}
1&0\\
\frac{\gamma(\z)}{\beta(\z)}\s^+_j&1
\end{pmatrix}_a
\begin{pmatrix}L_{A,j}(q\z)q^{-\s^3_j/2}&0\\
0&L_{A,j}(q^{-1}\z)q^{\s^3_j/2}
\end{pmatrix}_a\,,\nn
\ena
where $F_{a,A}=1-\ao_A\sigma^+_a$.
This is called the fusion relation. The monodromy matrix is triangular.
\be
&&T_{\{a,A\},[k,l]}(\z):=L_{\{a,A\},l}(\z/\xi_l)\cdots L_{\{a,A\},k}(\z/\xi_k)
=\begin{pmatrix}
A_{A,[k,l]}(\z)&0\\
C_{A,[k,l]}(\z)&D_{A,[k,l]}(\z)
\end{pmatrix},\\
&&A_{A,[k,l]}(\z)=T_{A,[k,l]}(q\z)q^{-S_{[k,l]}},\quad
D_{A,[k,l]}(\z)=T_{A,[k,l]}(q^{-1}\z)q^{S_{[k,l]}}.
\en
The triangular structure descends to
the adjoint action if the operand $X_{[k,l]}$ commutes with $F_{a,A}$
(it does if $X_{[k,l]}\in M_{[k,l]}$):
\bea
\bT_{\{a,A\}}(\z,\al)(X_{[k,l]})
&:=&(F_{a,A})^{-1}\left(\bT_{a}(\z,\al)\bT_{A}(\z,\al)(X_{[k,l]})\right)
F_{a,A}\label{eq:Tfusp}\\
&=&
\begin{pmatrix}
\A_A(\z,\al)(X_{[k,l]})&0\\
\C_A(\z,\al)(X_{[k,l]})&\D_A(\z,\al)(X_{[k,l]})
\end{pmatrix}_a,\nn
\ena
where
\bea
&&\A_A(\z,\al)(X_{[k,l]})
=\bT_A(q\z,\al)q^{\al-\bS}(X_{[k,l]}),\label{AFAT}\\
&&\D_A(\z,\al)(X_{[k,l]})
=\bT_A(q^{-1}\z,\al)q^{-\al+\bS}(X_{[k,l]}).\label{DFAT}
\ena
The Baxter relation \eqref{TQ} follows from
the diagonal part, $\A_{A,[k,m]}(\z,\al)$ and $\D_{A,[k,m]}(\z,\al)$
of this relation. They have no poles at $\z^2=\xi_j^2$, while 
the off-diagonal part $\C_{A,{[k,l]}}(\z,\al)$ does.
Now we use the latter. Namely, the following object will be basic 
for the construction of various other operators. 
For $X_{[k,l]}\in M_{[k,l]}$ we define
\bea
&&\kb(\z,\al)(X_{[k,l]})
:=\Tr_A\Bigl\{\C_A(\z,\al)\z^{\al-\bS}
\bigl(q^{-2 S_{[k,l]}}\,X_{[k,l]}\bigr)\Bigr\}\,.
\label{kb}
\ena
Since $[F_{a,A},\s^+_a]=0$, we have
\be
\kb(\z,\al)(X_{[k,l]})
=\Tr_{A,a}\left\{\s^+_a\bT_a(\z,\al)\bT_A(\z,\al)
\z^{\al-\bS}(q^{-2S_{[k,l]}}X_{[k,l]})\right\}.
\en
\subsection{Analytic structure of the twisted transfer matrices}\label{ANALY}
In order to read the behavior of the operators $\tb^*_{[k,l]}(\z,\al)$,
$\qb_{[k,l]}(\z,\al)$
and $\kb_{[k,l]}(\z,\al)$ in $\z$, it is useful to rewrite \eqref{kb} by using
\be
&&\tilde{L}^\circ_{a,j}(\z^2)
:=\z^{-\sigma^3_j/2}L^\circ_{a,j}(\z)\z^{\sigma^3_j/2}\,,\\
&&\tilde{L}^\circ_{A,j}(\z^2):=\z^{-\sigma^3_j/2-1}
L^\circ_{A,j}(\z)\z^{-\sigma^3_j/2}.
\en
Note that the second line is not a similarity transformation.
The matrices $\tilde{L^\circ}_{a,j}(\z^2)$
and $\tilde{L^\circ}_{a,j}(\z^2)^{-1}$ are rational functions in
$\z^2$; in the finite plane, they have poles only at
$\z^2=q^{-2}$ or $\z^2=q^{2}$, respectively.
At $\z^2=\infty$, they are regular and upper triangular in $M_a$.
The operators $L^\circ_{A,j}(\z)$ and
$(1-\z^2)\left(L^\circ_{A,j}(\z)\right)^{-1}$ are polynomials in $\z$.
The modified operators $\tilde L^\circ_{A,j}(\z^2)$
and $\tilde L^\circ_{A,j}(\z^2)^{-1}$ are 
rational functions in $\z^2$. In $\C^\times$,
$\tilde L^\circ_{A,j}(\z^2)$ has no pole, and 
$\tilde L^\circ_{A,j}(\z^2)^{-1}$ has poles only at $\z^2=1$.
At $\z^2=\infty$, they are regular.
We denote by $\widetilde\bT_{a,{[k,l]}}(\z^2,\al)$,
$\widetilde\bT_{A,{[k,l]}}(\z^2,\al)$
the modifications of $\bT_{a,{[k,l]}}(\z,\al)$, $\bT_{A,{[k,l]}}(\z,\al)$ where
$L_{a,{[k,l]}}(\z/\xi_j)$, $L_{A,{[k,l]}}(\z/\xi_j)$ are replaced with
$\tilde L^\circ_{a,{[k,l]}}(\z^2/\xi_j^2)$,
$\tilde L^\circ_{A,{[k,l]}}(\z ^2/\xi_j ^2)$, respectively. Namely, we have
\be
&&\bT_a(\z,\al)(X_{[k,l]})=\z^{\bS}\G^{-1}
\tilde \bT_a(\z^2,\al)\z^{-\bS}\G(X_{[k,l]}),\\
&&\bT_A(\z,\al)(X_{[k,l]})=\z^{\bS}\G^{-1}
\tilde \bT_A(\z^2,\al)\z^{\bS}\G ^{-1}(X_{[k,l]}),
\en
where $\G(X_{[k,l]})=G_{[k,l]}X_{[k,l]}G^{-1}_{[k,l]}$,
$G_{[k,l]}=\prod_{j\in[k,l]}\xi_j^{\s^3_j/2}$.
They are rational functions of $\z^2$, and the poles
in $\C^\times$ are only at $\z^2=q^{\pm2}\xi^2_j$ and $\z^2=\xi^2_j$,
respectively.

The operator $\tb^*_{[k,l]}(\z,\al)$ is a rational function in $\z^2$.
Its singularities in the finite plane are  poles
at $\z^2=q^{\pm2}\xi_j^2$. It is regular at $\z^2=\infty$.
The operator $\qb_{[k,l]}(\z,\al)$ has an overall factor
$\z^\al$. If $X_{[k,l]}$ is of spin $s$,
$\z^{-\al+s}\qb(\z,\al)(X_{[k,l]})$ is a rational function in $\z^2$.
Its poles in the finite plane are at $\z^2=\xi^2_j$, and
$\z^{-\al-s}\qb(\z,\al)(X_{[k,l]})$ is regular at $\z^2=\infty$.

Set
\be
\Tr_{A,a}:=\Tr_A\Tr_a.
\en
In later sections we will use similar notations such as
$\Tr_{A,B,a,b,c}$, {\it etc.}. The behavior of $\kb_{[k,l]}(\z,\al)$
easily follows from
\be
\kb(\z,\al)(X_{[k,l]})
&=&\z^{\al+s+1}\G^{-1}\Tr_{A,a}\left\{\s^+_a\widetilde\bT_a(\z^2,\al)
\widetilde\bT_A(\z^2,\al)\G^{-1}(q^{-2S_{[k,l]}}X_{[k,l]})\right\}.
\en
If $X_{[k,l]}$ is of spin $s$, $\z^{-\al+s-1}\kb(\z,\al)(X_{[k,l]})$
is a rational
function in $\z^2$. Its singularities in the finite plane are  poles at
$\z^2=\xi^2_j,q^{\pm2}\xi^2_j$, and
$\z^{-\al-s+1}\kb(\z,\al)(X_{[k,l]})$ is regular at $\z^2=\infty$.

Hereafter we assume 
$
\xi_i^2\neq q^2\xi_j^2,\ q^4\xi_j^2 \quad (i,j\in[k,l])$,
so that the three series of poles in $\kb_{[k,l]}(\z,\al)$
have no intersection. On the other hand, we do not require
$\xi^2_i\not=\xi^2_j$ $(i\not=j)$ unless otherwise stated.
Our construction goes as well for the homogeneous chain
as the inhomogeneous chain.
\subsection{$q$-exact forms, cycles and primitives}\label{subsec:exact}
Later we shall extract three kinds of operators out of the operator
$\kb_{[k,l]}(\z,\al)$. In this subsection we motivate this construction.
The key is its analytic structure
in $\z$: the poles are located in three series $\z^2=\xi^2_j,q^{\pm2}\xi^2_j$.

Let $\Delta_{\zeta}$ denote the $q$-difference operator with respect
to the variable $\zeta$:
\be
\Delta_{\zeta}f(\zeta):=f(q\zeta)-f(q^{-1}\zeta).
\en
In Section \ref{sec:4} we will establish the commutation relations
of the form
\be
&&\kb_{[k,l]}(\z_1,\al)\kb_{[k,l]}(\z_2,\al+1)
+\kb_{[k,l]}(\z_2,\al)\kb_{[k,l]}(\z_1,\al+1)\\
&&\quad=\Delta_{\z_1}\mathbf{m}^{(++)}_{[k,l]}(\z_1,\z_2,\al)+
\Delta_{\z_2}\mathbf{m}^{(++)}_{[k,l]}(\z_2,\z_1,\al),\\
&&\kb_{[k,l]}(\z_1,\al)\phi(\kb)_{[k,l]}(\z_2,\al+1)
+\phi(\kb)_{[k,l]}(\z_2,\al)\kb_{[k,l]}(\z_1,\al-1)\\
&&\quad=\Delta_{\z_1}\mathbf{m}^{(+-)}_{[k,l]}(\z_1,\z_2,\al)+
\Delta_{\z_2}\mathbf{m}^{(-+)}_{[k,l]}(\z_2,\z_1,\al).
\en
The right hand sides of these relations are `$q$-exact $2$ forms'.
Let us consider an analogy in differential calculus. If we have an exact
$1$ form $df(\z)$, its integral over a cycle $C$ is zero,
\be
\int_Cdf(\z)=0
\en
and the function $f(\z)$ is called the primitive integral.

In the context of our working,
we call an operator of the form $\mathbf{g}_{[k,l]}(\z,\al)=\Delta_\z 
\mathbf{h}_{[k,l]}(\z,\al)$
a $q$-exact 1 form
if $\mathbf{h}_{[k,l]}(\z,\al)=\z^{\al+\bS}\left(f_{[k,l]}(\z^2)\right)$,
and $ f_{[k,l]}(\z^2)$ is a rational
function in $\z^2$ whose poles in $\C^\times$ are
only at $\z^2=\xi_j^2$ for $j\in[k,l]$. 
We call $\mathbf{h}_{[k,l]}(\z,\al)$ a $q$-primitive integral of $\mathbf{g}_{[k,l]}(\z,\al)$
and denote it by $\Delta_\z^{-1}\mathbf{g}_{[k,l]}(\z,\al)$. We can take two kinds of cycles
$C=C_j,\tilde C_j$ on which the integrals are zero. The first kind of cycles
$C_j$ are ones which encircle the point $\z^2=\xi_j^2$, and the second kind
$\tilde C_j$ are those which encircle two points $\z^2=q^2\xi_j^2$
and $\z^2=q^{-2}\xi_j^2$. The integral
$\int_{C_j}\mathbf{g}_{[k,l]}(\z,\al)\frac{d\z^2}{\z^2}$
is zero because there is no pole at $\z^2=\xi_j^2$, and the integral
$\int_{\tilde C_j}\mathbf{g}_{[k,l]}(\z,\al)\frac{d\z^2}{\z^2}$ is also zero because
two residues at  $\z^2=q^{\pm2}\xi_j^2$ cancel each other.

In the above commutation relations, the singularity structure of
the operators $\mathbf{m}^{(++)}(\z_1,\z_2,\al)$, $\mathbf{m}_{[k,l]}^{(+-)}(\z_1,\z_2,\al)$
and $\mathbf{m}_{[k,l]}^{(-+)}(\z_1,\z_2,\al)$
are much improved compared to each term in the left hand side.
The right hand sides are $q$-exact in the above sense.
For example, we have
\be
\mathbf{m}_{[k,l]}^{(++)}(\z_1,\z_2,\al)=(\z_1\z_2)^{\al+\bS}
\widetilde{\mathbf{m}}_{[k,l]}^{(++)}(\z_1 ^2,\z_2^2,\al),
\en
the function $\widetilde{\mathbf{m}}_{[k,l]}^{(++)}(\z_1 ^2,\z_2^2,\al)$ is rational 
in $\z_1^2,\z_2^2$ such that the poles in $\z_1^2\in\C^\times$
are only at $\z_1^2=\xi_j^2$ for $j\in[k,l]$.
Similar statements hold for
$\mathbf{m}^{(+-)}_{[k,l]}(\z_1,\z_2,\al)$
and $\mathbf{m}^{(-+)}_{[k,l]}(\z_1,\z_2,\al)$
except that
there are simple poles at $\z_1^2=\z_2^2$ with
residues proportional to the identity
operator. This much will be proved in Section \ref{sec:4}.

Now, we integrate the above identities for $\kb_{[k,l]},\phi(\kb)_{[k,l]}$
over the cycles $C_j$ and $\tilde C_j$.
We denote the operators obtained as residues
by $\bar\cb^{(j)}_{[k,l]},\bar\bb^{(j)}_{[k,l]}$
and $\cb^{(j)}_{[k,l]},\bb^{(j)}_{[k,l]}$.
If we integrate the commutation relations
in both $\z_1$ and $\z_2$, in the left hand sides we obtain
anti-commutators of the operators (except that the value of $\alpha$
changes.) The right hand sides are zero. We get the Grassmann relations.

We can modify the operator $\kb_{[k,l]}(\z,\al)$ by subtracting these Grassmann
operators so that we get a $q$-exact operator. We
define the third kind of operator $\fb_{[k,l]}(\z,\al)$ as 
the $q$-primitive integral
of the modified operator. The commutation relations of $\fb_{[k,l]}$
with $\cb^{(j)}_{[k,l]},\bb^{(j)}_{[k,l]},
\bar\cb^{(j)}_{[k,l]},\bar\bb^{(j)}_{[k,l]}$ follow from those
for $\kb_{[k,l]},\phi(\kb)_{[k,l]}$. In the next subsection,
we define three kinds of operators in this way.
\subsection{Decomposition of $\kb_{[k,l]}$}\label{ccf}
We introduce operators $\ctb_{[k,l]}$, $\cb_{[k,l]}$, $\fb_{[k,l]}$ by
decomposing the operator $\kb_{[k,l]}$ in accordance with the poles
$\z^2=\xi_j^2,q^{\pm2}\xi_j^2$.
\bea
&&\quad\kb(\z,\al)(X_{[k,l]})\label{eq:kfcc}\\
&&=\left(\ctb(\z,\al)+\cb(q\z,\al)+\cb(q^{-1}\z,\al)
+\fb(q\z,\al)\ -\fb(q^{-1}\z,\al)\right)(X_{[k,l]}),\nn
\ena
or equivalently,
\be
\fb(\z,\al)(X_{[k,l]})
=\Delta_\z^{-1}\left(\left\{\kb(\z,\al)-\ctb(\z,\al)
-\cb(q\z,\al)-\cb(q^{-1}\z,\al)\right\}(X_{[k,l]})\right).
\en
We demand, for any element $X_{[k,l]}\in M_{[k,l]}$ with spin $s$, that
$\ctb(\z,\al)(X_{[k,l]})$, $\cb(\z,\al)(X_{[k,l]})$,
and $\fb(\z,\al)(X_{[k,l]})$
all have the form $\z^{\al-s+1}f_{[k,l]}(\z^2)$, where $f_{[k,l]}(\z^2)$
is a rational function
in $\z^2$ whose only poles are $\xi_j^2$ ($j\in[k,l]$) and $\infty$.  
Clearly such a decomposition is possible, and is unique modulo terms of
the form $\z^{\al-s+1}p(\z^2)$ where $p(\z^2)$ is a polynomial in $\z^2$
of degree $s$. We fix this ambiguity, which occurs when $s\geq0$,
by making the following choice.
\begin{align}
&\ctb(\z,\al)(X_{[k,l]})
\textstyle:=\frac{1}{2\pi i}\oint_\Gamma
\psi(\z/\xi,\al+s+1)
\kb(\xi,\al)(X_{[k,l]})\frac{d\xi^2}{\xi^2}\,,\label{eq:choice-ctilde} \\
&\cb(\z,\al)(X_{[k,l]})
\textstyle:=\frac{1}{4\pi i}\oint_\Gamma
\psi(\z/\xi,\al+s+1)
\left\{\kb(q\xi,\al)+\kb(q^{-1}\xi,\al)\right\}(X_{[k,l]})
\frac{d\xi^2}{\xi^2}\,,\label{eq:choice-c} \\
&\textstyle
\fb(\z,\al)(X_{[k,l]}):=\left\{\fb^{\rm sing}(\z,\al)+
\fb^{\rm reg}(\z,\al)\right\}(X_{[k,l]})\,,\label{eq:fbar}
\end{align}
where
\bea
\fb^{\rm sing}(\z,\al)(X_{[k,l]})
\textstyle:=\frac{1}{4\pi i}\oint_\Gamma
\psi(\z/\xi,\al+s+1)\{-\kb(q\xi,\al)
+\kb(q^{-1}\xi,\al)\}(X_{[k,l]})\frac{d\xi^2}{\xi^2}.\nn
\ena
Here
\bea
\psi(\z,\al):=\frac{1}{2}\frac{\z^2+1}{\z^2-1}\z^\al\,\label{psi}
\ena
and $p(\z ^2)=\z^{-\al+s-1}\fb^{\rm reg}(\z,\al)(X_{[k,l]})$
is a polynomial in $\z^2$ to be determined.
The integrands are rational functions in $\xi^2$ with possible poles at 
$\xi^2=\z^2,0,\xi^2_j,q^{\pm 2}\xi_j^2,q^{\pm 4}\xi_j^2$ ($j\in[k,l]$). 
The integrals are taken along a simple closed curve $\Gamma$ 
such that $\xi^2_j$ ($j\in[k,l]$) are inside, while
$q^{\pm 2}\xi^2_j,q^{\pm 4}\xi_j^2$ ($j\in[k,l]$), $0$ and $\z^2$ are outside.

Now we determine $p(z)$. Consider the rational one form in $\xi^2$
\be
\psi(\z/\xi,\al+s+1)
\kb(\xi,\al)(X)\frac{d\xi^2}{\xi^2}.
\en
We use the analytic behavior of $\z^{-\al+s-1}\kb(\z,\al)(X)$ in $\z^2$,
which was discussed in Subsection \ref{ANALY}.
First, the above one form has no pole at $\xi^2=\infty$.
Collecting the residues using \eqref{eq:kfcc}, we obtain that
\be
&&\left(\kb(\z,\al)-\ctb(\z,\al)-\cb(q\z,\al)-\cb(q^{-1}\z,\al)
-\fb^{\rm sing}(q\z,\al)+\fb^{\rm sing}(q^{-1}\z,\al)\right)(X_{[k,l]})\\
&&=\res_{\xi^2=0}\psi(\z/\xi,\al+s+1)\kb(\xi,\al)(X_{[k,l]})
\frac{d\xi^2}{\xi^2}.
\en
Then, the right hand side is $\z^{\alpha-s+1}$ times
a polynomial in $\z^2$ of degree at most $s$. Therefore,
for generic $\alpha$, $p(z)$ is uniquely determined by the equation
\be
\Delta_\z\left(\fb^{\rm reg}(\z,\al)(X_{[k,l]})\right)
=\res_{\xi^2=0}\psi(\z/\xi,\al+s+1)
\kb(\xi,\al)(X_{[k,l]})\frac{d\xi^2}{\xi^2}\,.
\en
In particular, $p(z)=0$ if $s\leq-1$. The decomposition \eqref{eq:kfcc}
follows from this.

Using this notation, we define
\be
&&\bar\bb_{[k,l]}(\z,\al):=\phi(\ctb)_{[k,l]}(\z,\al),
\qquad\bb_{[k,l]}(\z,\al):=\phi(\cb)_{[k,l]}(\z,\al).
\en

The operators $\bb_{[k,l]}(\z,\al)$, $\cb_{[k,l]}(\z,\al)$,
$\bar\bb_{[k,l]}(\z,\al)$, $\bar\cb_{[k,l]}(\z,\al)$
are called annihilation operators
because they annihilate the ``vacuum state'', the identity operator
$\id_{[k,l]}\in M_{[k,l]}$. (Recall that we fix the interval $[k,l]$
and are discussing operators acting on $M_{[k,l]}$.)
\begin{rem}
The operators $\z^{-(\al+s+1)}\xb(\z,\al)(X_{[k,l]})$ $(\xb=\ctb,\cb,\fb)$
are rational in $\z^2$. In the homogeneous case, they have a pole at
$\z^2=1$, while in the inhomogeneous case, if we assume that $\xi_j$'s
are distinct, their poles are
simple poles only at $\z^2=\xi_j^2$ $(j\in[k,l])$ in $\C^\times$.
\end{rem}
Until the end of this subsection we consider the inhomogeneous case
with distinct spectral parameters.
Let $f(\z^2)$ be a rational function in $\z^2$.
In order to unburden the formulas we use the residues of functions
of the form $\z^{\al+m}f(\z^2)$, e.g.,
\be
\textstyle\res_{\z=\xi_j}\ctb(\z,\al)(X_{[k,l]})\frac{d\z^2}{\z^2}
=\xi_j^{\al+s+1}
\textstyle\res_{\z^2=\xi_j^2}\z^{-(\al+s+1)}\ctb(\z,\al)(X_{[k,l]})\frac{d\z^2}{\z^2}.
\en
In this notation, by the definition we have
\be
&&\textstyle\res_{\z=\xi_j}\ctb(\z,\al)(X_{[k,l]})\frac{d\z^2}{\z^2}
=\res_{\z=\xi_j}\kb(\z,\al)(X_{[k,l]})\frac{d\z^2}{\z^2},\\
&&\textstyle
\res_{\z=\xi_j}\cb(\z,\al)(X_{[k,l]})\frac{d\z^2}{\z^2}
=\textstyle{\frac12}\left(\res_{\z=q^{-1}\xi_j}+\res_{\z=q\xi_j}\right)
\kb(\z ,\al)(X_{[k,l]})\frac{d\z^2}{\z^2},\\
&&\textstyle
\res_{\z=\xi_j}\fb(\z,\al)(X_{[k,l]})\frac{d\z^2}{\z^2}
=\textstyle{\frac12}\left(\res_{\z=q^{-1}\xi_j}-\res_{\z=q\xi_j}\right)
\kb(\z,\al)(X_{[k,l]})\frac{d\z^2}{\z^2}.
\en

The following is less obvious. We consider the residue at the right end,
$\z=\xi_l$.
\begin{lem}\label{FQ}
The residues of $\kb_{[k,l]}(\z,\al)$ at $\z^2=q^{\pm2}\xi^2_l$ are given
in terms of the residue of $\qb_{[k,l]}(\z,\al):$
\be
&&\textstyle
\res_{\z=q^{-1}\xi_l}\kb(\z,\al)(X_{[k,l]})\frac{d\z^2}{\z^2}=-\res_{\z=\xi_l}
\Bigl(\sigma^+_l\ \qb(\z,\al)(X_{[k,l]})\Bigr)\frac{d\z^2}{\z^2}\,,\\
&&\textstyle
\res_{\z=q\xi_l}\kb(\z,\al)(X_{[k,l]})\frac{d\z^2}{\z^2}
=-\res_{\z=\xi_l}\Bigl(\qb(\z,\al)(X_{[k,l]})\ \sigma^+_l\Bigr)
\frac{d\z^2}{\z^2}\,.
\en
In particular, we have
\be
\textstyle
\res_{\z=\xi_l}\fb(\z,\al)(X_{[k,l]})\frac{d\z^2}{\z^2}
=-\frac12\res_{\z=\xi_l}
\Bigl[\sigma^+_l,\qb(\z,\al)(X_{[k,l]})\Bigr]
\frac{d\z^2}{\z^2}.
\en
\end{lem}
\begin{proof}
We prove the first formula. The other one is similar. We start from
\be
&&\kb(\z,\al)(X_{[k,l]})=
\Tr_{A,a}\Bigl\{\s^+_aL_{\{a,A\},l}(\z/\xi_l)\\
&&\quad\times\bT_{\{a,A\},[k,l-1]}(\z,\al)\z^{\al-\bS}
(q^{-2S_{[k,l]}}X_{[k,l]})
L_{\{a,A\},l}(\z/\xi_l)^{-1}\Bigr\}.
\en
We use \eqref{eq:Lfusp} for $L_{\{a,A\},l}(\z)$ and
\eqref{eq:Tfusp} for $\bT_{\{a,A\},[k,l-1]}(\z,\al)$.
We must be careful about $\bS$ in $\eqref{AFAT}$ and $\eqref{DFAT}$.
When the formulas are used in
$\bT_{\{a,A\},[k,l-1]}(\z,\al)$ this means $\bS_{[k,l-1]}$.
To see if there is a pole at $\z=q^{-1}\xi_j$ we use
\eqref{Lcirc}, and that $\overline L^\circ(\z):=(1-\z^2)L^\circ_{A,l}(\z)^{-1}$
is regular at
$\z=1$. For the normalization factor we use
\be
\frac{\s(q\z)}{\s(q^{-1}\z)}=\frac{1-\z^2}{1-q^2\z^2}.
\en
Taking the residue at $\z=q^{-1}\xi_j$ we obtain
\be
&&\res_{\z=q^{-1}\xi_j}\kb_{[k,l]}(\z,\al)(X_{[k,l]})\\&&\quad=
\s^+_lL^\circ_{A,l}(1)\Tr_A\bT_{A,[k,l-1]}(\xi_l,\al)\xi_l^{\al-\bS_{[k,l]}}
(q^{-2S_{[k,l]}}X_{[k,l]})
\overline L^\circ_{A,l}(1)^{-1}.
\en
By a similar calculation for $\qb_{[k,l]}$ we obtain the first formula.
\end{proof}
\subsection{Creation operators $\bb^*_{[k,l]}$ and $\cb^*_{[k,l]}$}
Our main objects in this paper are the creation operators.
We define them in terms of $\fb_{[k,l]}$ and $\tb^*_{[k,l]}$ as
\bea
&&\bb^*(\z,\al)(X_{[k,l]}):=\left(\fb(q\z,\al)+\fb(q^{-1}\z,\al)
-\tb^*(\z,\al)\fb(\z,\al)\right)(X_{[k,l]})\,.\label{cstar}\\
&&\cb^*(\z,\al)(X_{[k,l]}):=-\phi(\bb^*)(\z,\al)(X_{[k,l]}).\label{CSTAR-}
\ena
Notice the similarity of this definition with Baxter's TQ relation \eqref{TQ}:
the same second order linear difference operator is
used in the right hand side of \eqref{cstar}. This particular combination of
the operators enjoys several miraculous properties such as
the regularity, the reduction and the commutation relations.
In the following sections we shall establish them.
\begin{rem}
The construction of operators in this section goes equally well
when $q^2$ is a root of unity other than $-1$. With a little more care
the case $q^2=-1$ can be also treated. We hope to discuss these in a separate
publication.
\end{rem}

\section{Reduction relations and extension to infinite volume}
\label{sec:3}

In this section, we discuss certain stability of the operators $\tb^*_{[k,l]}$,
$\kb_{[k,l]}$, {\it etc.}, when the interval $[k,l]$ are enlarged
to $[k',l']$. It is called the reduction relation. The reduction relation
will be used essentially in the definition of operators on the infinite
chain.
\subsection{Space of quasi-local operators}
Set
\be
S(k):=\frac12\sum_{j=-\infty}^k\s^3_j,
\en
and consider formal expressions
$q^{2(\al-s)S(0)}\mathcal O$
where $\mathcal O$ is a local operator of spin $s$,
i.e., $\mathcal O$ is an element of $M_{[k,l]}$ for some interval $[k,l]$
such that $\bS(\mathcal O)=s\mathcal O$. We call them `quasi-local operators'.
We denote by $\Wal$ the space spanned by quasi-local operators
of the form above, where  $\al$ is
fixed and $s$ can be any integer. Mathematically, one can define the space $\Wal$ as an inductive limit.
Physically, we want to compute the vacuum expectation values
\be
\langle q^{2\al S(0)}\mathcal O\rangle=
\frac{\langle{\rm vac}|q^{2\al S(0)}\mathcal O|{\rm vac}\rangle}
{\langle{\rm vac}|q^{2\al S(0)}|{\rm vac}\rangle}.
\en
Therefore, we are interested in the spin zero case, i.e., $s=0$.
The multiplication of $q^{2\al S(0)}$ is the insertion of a disorder
field in the infinite chain.

As we did in Introduction, we set
\be
\W:=\oplus_{\al\in\C}\W^{(\al)}.
\en
The subspace of quasi-local operators, stable outside the interval
$[k,l]$, will be denoted by $(\W )_{[k,l]}$.
We will define operators acting on $\W$. All the operators
we discuss in this paper are block diagonal
\be
\xb:\W^{(\al)}\rightarrow\W^{(\al)}.
\en
From now on, we fix $\al$ and discuss the restriction of $\xb$ on $\W^{(\al)}$
denoting it by the same symbol $\xb$ without specifying $\al$.
We note that in Section 2 the symbol $\al$ was used 
as a `dummy variable' rather than a fixed parameter. 
We shall keep using $\al$ in these two different ways, but
there should be no fear of confusion. 
In formulas containing infinite interval, 
$\al$ is a fixed 
parameter specifying the subspace we work with,    
while in formulas containing only finite intervals 
it is used as a dummy variable.

The operator $\bS$ defines the spin $s$ on $\Wal$.
We have the decomposition
\be
\Wal=\oplus_{s\in\Z}\Wal_s,\quad\Wal_s:=\{X\in\Wal\,|\,\bS X=sX\}.
\en
The creation and annihilation operators change spin $s$.
Accordingly, they change the semi-infinite tail $q^{2(\al-s)\bS(0)}$.
To follow up this change it is convenient to introduce an operator
\be
\alb=\al-\bS.
\en
Note that $\alb+\bS$ is a $c$-number on $\W^{(\al)}$, i.e.,
it commutes with all kind of operators.

In Section 2 we constructed the creation and annihilation operators
$\xb_{[k,l]}(\z,\al)$ acting on $M_{[k,l]}$. Recall that we denoted by $s(\xb)$
the spin of the operator $\xb_{[k,l]}(\z,\al)$.
In this section we will define operators $\xb(\z)$
acting on $\W^{(\al)}$ in such a way that for all $s\in\Z$
\be
\xb(\z):\W^{(\al)}_{s-s(\xb)}\rightarrow\W^{(\al)}_s.
\en
In the homogeneous case, we construct
$\xb(\z)$ as the inductive limit of $\xb_{[k,l]}(\z,\al)$:
\be
\xb(\z)|_{\W^{(\al)}_{s-s(\xb)}}=
\lim_{k\rightarrow-\infty\atop l\rightarrow\infty}
\xb_{[k,l]}(\z,\al-s).
\en
To be precise this means for $X_{[k,m]}\in \W^{(\al)}_{s-s(\xb)}$
\begin{align}
&\xb(\z)\(q^{2(\al-s+s(\xb))S(k-1)}X_{[k,m]}\)\label{INDUC}\\
&=
\begin{cases}
q^{2(\al-s)S(k-1)}\xb_{[k,l]}(\z,\al-s)(X_{[k,m]})\ \hbox{ for $l\geq m$}&
\hbox{if $\xb$ is annihilation};\\
q^{2(\al-s)S(k-1)}\xb_{[k,l]}(\z,\al-s)(X_{[k,m]})\ \bmod(\z^2-1)^{l-m}&
\hbox{if $\xb$ is creation}.
\end{cases}\nn
\end{align}
In the inhomogeneous case, for the annihilation operators, the inductive
construction is the same and we obtain operators $\xb(\z)$
acting on $\W^{(\al)}$.
On the other hand, for the creation operators, the inductive construction
leads to operators whose domains are restricted in such a way that
$\xb(\xi_j)$ is defined only on the operators of the form
$q^{2(\al-s)S(k-1)}X_{[k,m]}$ with $X_{[k,m]}\in M_{[k,m]}$ where $m<j$.

\subsection{Left reduction relation}
The definition of the twisted transfer matrix \eqref{TTM}
has a general feature, which we call the left reduction property.
Suppose that the quantum space is a tensor product of two representations
$\pi^{(i)}_{\rm qua}:U_q\mathfrak{b}^-\rightarrow V^{(i)}_{\rm qua}$ $(i=1,2)$,
\be
V_{\rm qua}=V^{(1)}_{\rm qua}\otimes V^{(2)}_{\rm qua}.
\en
Then, if $Y\in\End(V^{(2)}_{\rm qua})$, we have
\bea
\tb_{V_{\rm qua}}(\al)(q^{\al\pi^{(1)}_{\rm qua}(h_1)}\otimes Y)
=q^{\al\pi^{(1)}_{\rm qua}(h_1)}\otimes\tb_{V^{(2)}_{\rm qua}}(\al)(Y).
\label{LEFT}
\ena
This is obvious from \eqref{H} and
\be
L_{V_{\rm aux}\otimes V_{\rm qua}}
=L_{V_{\rm aux}\otimes V^{(2)}_{\rm qua}}L_{V_{\rm aux}\otimes
V^{(1)}_{\rm qua}}.
\en
Applying \eqref{LEFT} to the operator $\tb^*_{[k,l]}(\z,\al)$ we obtain
\be
\tb^*(\z,\al)(q^{\al\s^3_{k-1}}X_{[k,l]})=
q^{\al\s^3_{k-1}}\tb^*(\z,\al)(X_{[k,l]}),
\en
where $X_{[k,l]}\in M_{[k,l]}$.
This is called the left reduction relation for $\tb^*$.
Note that we keep the convention on suffixes; e.g.,
$\tb^*(\z,\al)(q^{\al\s^3_{k-1}}X_{[k,l]})=
\tb^*_{[k-1,l]}(\z,\al)(q^{\al\s^3_{k-1}}X_{[k,l]})$.

By a similar argument we obtain
\be
\kb(\z,\al)(q^{(\al+1)\s^3_{k-1}}X_{[k,l]})=
q^{\al\s^3_{k-1}}\kb(\z,\al)(X_{[k,l]}).
\en
This is also called the left reduction relation.
The shift of the parameter $\al$ occurs because of the factor $q^{-2S_{[k,l]}}$
in the definition of $\kb_{[k,l]}(\z,\al)$. By the spin reversal we obtain
\be
\phi(\kb)(\z,\al)(q^{(\al-1)\s^3_{k-1}}X_{[k,l]})=
q^{\al\s^3_{k-1}}\phi(\kb)(\z,\al)(X_{[k,l]}).
\en
In all cases, the action of the operator $\xb$ changes
the coefficient in front of $\s^3_{k-1}$ by $-s(\xb)$.
The left reduction relation enables us to define the action of $\xb$
on the semi-infinite interval $(-\infty,l]$ by changing $\s^3_{k-1}$
to $2S(k-1)$. For example, we define
\be
\tb^*(\z,\al)(q^{2\al S(k-1)}X_{[k,l]})=
q^{2\al S(k-1)}\tb^*(\z,\al)(X_{[k,l]}),
\en
The real question is how to extend it to the complete
infinite chain. We want to obtain something independent of $l$ when
$l\rightarrow\infty$ out of the operator on the interval $(-\infty,l]$.
\subsection{Locality of annihilation operators
and quasi-locality of creation operators}
\label{subsec:Quasi}
We define the support of a quasi-local operator $X\in\W^{(\al)}$
to be the minimal interval $[k,l]$ such that
\be
X=q^{2\al S(k-1)}X_{[k,l]}
\en
for some $X_{[k,l]}\in M_{[k,l]}$ holds. Then, we define its length by
\be
{\rm length}(X)=l-k+1
\en
Note that $M_{[k,k-1]}=\C (q^\al)\cdot I$ by definition, where $I$
is the identity operator; the support of $q^{2\al S(0)}$ is
the virtual interval $[k,k-1]$, and ${\rm length}(q^{2\al S(0)})=0$.
The operator $q^{2\al S(0)}$ belongs to $\W^{(\al)}_0$.

We will define two kinds of operators on $\W^{(\al)}$:
creation and annihilation operators. Let us discuss them separately.

\medskip\noindent
{\it Annihilation operators.}\quad
We have already defined annihilation operators 
$\xb_{[k,l]}(\z,\al)$ ($\xb=\bb,\cb,\overline\bb,\overline\cb$) on the finite
interval $[k,l]$. We will define the operator $\xb(\z)$
acting on $\W^{(\al)}$. There is not much difference in the homogeneous and
inhomogeneous cases except for the analytic structure. The singularity in
$\z$ is at $\z^2=1$ in the homogeneous case, while the singularities
in the inhomogeneous case are at $\z^2=\xi_j^2$.
The definition goes as follows.

The right reduction relation for annihilation operators is exactly
the same as the left reduction. Suppose that $k\leq m\leq l$.
We will prove that if $X_{[k,m]}\in M_{[k,m]}$ then
we have an equality
\be
\xb_{[k,l]}(\z,\al)(X_{[k,m]})=\xb_{[k,m]}(\z,\al)(X_{[k,m]}).
\en
Therefore, we can define $\xb(\z):\W^{(\al)}\rightarrow\W^{(\al)}$.
Namely, for $X_{[k,m]}\in M_{[k,m]}$ of spin $s-s(\xb)$, we define
\be
\xb(\z)(q^{2(\al-s+s(\xb))S(k-1)}X_{[k,m]}):=q^{2(\al-s)S(k-1)}
\xb(\z,\al-s)(X_{[k,m]}).
\en
We call this property of the annihilation operators the locality.

In the homogeneous case, one can define the annihilation operators $\xb_p$
where $\xb=\bb,\cb,\overline\bb,\overline\cb$, by the series expansion
\be
&&\xb(\z)=\begin{cases}
\z^{-\al}\sum_{p=0}^\infty(\z^2-1)^{-p}\xb_p
&\hbox{ if $\xb=\bb,\overline\bb$};\\[5pt]
\z^\al\sum_{p=0}^\infty(\z^2-1)^{-p}\xb_p
&\hbox{ if $\xb=\cb,\overline\cb$}.
\end{cases}
\en
In the inhomogeneous case, the corresponding objects are the residues
$\res_{\z=\xi_j}\xb(\z)$ at the simple poles $\z=\xi_j$.

\medskip\noindent
{\it Creation operators.}\quad
Creation operators may enlarge the support of quasi-local operators to the
right. There is some difference how much the support is
enlarged to the right in the homogeneous and inhomogeneous cases.

\medskip\noindent
{\it Homogeneous Case.}\quad
In the homogeneous case, we will define operators $\xb_p$ acting on
$\W^{(\al)}$, where $\xb=\tb^*,\bb^*,\cb^*$ and $p\in\Z_{\geq1}$,
such that if the support of $X\in\W^{(\al)}$ is contained in $[k,m]$
then the support of $\xb_p(X)$ is contained in $[k,m+p]$. Namely,
the length of quasi-local operators
is incremented by at most $p$. Then, we define the operator
$\xb(\z)$ as the formal power series with the coefficients $\xb_p$.
\be
&&\tb^*(\z):=\sum_{p=1}^\infty(\z^2-1)^{p-1}\tb^*_p,\\
&&\bb^*(\z):=\z^{\al+2}\sum_{p=1}^\infty(\z^2-1)^{p-1}\bb^*_p,\\
&&\cb^*(\z):=\z^{-\al-2}\sum_{p=1}^\infty(\z^2-1)^{p-1}\cb^*_p.
\en

\medskip\noindent
{\it Inhomogeneous Case.}\quad
We assume $\xi_i\not=\xi_j$ for $i\not=j$.
We have already defined operators $\xb_{[k,l]}(\z,\al)$ where
$\xb=\tb^*,\bb^*,\cb^*$. We will prove that they satisfy the properties that
if $X_{[k,m]}\in M_{[k,m]}$ and $m<j\leq l$ then
$\xb_{[k,l]}(\z,\al)(X_{[k,m]})$ is regular at $\z=\xi_j$,
the specialization $\xb_{[k,l]}(\xi_j,\al)(X_{[k,m]})$
belongs to $M_{[k,j]}$ and it is independent of $l$.

When $X\in\bigl(\W^{(\al)}_{s-s(\mathbf{x})}\bigr)_{[k,m]}$ we denote
\be
X=q^{2(\al-s+s(\mathbf{x}))S(k-1)}X_{[k,m]},
\en
If $X\in\bigl(\W^{(\al)}_s\bigr)_{(-\infty,j-1]}$ we can define $\xb(\xi_j)(X)$ by the inductive limit
\be
\xb(\xi_j)(X):=
\lim_{k\rightarrow-\infty\atop l\rightarrow\infty}
q^{2(\al-s)S(k-1)}\xb_{[k,l]}(\xi_j,\al-s)(X_{[k,j-1]}).
\en

We call these properties of the creation operators the quasi-locality.
In the following subsections we will prove the quasi-locality of the creation
operators.
\subsection{Creation operator $\tb^*$ and local integrals of motion}\label{COPT}
In this subsection, we clarify the quasi-locality in details in the case
of the creation operator $\tb^*$. In particular, in the homogeneous case,
we show that the action of $\tb^*(\z)$ is given in terms of the shift
operator and the exponential adjoint action of the local integrals of motion.

We define the shift operator
$\taub_{[k,l]}:M_{[k,l]}\rightarrow M_{[k+1,l+1]},
\ X_{[k,l]}\mapsto\taub(X_{[k,l]})$ by
\be
\taub(X_{[k,l]})=K_{k,k+1}\cdots K_{l,l+1}
\mathbb P_{k,k+1}\cdots\mathbb P_{l,l+1}(X_{[k,l]}),
\en
where $K_{i,j}$ is the exchange of the inhomogeneous parameters
$\xi_i$ with $\xi_j$.
The shift operator $\taub$ is also defined on $\W^{(\al)}$:
\be
\taub(q^{\al S(k-1)}X_{[k,m]}):=q^{\al S(k)}\taub(X_{[k,m]}).
\en

The key observation is that
\be
\bL_{a,j}(1)=\bP_{a,j}.
\en

\medskip\noindent
{\it Homogeneous Case.}\quad
In the homogeneous case, it leads to a simple fact that if
$X_{[k,m]}\in M_{[k,m]}$ and $m<l$ then the operator
$\tb^*_{[k,l]}(1,\al)$ satisfies
\be
{\textstyle\frac12}\tb^*_{[k,l]}(1,\al)(X_{[k,m]})
=q^{\al\s^3_k}\taub(X_{[k,m]}).
\en
It means in the inductive limit we have the shift operator
\be
\lim_{k\rightarrow-\infty\atop l\rightarrow\infty}
\textstyle\frac12\tb^*_{[k,l]}(1,\al)=\taub.
\en

Let us expand $\frac12\tb^*_{[k,l]}(\z,\al)(X_{[k,m]})$
in $\z^2-1$. Set
\be
\widetilde R^\vee_{i,j}(\z^2):=\z^{\s^3_i/2} R_{i,j}(\z)P_{i,j}\z^{-\s^3_j/2},\quad
\widetilde\R^\vee_{i,j}(\z^2):=\z^{\bS_i}\R_{i,j}(\z)\bP_{i,j}\z^{-\bS_j}.
\en
We have
\be
\tb^*_{[k,l]}(\z,\al)(X_{[k,m]})=\Tr_a\{
\widetilde\R^\vee_{a,l}(\z^2)\widetilde\R^\vee_{l,l-1}(\z^2)\cdots
\widetilde\R^\vee_{k+1,k}(\z^2)(q^{\al\s^3_k}\taub(X_{[k,m]}))\}.
\en
 Define an operator $\rb_{i,j}(\z^2)$ by
\be
\widetilde\R^\vee_{i,j}(\z^2)=1+(\z^2-1)\rb_{i,j}(\z^2).
\en
Note that $\rb_{i,j}(\z^2)$ is regular at $\z^2=1$ and that
$\rb_{i,j}(\z^2)(Z)=0$ if $Z$ is a local operator such that
its action on the $i$-th and the $j$-th components
is proportional to the identity operator or $q^{\al(\s^3_i+\s^3_j)}$.
We define $\tilde \R^\vee_{[k,l]}(\z^2)$ acting on $M_{[k,l]}$ by
\be
\tilde \R^\vee(\z^2)(X_{[k,l]}):=\widetilde\R^\vee_{l,l-1}(\z^2)
\cdots\widetilde\R^\vee_{k+1,k}(\z^2)(X_{[k,l]}).
\en
We have
\be
&&\tb^*_{[k,l]}(\z,\al)(X_{[k,m]})\\
&&\quad=2\sum_{j=m}^{l-1}(\z^2-1)^{j-m}
\rb_{j+1,j}(\z^2)\cdots\rb_{m+2,m+1}(\z^2)
\tilde \R^\vee(\z^2)(Y_{[k,m+1]})\\
&&\quad+(\z^2-1)^{l-m}\Tr_a\left\{\rb_{a,l}(\z^2)\rb_{l,l-1}(\z^2)\cdots
\rb_{m+2,m+1}(\z^2)\tilde\R^\vee(\z^2)(Y_{[k,m+1]})\right\}.
\en
where $Y_{[k,m+1]}:=q^{\al\s^3_k}\taub(X_{[k,m]})$.
Therefore, the inductive limit is well-defined as a formal power series
in $\z^2-1$. Namely, for $X\in\W^{(\al)}_s$ such that the support of $X$
is contained in $[k,m]$ we define
\be
\tb^*(\z)(X)
&=&\lim_{l\rightarrow\infty}
q^{2(\al-s)S(k-1)}\tb^*_{[k,l]}(\z,\al-s)(X_{[k,m]})\\
&=&2q^{2\al S(k-1)}\sum_{j=m}^\infty(\z^2-1)^{j-m}\rb_{j+1,j}(\z^2)
\cdots\rb_{m+2,m+1}(\z^2)\tilde \R^\vee(\z^2)(Y_{[k,m+1]}).
\en
The operators $\tb^*_p$ are the coefficients of $\tb^*(\z)$.
\be
\tb^*(\z)=\sum_{p=1}^\infty(\z^2-1)^{p-1}\tb^*_p.
\en
From this definition it is clear that the operator $\tb^*_p$
enjoys the quasi-locality discussed in Subsection \ref{subsec:Quasi}.

Later we use
\begin{lem}\label{EXPANSION}
Suppose that $k\leq m<l$, and let $Y_{[k,m],c}\in M_{[k,m]}\otimes M_c$.
Set $Y_{[k,m],m+1}:=\bP_{c,m+1}(Y_{[k,m],c})$.
Then, we have
\be
&&\bT_{c,[m+1,l]}(\z)(Y_{[k,m],c})\\
&&\quad=\sum_{j=m}^{l-1}(\z^2-1)^{j-m}
\rb_{j+1,j}(\z^2)\cdots\rb_{m+2,m+1}(\z^2)\z^{\bS_{m+1}}(Y_{m+1,[k,m]})\\
&&\quad+(\z^2-1)^{l-m}\z^{-\bS_c}\rb_{c,l}(\z^2)\rb_{l,l-1}(\z^2)\cdots
\rb_{m+2,m+1}(\z^2)\z^{\bS_{m+1}}(Y_{m+1,[k,m]}).
\en
\end{lem}

Let us return to $\tb^*(\z)$. We have
\be
\widetilde R^\vee_{i,j}(\z^2)
=\exp\left\{\log\left(1+\frac{(1+q^2)(\z^2-1)}{1-q^2\z^2}\right)
\cdot h^{(1)}_{i,j}\right\},
\en
where 
\be
h^{(1)}_{i,j}
:=-\frac1{q+q^{-1}}\left(\s^+_i\s^-_j+\s^-_i\s^+_j+\frac{q+q^{-1}}4
(\s^3_i\s^3_j-1)+\frac{q-q^{-1}}4(\s^3_i-\s^3_j)\right)
\en
is the local density of the Hamiltonian.

The local integrals of motion $I_p$ $(p\geq1,\,I_1=\frac 1 {q-q^{-1}}H_{XXZ})$
are defined as coefficients of the formal series in $\z^2-1$:
\be
\lim_{N\rightarrow\infty}\log\left(
\widetilde R^\vee_{N,N-1}(\z^2)\cdots\widetilde R^\vee_{-N+1,-N}(\z^2)\right)
=\sum_{p=1}^\infty(\z^2-1)^pI_p.
\en
By the Campbell-Hausdorff formula, each $I_p$ is a sum of local densities 
$h^{(p)}_{[j,j+p]}\in M_{[j,j+p]}$, which commute with $X\in\W^{(\al)}$
if the support of $X$ does not intersect with $[j,j+p]$.
Thus, $\mathbb I_p=[I_p,\cdot]$ is well-defined on $\W^{(\al)}$
and we have
\be
\textstyle\frac12\tb^*(\z)
=\exp\left(\sum_{p=1}^\infty(\z^2-1)^p\mathbb I_p\right)\taub.
\en

\medskip\noindent
{\it Inhomogeneous Case.}\quad
We prove
\begin{lem}\label{VACCRE}
Suppose that $k\leq m<j\leq l$. Then, we have
\be
\textstyle{\frac12}\tb^*_{[k,l]}(\xi_j,\al)(X_{[k,m]})
=\sb_{j-1}\cdots\sb_k(q^{\al\s^3_k}\cdot\taub(X_{[k,m]})).
\en
\end{lem}
\begin{proof}
We have
\be
\textstyle{\frac12}\tb^*_{[k,l]}(\xi_j,\al)(X_{[k,m]})
&=&\half\tr_a\left\{\mathbb T_{a,[j+1,l]}(\xi_j)
\mathbb P_{a,j}\mathbb T_{a,[k,j-1]}(\xi_j)(q^{\al\s^3_a}X_{[k,m]})\right\}\\
&=&\mathbb T_{j,[k,j-1]}(\xi_j)(q^{\al\s^3_j}X_{[k,m]})\\
&=&\mathbb R_{j,j-1}(\xi_j/\xi_{j-1})\cdots\mathbb R_{j,k}(\xi_j/\xi_k)
\bigl(q^{\al\s^3_j}X_{[k,m]}\bigr)\\
&=&\check{\mathbb R}_{j,j-1}(\xi_j/\xi_{j-1})\cdots\check{\mathbb R}_{k+1,k}(\xi_j/\xi_k)
q^{\al\s^3_k}\mathbb P_{k+1,k}\cdots\mathbb P_{j,j-1}(X_{[k,m]})\\
&=&\check{\mathbb R}_{j,j-1}(\xi_j/\xi_{j-1})\cdots\check{\mathbb R}_{k+1,k}(\xi_j/\xi_k)q^{\alpha\sigma^3_k}K_{j-1,j}\cdots K_{k,k+1}(\taub(X_{[k,m]}))\\
&=&\sb_{j-1}\cdots\sb_k(q^{\al\s^3_k}\cdot\taub(X_{[k,m]})).
\en
\end{proof}
\begin{cor}
If $X\in\bigl(\W^{(\al)}_s\bigr)_{(-\infty,j-1]}$,
we can define $\tb^*(\xi_j)(X)$ by the inductive limit
\be
\tb^*(\xi_j)(X):=
\lim_{k\rightarrow-\infty\atop l\rightarrow\infty}
q^{2(\al-s)S(k-1)}\tb^*_{[k,l]}(\xi_j,\al-s)(X_{[k,j-1]}),
\en
and we have
\be
\half \tb^*(\xi_j)(X)=\lim_{k\rightarrow-\infty}\sb_{j-1}\cdots\sb_k\taub(X).
\en
\end{cor}
\subsection{Right reduction relation for $\kb_{[k,l]}(\z,\al)$}
In this subsection we prove the right reduction relation for the operator
${\bf k}_{[k,l]}(\zeta, \alpha)$. It implies
the right reduction relation for the annihilation operators 
$\bb,\cb,\overline\bb,\overline\cb$ discussed in 
Subsection \ref{subsec:Quasi}.

We use the anti-automorphism $\theta_j$ of $M_j$:
\be
\theta_j(x):=\s^2_jx_j^{t_j}\s^2_j\ \hbox{ for $x_j\in M_j$.} 
\en
In general, we denote $\theta_{[k,l]}:=\prod_{j=k}^l\theta_j$.
It has the property (crossing symmetry):
\be
\theta_j(L_{a,j}(\z))=L_{a,j}(q\z)^{-1},\quad
\theta_j(L_{a,j}(\z)^{-1})=L_{a,j}(q^{-1}\z).
\en
This property is universal, i.e., valid for
$L_{A,j}(\z)$, $L_{\{a,A\},j}(\z)$, $T_{a,[k,l]}(\z)$, {\it etc.}.
\begin{lem}
Suppose that $k\leq m<l$. Let $X_{[k,m]}\in M_{[k,m]}$. Then we have
\bea
&& 
{\bf k}_{[k,l]}(\zeta, \alpha)(X_{[k,m]})=
{\bf k}(\zeta, \alpha)(X_{[k,m]})+\Delta_\z \mathbf{v}_{[k,l]}(\zeta,\al)
(X_{[k,m]}),
\label{eq:k*-reduction2} \\[5pt]
&\hbox{where}&\nonumber\\[5pt]
&&\mathbf{v}_{[k,l]}(\zeta,\al)
(X_{[k,m]})=\Tr_A\left( V_{A, [m+1, l]}(\z)
\bT_A(\z,\al)\z^{\alpha-\bS} 
(q^{-2S_{[k,m]}}X_{[k,m]})\right),\label{DELTAG}\\ 
&& V_{A,[m+1,l]}(\zeta)=-\theta_{[m+1,l]}\left(C_{A,[m+1,l]}(\zeta)
q^{S_{[m+1,l]}}T_{A,[m+1,l]}(q\zeta)^{-1}\right).
 \label{eq:def-V}
\ena
\end{lem}
\begin{proof}
We use $J=[k,m],K=[m+1,l]$. Write the operator $\kb_{[k,l]}$ separating
the $K$ part from the $J$ part:
\be
\kb_{[k,l]}(\z,\al)(X_{J})=\Tr_{a,A}\left[
\s^+_aT_{\{a,A\},K}(\z)q^{-2S_K}\bT_{\{a,A\}}(\z,\al)
\z^{\al-\bS}(q^{-2S_J}X_J)T_{\{a,A\},K}(\z)^{-1}\right]
\en
We want to bring $Q=T_{\{a,A\},K}(\z)^{-1}$ together with
$P=\s^+_aT_{\{a,A\},K}(\z)q^{-2S_K}$. Using the cyclicity of trace we obtain
\be
\kb_{[k,l]}(\z,\al)(X_{J})
=\Tr_{a,A}\left\{\theta_K\left(\theta_K(Q)\theta_K(P)\right)
\bT_{\{a,A\}}(\z,\al)\z^{\al-\bS}(q^{-2S_J}X_J)\right\}.
\en
The expression $\theta_K\left(\theta_K(Q)\theta_K(P)\right)$ is used to keep
the order of the product $PQ$ with respect to the quantum space $K$
but reverse the order to $QP$ with respect to the auxiliary space $a,A$.
The rest of the proof is straightforward.
\end{proof}
The right reduction property of the annihilation operators, which
was discussed in Subsection \ref{subsec:Quasi}, follows from
\eqref{eq:k*-reduction2} and the following
\begin{rem}
The operator $\Delta_\z \mathbf{v}_{[k,l]}(\zeta,\al)$ is $q$ exact in the sense of Subsection
\ref{subsec:exact}.
\end{rem}
To see this one can rewrite
$$V_{A,[m+1,l]}(\zeta)=-\theta_{[m+1,l]}\Bigl(
\sum_{j=m+1}^l\frac{q-q^{-1}}{\z/\xi_j-\xi_j/\z}
T_{A,[j+1,l]}(q^{-1}\z)q^{2S_{[j+1,l]}}\s^+_jT_{A,[j+1,l]}(q\z)^{-1}\Bigr).
$$

In particular, we have
\be
V_{A,l}(\z)=\frac{q-q^{-1}}{\z/\xi_l-\xi_l/\z}\s^+_l.
\en

The right reduction relation for $\fb_{[k,l]}(\z,\al)$ reads
\begin{cor}
\be
&& 
{\bf f}_{[k,l]}(\zeta,\alpha)(X_{[k,m]})=
{\bf f}(\zeta,\alpha)(X_{[k,m]})+\mathbf{v}_{[k,l]}(\zeta,\al)(X_{[k,m]})\,.\label{eq:f*-reduction2}
\en
\end{cor}
\subsection{Right reduction for $\bb^*_{[k,l]}(\z,\al)$ and its regularity}
The right reduction relation for the creation operator $\bb^*_{[k,l]}(\z,\al)$ reads
\begin{lem}\label{APPB}
Suppose that $k\leq m<l$. For $X_{[k,m]}\in M_{[k,m]}$ we have
\bea
&&\bb^*_{[k,l]}(\z,\al)(X_{[k,m]})=
\Tr_c\left\{\bT_{c,[m+1,l]}(\z)\gb_c(\z,\al)(X_{[k,m]})\right\},
\label{BBTU}
\ena
where
\begin{align}
\gb_c(\z,\al)(X_{[k,m]})&=\Bigl(\textstyle{\frac12}\fb(q\z,\al)
+\textstyle{\frac12}\fb(q^{-1}\z,\al)
-\bT_c(\z,\al)\fb(\z,\al)+\ub_c(\z,\al)\Bigr)(X_{[k,m]}),\nn\\
\ub_c(\z,\al)(X_{[k,m]})&=\Tr_{A,a}\left\{Y_{a,c,A}
\bT_{\{a,A\}}(\z,\al)\z^{\al-\bS}\left(q^{-2S_{[k,m]}}X_{[k,m]}\right)
\right\},\nn\\
Y_{a,c,A}&=-{\textstyle\frac12}\s^3_c\s^+_a+\s^+_c\s^3_a-\ab_A\s^+_c\s^+_a.
\label{Y_{a,c,A}}
\end{align}
\end{lem}
The proof will be given in Appendix B. In this formula the automorphism
$\theta_{[m+1,l]}$ does not appear. By introducing the auxiliary space
indexed by $c$, we have eliminated
the $\theta_{[m+1,l]}$ used in $V_{A,[m+1,l]}(\z)$.


An immediate consequence of the right reduction relation is the regularity
of $\bb^*_{[k,l]}(\z,\al)({X}_{[k,m]})$. In the homogeneous case
it is regular at $\z=1$, and in the inhomogeneous case it is regular at
$\z=\xi_j$ where $j\in[k,l]$. For the proof it is enough to consider
the latter case with distinct spectral parameters since other cases are
obtained by specialization.
\begin{lem}\label{regularity}
Suppose $k\le m<l$. Then $\bb^*_{[k,l]}(\z,\al)({X}_{[k,m]})$
is regular at $\z^2=\xi_j^2$ for any $j\in[k,l]$.
\end{lem}
\begin{proof}
Since $\bT_{c,[m+1,l]}(\z)$, $\fb_{[k,m]}(q\z,\al)$ and
$\fb_{[k,m]}(q^{-1}\z,\al)$ are regular at $\z^2=\xi_j^2$, it is enough to
show that
\be
\textstyle
\res_{\z=\xi_j}\bT_c(\z,\al)\fb(\z,\al)(X_{[k,m]})\frac{d\z^2}{\z^2}=
\res_{\z=\xi_j}\ub_c(\z,\al)(X_{[k,m]})\frac{d\z^2}{\z^2}.
\en
By the $R$ matrix symmetry without loss of generality we assume $j=m$.
We have
\be
(LHS)&=&-\half\res_{\z=\xi_m}\bT_c(\z,\al)
\left[\s^+_m,\qb(\z,\al)(X_{[k,m]})\right]\textstyle\frac{d\z^2}{\z^2}\\
&=&-\half\res_{\z=\xi_m}\left[\s^+_c,\bT_c(\z,\al)
\qb(\z,\al)(X_{[k,m]})\right]\textstyle\frac{d\z^2}{\z^2}\\
&=&-\half\res_{\z=\xi_m}\left[\s^+_c,\Tr_A
\left\{F_{c,A}\C_A(\z,\al)\s^-_c\z^{\al-\bS}(q^{-2S_{[k,m]}}X_{[k,m]})
F_{c,A}^{-1}\right\}\right]\textstyle\frac{d\z^2}{\z^2}\\
&=&-\res_{\z=\xi_m}\Tr_A\left\{(\half\s^3_c+\ab_A\s^+_c)
\C_A(\z,\al)\z^{\al-\bS}(q^{-2S_{[k,m]}}X_{[k,m]})
\right\}\textstyle\frac{d\z^2}{\z^2}\\&=&(RHS),
\en
where in the first line we used Lemma \ref{FQ},
going to the second line we used the
fact that $\bT_{c,[k,m]}(\xi_m,\al)$ contains the permutation $\bP_{c,m}$,
going to the third line we did the fusion, and dropped the diagonal terms
since they are regular at $\z=\xi_m$. For the same reason we dropped the term
containing $\s^+_c\s^3_a$ to obtain
$\res_{\z=\xi_m}\ub_c(\z,\al)(X_{[k,m]})\frac{d\z^2}{\z^2}$.
\end{proof}
\subsection{Creation operator $\bb^*(\z)$}
Now we define the creation operators $\bb^*(\z)$ on the space $\W^{(\al)}$.
We discuss the homogeneous and inhomogeneous cases separately.
There is a crucial difference in the two cases: the operators $\bb^*_p$
are defined on the whole space $\W^{(\al)}$ in the homogeneous case,
while the operators $\bb^*(\xi_j)$ is defined only on a certain subspace of
$\W^{(\al)}$ in the inhomogeneous case.

\medskip\noindent
{\it Homogeneous Case.}\quad
Let $k\leq m<l$.
The operator $\z^{-\al}\bb^*_{[k,l]}(\z,\al)$ is a rational function
in $\z^2$ and regular at $\z^2=1$ when it acts on ${X}_{[k,m]}\in M_{[k,m]}$.
Lemma \ref{APPB} shows that the dependence of
$\bb^*_{[k,l]}(\z,\al)({X}_{[k,m]})$ on $l$ comes only from
$\bT_{c,[m+1,l]}(\z)$. Therefore, from Lemma \ref{EXPANSION} we see that the
coefficients in the expansion
\be
\bb^*_{[k,l]}(\z,\al)({X}_{[k,m]})=\z^{\al+2}\sum_{p=1}^\infty(\z^2-1)^{p-1}
(\bb^*_p)_{[k,l]}({X}_{[k,m]})
\en
stabilizes when $l\rightarrow\infty$. From this one can define $\bb^*_p$ on
$\W^{(\al)}$.

\medskip\noindent
{\it Inhomogeneous Case.}\quad
By exactly the same argument as in Subsection \ref{COPT},
we can show that
\be
\bb^*_{[k,l]}(\xi_j,\al)({X}_{[k,m]})=\bb^*_{[k,j]}(\xi_j,\al)({X}_{[k,m]}).
\en
The above relation implies that 
the operator $\bb^*(\xi_j)$ is well-defined on $\bigl(\W^{(\al)}_s\bigr)_{(-\infty,j-1]}$. 
\section{Commutation relations}\label{sec:4}

It this section we shall find the commutation relations
of the annihilation operators $\bb $, $\cb$ with the
creation operators  $\tb ^*$, $\cb ^*$ and $\bb ^*$.
We shall also comment on the known commutation
relations \cite{HGSI}
of the annihilation operators among themselves.
We shall restrict our consideration to the more complicated
homogeneous case. 
The commutation relations for the 
inhomogeneous case will be presented at the end of the section
with necessary comments on their derivation. 

Before starting, recall the connection bewteen operators in infinite volume
and those on finite intervals \eqref{INDUC}. On the basis of these relations,
we drive commutation relations for the operators in infinite volume
from those for finite intervals.

\subsection{ Commutation relations of  $\cb$,
$\bar{\cb}$ with $\tb ^*$.}

The derivation of the commutation relations is a complicated problem, so, this section will
be rather technical.

We shall act by operators $\bb $, $\bb^*$,
{\it etc.} on the quasi-local
operators of the form $q^{2\al S(k-1)}X_{[k,m]}$. It is
clear from left reduction relations that in that case they can be
reduced to $\bb _{[k,\infty)}$, $\bb^* _{[k,\infty)}$, {\it etc.}
acting on $X_{[k,m]}$. Let us take $l\gg m$. 
Then the construction of the operators $\bb ^*$, $\cb ^*$,
implies for the homogeneous case:
$$\bb _{[k,\infty)}^*(\z)(X_{[k,m]})\equiv\bb _{[k,l]}^*(\z)(X_{[k,m]})
\quad \mod\,.$$
We shall consider  the commutation relations of $\cb$,  $\bar{\cb}$ with $\tb ^*$, $\cb ^*$ and $\bb ^*$. 
The operators $\cb (\z)$ and $\bar{\cb}(\z)$ are defined by (\ref{eq:choice-ctilde}), (\ref{eq:choice-c})
via $\kb (\xi)$. So, in order to treat them simultaneously we 
shall actually consider the commutation relations with $\kb (\xi)$ 
computing them up to $q$-exact forms
defined in Subsection \ref{subsec:exact}. Equality up to $q$-exact forms
in $\xi$ will be denoted by $\simeq _{\xi}$.

We begin with the following technical Lemma.
In the statement and the proof, we use a $2\times2$ matrix 
algebra $M_c$ with spectral parameter $\z$ in two ways: 
as an auxiliary space, and as an additional 
quantum space. In the right hand side of (\ref{auxhom}) below, 
the inhomogeneous parameter corresponding to $c$ is 
to be understood as $\z$.   
\begin{lem}\label{auxlemma}
Suppose that $k\leq m<l$ and $Y_{[k,m],c}\in M_{[k,l]}\otimes M_c$. We have
\bea
&&\kb_{[k,l]}(\xi,\al)\Tr_c\bT_{c,[m+1,l]}(\z)(Y_{[k,m],c})
\label{auxhom}\\
&&\quad\simeq_\xi\Tr_c\bT_{c,[m+1,l]}(\z)
\kb_{[k,m]\sqcup c}(\xi,\al)(Y_{[k,m],c})\bmod(\z^2-1)^{l-m}.\nn
\ena
\end{lem}
\begin{proof}
Consier the following expression:
\be
X_{[k,l]}:=\Tr_c\kb_{[k,l]\sqcup c}(\xi,\al)\bT_{c,[m+1,l]}(\z)(Y_{[k,m],c}).
\en
There are two ways to compute $X_{[k,l]}$. 
First write
\be
X_{[k,l]}=\Tr_{c,b,B}\Bigl\{Z_{c,b,B}
\bT_{\{b,B\},[k,l]}(\xi,\al)\xi^{\al-\bS_{[k,l]}}\left(q^{-2S_{[k,l]}}
\bT_{c,[m+1,l]}(\z)(Y_{[k,m],c})\right)\Bigr\},
\en
where
\be
Z_{c,b,B}:=\xi^{\bS_c}\bL_{\{b,B\},c}(\xi/\z)^{-1}(\s^+_b)q^{-\s^3_c}.
\en
Lemma \ref{EXPANSION} says modulo $(\z^2-1)^{l-m}$ the $c$-dependence in
$\bT_{c,[m+1,l]}(\z)(Y_{[k,m],c})$ disappears:
\be
\bT_{c,[m+1,l]}(\z)(Y_{[k,m],c})\equiv W_{[k,l]}\quad\bmod(\z^2-1)^{l-m}.
\en
Denoting by $\equiv_{l-m}$ equalities modulo $(\z^2-1)^{l-m}$, we have
\be
X_{[k,l]}&\equiv_{l-m}&\Tr_{c,b,B}\Bigl\{Z_{c,b,B}\bT_{\{b,B\},[k,l]}(\xi,\al)
\xi^{\al-\bS_{[k,l]}}(q^{-2S_{[k,l]}}W_{[k,l]})\Bigr\}\\
&=_{\phantom{l-m}}&
\Tr_{b,B}\Bigl\{(\Tr_cZ_{c,b,B})\bT_{\{b,B\},[k,l]}(\xi,\al)\xi^{\al-\bS_{[k,l]}}
\half\Tr_c(q^{-2S_{[k,l]}}W_{[k,l]})\Bigr\}\\
&\equiv_{l-m}&
\Tr_{b,B}\Bigl\{(\Tr_cZ_{c,b,B})\bT_{\{b,B\},[k,l]}(\xi,\al)\xi^{\al-\bS_{[k,l]}}
\half\Tr_c(q^{-2S_{[k,l]}}\bT_{c,[m+1,l]}(\z)(Y_{[k,m],c}))\Bigr\}\\
\en
It is easy to check $\Tr_cZ_{c,b,B}=2\s^+_b$. So, we obtain
\be
X_{[k,l]}\equiv_{l-m}
\kb_{[k,l]}(\xi,\al)\Tr_c\bT_{c,[m+1,l]}(\z)(Y_{[k,m],c}).
\en
Now write
\be
&&X_{[k,l]}=\Tr_{c,b,B}\s^+_b\bL_{\{b,B\},c}(\xi/\z)\bT_{\{b,B\},[m+1,l]}(\xi)
\bT_{\{b,B\},[k,m]}(\xi,\al)\\
&&\quad\times\bT_{c,[m+1,l]}(\z)
\xi^{\al-\bS_{[k,l]}-\bS_c}(q^{-\s^3_c-2S_{[k,l]}}Y_{[k,m],c}).
\en
Using the Yang-Baxter equation and then the right reduction relation we obtain
\begin{align}
&X_{[k,l]}=\Tr_{c,b,B}\s^+_b\bT_{c,[m+1,l]}(\z)\bT_{\{b,B\},[m+1,l]}(\xi)
\bL_{\{b,B\},c}(\xi/\z)\bT_{\{b,B\},[k,m]}(\xi,\al)\nn\\
&\ \times\xi^{\al-\bS_{[k,l]}-\bS_c}(q^{-\s^3_c-2S_{[k,l]}}Y_{[k,m],c})
=\Tr_c\bT_{c,[m+1,l]}(\z)\kb_{[k,m]\sqcup c\sqcup[m+1,l]}(\xi,\al)
(Y_{[k,m],c})\nn\\&
\ \simeq_\xi\Tr_c\bT_{c,[m+1,l]}(\z)\kb_{[k,m]\sqcup c}(\xi,\al)
(Y_{[k,m],c}).\nn
\end{align}
\end{proof}

\noindent
{\bf Remark.} {\it We can allow $Y_{c,[k,m]}$ to be a function
of $\z ^2$ regular
at $\z ^2=1$.}
\vskip .3cm
\noindent
From this Lemma we get the first couple of commutation relation.
\begin{cor}
The operators $\cb$, $\bar{\cb}$ commute with $\tb ^*$ 
\begin{align}
[\cb (\z'),\tb ^*(\z)]=0,\quad [\bar{\cb } (\z'),\tb ^*(\z)]=0\,.
\label{ct*}
\end{align}
\end{cor}
\begin{proof}
From the general remarks given at the beginning of Section 4,
it is enough to deduce the following equality from Lemma \ref{auxlemma}:
\bea
&&\kb_{[k,l]}(\xi,\al)\tb^*(\z,\al+1)(X_{[k,m]})
\simeq_\xi\tb^*_{[k,l]}(\z,\al)\kb(\xi,\al)(X_{[k,m]})\
\bmod(\z^2-1)^{l-m}\label{KT^*}
\ena
Set $Y_{[k,m],c}=\Tb_{c,[k,m]}(\z,\al)(q^{\sigma ^3_c}X_{[k,m]})$ in \eqref{auxhom}.
The (LHS) immediately gives the (LHS) of \eqref{KT^*}. For the (RHS),
move $ q^{-\sigma ^3_c-2S_{[k,m]}}$ through 
$\mathbb{T}_{c,[k,m]}(\z,\al)$ and use the Yang-Baxter equation
\be
\bL_{\{b,B\},c}(\xi/\z)\Tb_{\{b,B\},[k,m]}(\xi,\al)\Tb_{c,[k,m]}(\z,\al)
=\Tb_{c,[k,m]}(\z,\al)\Tb_{\{b,B\},[k,m]}(\xi,\al)\bL_{\{b,B\},c}(\xi/\z).
\en
Finally $\bL_{\{b,B\},c}(\xi/\z)$ will disappear because of
\be
\mathbb{L}_{\{b,B\},c}(\xi/\z)(q^{-2S_{[k,m]}}X_{[k,m]})=
q^{-2S_{[k,m]}}X_{[k,m]},
\en
and we obtain the (RHS) of \eqref{KT^*}.
\end{proof}
\subsection {Commutation relations of  $\cb$,
$\bar{\cb}$ and $\bb ^*$.}

Now we want to consider the commutation relations among $\cb$,
$\bar{\cb}$, $\bb ^*$. They are based on the Yang-Baxter equation:
\begin{align}
 R_{\{a,A\},\{b,B\}}(\z _1/\z _2) \mathbb{T}_{\{a,A\}}(\z_1)
 \mathbb{T}_{\{b,B\}}(\z_2)= \mathbb{T}_{\{b,B\}}(\z_2)
 \mathbb{T}_{\{a,A\}}(\z_1)R_{\{a,A\},\{b,B\}}(\z _1/\z _2)\,,\label{Rmat}
\end{align}
See Appendix A for more details. We start from commutation relation of
$\mathbf{k}$ with itself. 
\begin{lem}\label{kk}
The commutation relation for $\kb_{[k,l]}(\z,\al)$ is given
by ``$q$-exact 2 forms''$:$
\bea
&&\kb_{[k,l]}(\z_1,\al)\kb_{[k,l]}(\z_2,\al+1)+\kb_{[k,l]}(\z_2,\al)\kb_{[k,l]}(\z_1,\al+1)
\label{2form++}\\
&&\qquad=\Delta_{\z_1}\mathbf{m}^{(++)}_{[k,l]}(\z_1,\z_2,\al)
+\Delta_{\z_2}\mathbf{m}^{(++)}_{[k,l]}(\z_2,\z_1,\al),\nn
\ena
where
\begin{align}
&\mathbf{m}^{(++)}(\z_1,\z_2,\al)(X_{[k,l]})\label{M}\\&=\Tr_{b,A,B}\left(M_{b,A,B}(\z_1/\z_2)\bT_A(\z_1,\al)
\bT_{\{b,B\}}(\z_2,\al)(\z_1\z_2)^{\al-\bS}(q^{-4S_{[k,l]}}X_{[k,l]})\right),\nn\\
&M_{b,A,B}(\z)=\frac{\z^{-1}q^{-1}}{\z-\z^{-1}}
\begin{pmatrix}
q^{2D_B+1}\ab^*_Aq^{-2D_A}\ab^*_A&-\z^{-1}q^{D_B}(1+\z u_{A,B})\ab^*_Aq^{D_B}\\
0&-q^{2D_B-1}\ab^*_Aq^{-2D_A}\ab^*_A
\end{pmatrix}_b\,,\nn
\end{align}
with $u_{A,B}=\ab^*_Aq^{-2D_A}\ab_B$.
\end{lem}
\begin{proof} A similar formula is proved in \cite{HGSI}, so the proof here
is brief. Denote $\A_A(\z_1,\al)$, $\C_A(\z_1,\al)$, $\D_A(\z_1,\al)$,
$\A_B(\z_2,\al)$, $\C_B(\z_2,\al)$, $\D_B(\z_2,\al)$ by $A_1$, $C_1$, $D_1$,
$A_2$, $C_2$, $D_2$.

First, consider (RHS) of \eqref{2form++}.
There are some cancellations. Namely, the term in $\mathbf{m}^{(++)}_{[k,l]}(q^{-1}\z_1,\z_2)$
which comes from the $(1,1)$ element in $M_{b,A,B}(\z)$ cancels with the one
in $\mathbf{m}^{(++)}_{[k,l]}(q\z_2,\z_1)$ from the $(2,2)$ element. This is a consequence
of the Yang-Baxter relation $R_{33}D_1A_2=A_2D_1R_{33}$.
Another cancellation come from $R_{22}A_1D_2=D_2A_1R_{22}$. So, we will
prove the equality  \eqref{2form++} for the rest.

From the Yang-Baxter relation (\ref{Rmat})
one finds that
\bea
&&C_1C_2-R_{44}^{-1}C_2C_1R_{11}\label{3.7}\\
&&\quad=
-R_{44}^{-1}R_{42}A_1C_2+D_1C_2R_{33}^{-1}R_{31}
 -R_{44}^{-1}R_{43}R_{33}^{-1}A_2C_1R_{11}+R_{44}^{-1}D_2C_1R_{21}\nn\\
&&\quad+R_{44}^{-1}(R_{43}R_{33}^{-1}R_{31}-R_{41})A_1A_2
 -D_1D_2R_{44}^{-1}(R_{43}R_{33}^{-1}R_{31}-R_{41}).\nn
\ena
We use \eqref{H} frequently in the calculation below.
Rewrite $(LHS)$ of (\ref{2form++}), e.g.,
\begin{align}
&\kb(\z_1,\al)\kb(\z_2,\al+1)(X_{[k,l]})\label{often1}\\
&\quad=
\Tr_{\scriptscriptstyle a,b,A,B}
\Bigl\{\s^+_a\s^+_b\bT_{\{a,A\}}(\z_1,\al)\z_1^{\al-\bS}q^{-2S_{[k,l]}}
\bT_{\{b,B\}}(\z_2,\al+1)\z_2^{\al-\bS+1}(q^{-2S_{[k,l]}}X_{[k,l]})
\Bigr\}\nn\\
&\quad={\textstyle\frac{\z_2}{q\z_1}}\Tr_{\scriptscriptstyle a,b,A,B}
\Bigl\{q^{2D_B}\s^+_a\s^+_b\bT_{\{a,A\}}(\z_1,\al)
\bT_{\{b,B\}}(\z_2,\al)(\z_1\z_2)^{\al-\bS}(q^{-4S_{[k,l]}}X_{[k,l]})\Bigr\}.\nn
\end{align}
Thus, we obtain
\be
(LHS)={\textstyle\frac{\z_2}{q\z_1}}\Tr_{\scriptscriptstyle a,b,A,B}
\Bigl\{q^{2D_B}(C_1C_2-R_{44}^{-1}C_2C_1R_{11})
(\z_1\z_2)^{\al-\bS}(q^{-4S_{[k,l]}}X_{[k,l]})\Bigr\}.
\en
In the right hand side evaluating, for example, the term with 
$A_2C_1$, one has to remember that $\A _B(\z_2,\al)=\mathbb{T }_B(\z _2q,\al)q^{\al-\mathbb{S}}$ and move $q^{-\mathbb{S}}$ trough $\C_A(\z_1,\al)$ using
\begin{align}
q^{-\mathbb{S}}\mathbb{T }_{\{a,A\}}(\z _1,\al)=
q^{D_A+\frac 1 2 \sigma ^3_a}
\mathbb{T } _{\{a,A\}}(\z _1,\al)q^{-\mathbb{S}-D_A-\frac 1 2 \sigma ^3_a}\,,
\label{often2}
\end{align}
and use the cyclicity of trace.
After some calculations using \eqref{often1}, \eqref{often2},
the equality \eqref{3.7} gives rise to (\ref{2form++}).
\end{proof}

\vskip .2cm
\noindent

\begin{lem}\label{nopole}
The singularity at $\z_1 ^2=\z_2 ^2$ which is present in (\ref{M}) cancels when $M_{b,A,B}(\z)$ is substituted into 
$\mathbf{m}^{(++)}_{[k,l]}(\z _1,\z _2,\al)$.
\end{lem}

\begin{proof}
Indeed, suppose 
$$\mathbf{m}^{(++)}_{[k,l]}(\z _1,\z _2,\al)=\frac 1 {\z _1^2-\z _2^2}f(\z _2^2)\z _2^{2\al}+\text{regular}\,,$$
where $f(\z _2^2)$ is a rational function.
The left hand side of (\ref{2form++}) is regular at $\z _1^2=\z _2^2q^{\pm 2}$. So, the poles at this point must cancel in the right hand side. This requirement leads to an equation for $f(\z _2^2)$ which has no
rational solutions. Hence $f(\z _2^2)=0$ and the singularity of
$\mathbf{m}^{(++)}_{[k,l]}(\z _1,\z _2,\al)$ is fictitious. 
\end{proof}

Integrating (\ref{2form++}) in $\z_1$ and $\z_2$, and using the commutativity
of two integrations assured by Lemma \ref{kk}, we have
\begin{thm}
In the homogeneous case we have the commutation relations:
\begin{align}
[\cb (\z),\cb(\z ')]_+=0,\quad [\bar{\cb} (\z),\cb(\z ')]_+=0,\quad[\bar{\cb} (\z),\bar{\cb}(\z ')]_+=0\,.\label{commc}
\end{align}
\end{thm}


Now we are ready to attack much more complicated case 
of the commutation relations between $\cb$ and $\bb ^*$.
The operator $\bb ^*(\z,\al)$ is constructed via the operator
$\fb (\z ,\al)$. 
First, we derive the commutation relations between $\fb$ and $\kb$.
\begin{lem}\label{lemmakf}
We have:
\begin{align}
\fb_{[k,l]}(\z,\al)\kb_{[k,l]}(\xi,\al+1)
+\kb_{[k,l]}(\xi,\al)\fb_{[k,l]}(\z,\al+1)\simeq _{\xi}\mb_{[k,l]}^{(++)}(\z,\xi,\al).
\label{comkf}
\end{align}
\end{lem}
\begin{proof}
Denote the difference (LHS)-(RHS) of (\ref{comkf}) by $\mathbf{x}_{[k,l]}(\z,\xi,\al)$, we want to show that it is
$q$-exact
in $\xi$. It is enough to prove
this statement in the inhomogeneous case where $\xi_j$ are distinct. Then,
because of Lemma \ref{nopole}, it is equivalent to the vanishing of the integrals
$\mathbf{y}_{[k,l]}(\z,\al;\Gamma)
=\int_{\Gamma}\mathbf{x}_{[k,l]}(\z,\xi,\al)\frac{d\xi^2}{\xi^2}$
for $\Gamma=C_j,\tilde C_j$. Let us prove $
\mathbf{y}_{[k,l]}(\z,\al;\Gamma)=0$.
Recall that
\be
\Delta_\z\fb_{[k,l]}(\z,\al)
=\kb_{[k,l]}(\z,\al)-\bar\cb_{[k,l]}(\z,\al)-\cb_{[k,l]}(\z q,\al)
-\cb_{[k,l]}(\z q^{-1},\al).
\en
We know already that $\cb_{[k,l]}(\z,\al),\bar\cb_{[k,l]}(\z,\al)$
anti-commute with $\kb_{[k,l]}(\xi,\al)$ up to $q$-exact form in $\xi$.
Therefore we have
$\Delta_\z\mathbf{x}_{[k,l]}(\z,\xi,\al)\simeq_\xi
\Delta_\xi\mb_{[k,l]}^{(++)}(\xi,\z,\al)$.
Hence
\be
\Delta_\z\mathbf{y}_{[k,l]}(\z,\al;\Gamma)
=\Delta_\z\int_{\Gamma}\mathbf{x}_{[k,l]}(\z,\xi,\al)\frac{d\xi^2}{\xi^2}
=\int_{\Gamma}\Delta_\xi\mb_{[k,l]}^{(++)}(\xi,\z,\al)\frac{d\xi^2}{\xi^2}=0.
\en
From this follows that $\mathbf{y}_{[k,l]}(\z,\al;\Gamma)=0$, and therefore
$\mathbf{x}_{[k,l]}(\z,\xi,\al)\simeq_\xi0$.
\end{proof}

\begin{thm}\label{thcb*}
In the homogeneous case  the operators $\cb$ and $\bb ^*$
anticommute:
\begin{align}
[\bb ^*(\z),\cb (\z ' )]_+=0\,. \label{cb*hom}
\end{align}
\end{thm}
\begin{proof}
Consider the intervals $[k,m]$, $[k,l]$ for $l>m$. We may drop the suffix
$[k,m]$ in the following formulas within the rules discussed in 
Subsection 2.1.
Use Lemma \ref{auxlemma} in
\begin{align}
&\kb _{[k,l]}(\xi ,\al) \bb _{[k,l]}^*(\z ,\al+1)\bigl(X_{[k,m]}\bigr)
=\kb _{[k,l]}(\xi ,\al) \tr _c\mathbb{T}_{c,[m+1,l]}
(\z)\gb _c(\z ,\al +1)\bigl(X_{[k,m]}\bigr)
&\nn\\&\simeq _{\xi}\Tr _c\mathbb{T}_{c,[m+1,l]}(\z)
\kb_{[k,m]\sqcup c}(\xi,\al)
\gb _c(\z ,\al +1)\bigl(X_{[k,m]}\bigr) 
\quad \bmod(\z^2-1)^{l-m}\,.\nn
\end{align}
On the other hand using the right reduction for $\kb$ we have
\begin{align}
& \bb _{[k,l]}^*(\z ,\al)\kb _{[k,l]}(\xi ,\al+1)\bigl(X_{[k,m]}\bigr)
\simeq _{\xi}\tr _c\mathbb{T}_{c,[m+1,l]}
(\z)\gb _c(\z ,\al )\kb(\xi ,\al+1)\bigl(X_{[k,m]}\bigr)\quad
\,.\nn
\end{align}
So, the anticommutator is of the form 
\begin{align}
&\{\kb_{[k,l]}(\xi ,\al)\bb_{[k,l]}^*(\z,\al+1)+\bb_{[k,l]}^*(\z,\al)\kb_{[k,l]}(\xi,\al+1)\}
\left(X_{[k,m]}\right)
\nn\\
&\simeq _\xi
\Tr_c\bT_{c,[m+1,l]}(\z)\mathcal{X}_{c,[k,m]}(\z,\xi)
\qquad \mod\,,\nn\\
&\mathcal{X}_{c,[k,m]}(\z,\xi)=\{\kb_{[k,m]\sqcup c}(\xi,\al)
\gb _{c}(\z ,\al +1)+\gb _{c}(\z ,\al )\kb(\xi ,\al+1)\}(X_{[k,m]})\,.
\label{jjj}
\end{align}
We want to show that $\mathcal{X}_{c,[k,m]}(\z,\xi)\simeq _{\xi}0$.
Recall that
\begin{align}
&\gb _{c}(\z ,\al)=
\half\fb (\z q,\al)+\half\fb (\z q^{-1},\al)- \mathbb{T}_{c}(\z,\al)
\fb (\z ,\al)
+
\ub  _{c}(\z ,\al)\,.
\nn
\end{align}
Substitute this into (\ref{jjj}). When 
$\gb _{c}(\z ,\al +1)$ is replaced with
$\half\fb (\z q,\al)+\half\fb (\z q^{-1},\al)$, we use the right reduction for
$\kb_{[k,m]\sqcup c}(\xi,\al)$ to drop $c$ from it.
When $\gb _{c}(\z ,\al +1)$ is replaced with
$-\mathbb{T}_{c}(\z,\al+1)\fb(\z,\al+1)$
we use the Yang-Baxter relation after rewriting
\begin{align}
\kb_{[k,m]\sqcup c}
&(\xi,\al)\mathbb{T}_{c}(\z,\al+1)\fb(\z,\al+1)(X_{[k,m]})\nn\\
=\Tr_{a,B}&\s^+_b\mathbb{L}_{\{b,B\},c}(\xi/\z)\mathbb{T}_{\{b,B\}}(\xi,\al)
\xi^{-\bS_c-\bS} q^{-\sigma ^3_c-2S_{[k,m]}}
\mathbb{T}_{c}(\z,\al+1)\fb (\z ,\al+1)(X_{[k,m]})\,,\nn\\
=\Tr_{a,B}&\s^+_b\mathbb{L}_{\{b,B\},c}(\xi/\z)\mathbb{T}_{\{b,B\}}(\xi,\al)
\mathbb{T}_{c}(\z,\al)
\xi^{-\bS} q^{-2S_{[k,m]}}\fb (\z ,\al+1)(X_{[k,m]})\,.\nn
\end{align}
Then the anticommutation relation (\ref{comkf}) gives
\begin{align}
&\mathcal{X}_{c,[k,m]}(\z,\xi)
=\bigl\{\half\mb^{(++)}(\z q,\xi,\al)+\half\mb^{(++)}(\z q^{-1},\xi,\al)
-\mathbb{T}_c(\z,\al)\mb^{(++)}(\z,\xi,\al)\label{idenb*c}\\
&+\ub_c(\z,\al)\kb(\xi,\al+1)+\kb_{[k,m]\sqcup c}(\xi,\al)
\ub_{c}(\z,\al+1)\bigr\}(X_{[k,m]})\,.\nn
\end{align}
Define $\eta=\z/\xi$. We write $\mathcal{X}_{c,[k,m]}(\z,\xi)$
in the following form.
\bea
&&\mathcal{X}_{c,[k,m]}(\z,\xi)=
\Tr_{a,b,,A,B}\Bigl\{W^{(1)}_{a,b,c,A,B}(\eta)
\bT_{\{a,A\}}(\z,\al)\bT_{\{b,B\}}(\xi,\al)\label{X_c}\\
&&+W^{(2)}_{a,b,c,A,B}(\eta)
\bT_{\{b,B\}}(\xi,\al)\bT_{\{a,A\}}(\z,\al)\Bigr\}
(\z\xi)^{\al-\bS}\left(q^{-4S_{[k,m]}}X_{[k,m]}\right).\nn
\ena
The term with $W^{(1)}_{a,b,c,A,B}(\eta)$ comes from the first four terms
in \eqref{idenb*c}. It reads
\begin{align}
W^{(1)}_{a,b,c,A,B}(\eta):=\half(q^{-1}\beta(\eta)\tau^+_a&+q\beta(\eta^{-1})
\tau^-_a)M_{b,A,B}(\eta)\label{four}\\
&-F_{a,A}^{-1}P_{a,c}M_{b,A,B}(\eta)F_{a,A}
+\sigma ^+_b\eta ^{-1}q^{2D_B-1} Y_{a,c,A}\,,\nn
\end{align}
where $Y_{a,c,A}$ is given by \eqref{Y_{a,c,A}}.
The only non-trivial point in this derivation is to understand that
\begin{align}
&\Tr _{a,b,A,B}M_{b,A,B}(\eta q)
\mathbb{T}_{A}(\z q,\al)
\mathbb{T}_{\{b,B\}}(\xi,\al)(\z q)^{\al -\mathbb{S}}
\label{uu}\\
&=\Tr _{a,bA,B} q^{-1}\beta (\eta)M_{b,A,B}(\eta )
\tau ^+_a\mathbb{T}_{\{a,A\}}(\z,\al)
\mathbb{T}_{\{b,B\}}(\xi,\al)\z^{\al  -\mathbb{S}}\,.\nn
\end{align}
To see that one has to use (\ref{often2}).
The last term in (RHS) of \eqref{idenb*c} gives rise to the second term
in \eqref{X_c} where
\begin{align}
&W^{(2)}_{a,b,c,A,B}(\eta)
:=\eta\theta _{c}\( L_{\{b,B\}c}(\eta^{-1}q^{-1})\sigma ^+_b
Y_{a,c,A}q^{\sigma ^3_c}
L_{\{b,B\}c}(\eta ^{-1}q)^{-1}\)q^{2D_A+\sigma ^3_a}.\nn
\end{align}

In \eqref{X_c} there are two kinds of ambiguities: first,
(RHS) of \eqref{comkf} does not change if we add terms independent of $c$ to
$\mathcal{X}_{c,[k,m]}(\z,\xi)$; second, (RHS) of \eqref{X_c} does not change
if we add terms proportional to $\s^-_a$ or $\s^-_b$ to
$W^{(i)}_{a,b,c,A,B}(\eta)$. In the following
we use $\equiv$ to mean equality modulo such quantities, and read it
``equal to modulo irrelevant terms''.

\noindent
Write
\begin{align}
&W^{(2)}_{a,b,c,A,B}(\eta)
\equiv\left(\sigma _b^++\gamma (\eta ^{-1})\sigma ^+_c\tau ^+_b\right)
N_{a,c,A,B}(\eta)+\gamma (\eta )\tau ^-_bN_{a,c,A,B}(\eta)\sigma ^+_c
\label{last}
\end{align}
where 
$$N_{a,c,A,B}(\eta):=\eta\theta_c\bigl(L_{B,c}(\eta^{-1})q^{-\sigma^3_c/2}
Y_{a,c,A}q^{\sigma^3_c/2}L_{B,c}(\eta^{-1})^{-1}\bigr)q^{\sigma^3_a+2D_A}\,.$$

Now we reverse the order of the product
$\mathbb{T}_{\{b,B\}}(\xi,\al)\mathbb{T}_{\{a,A\}}(\z,\al)$ in \eqref{X_c}
by using the Yang-Baxter relation \eqref{YBabAB}. However, before doing that
it is very convenient to subtract some $q$-exact forms in $\xi$.
Comparing (\ref{four}) and (\ref{last}) we see that
the structure of singularities is different:
(\ref{last}) contains poles at $\eta ^2 =q^{\pm 2}$ and $\eta ^2=1$,
while (\ref{four}) contains singularities at $\eta ^2=1$ only.
The unwanted singularities in (\ref{last}) will cancel modulo irrelevant terms
if we add the following term to $W^{(2)}_{a,b,c,A,B}(\eta)$:
\begin{align}
W^{(3)}_{a,b,c,A,B}(\eta)
:=W^{(2)}_{a,b,c,A,B}(\eta)+W^{(4)}_{a,b,c,A,B}(\eta)\,,\nn
\end{align}
where
\begin{align}
W^{(4)}_{a,b,c,A,B}(\eta)
:=\eta(1-\half\Tr_c)
\bigl\{q^{-1}\beta(\eta^{-1})\tau^+_b\sigma^+_cN_{a,c,A,B}(\eta)\bigr\}
+q&\beta(\eta)\tau^-_bN_{a,c,A,B}(\eta)\sigma^+_c\bigr\}\nn\\
&+\half q^{2D_A}\eta^2q^{-1}\sigma^3_b\sigma^+_a\sigma^+_c\,.\nn
 \end{align}
This term can be added to $W^{(2)}_{a,b,c,A,B}(\eta)$ because
\begin{align}
&\Tr _{a,b,A,B} W^{(4)}_{a,b,c,A,B}(\z/\xi)\mathbb{T}_{\{b,B\}}(\xi,\al)\mathbb{T}_{\{a,A\}}(\z,\al)(\z\xi)^{\al -\mathbb{S}}\label{exactaddition}\\
&=\Delta _{\xi}\Tr _{a,A,B} S_{a,c,A,B}(\z/\xi)\mathbb{T}_{B}(\xi,\al)\mathbb{T}_{\{a,A\}}(\z,\al)(\z\xi)^{\al -\mathbb{S}}\,,\nn
\end{align}
 where 
\begin{align}
S_{a,c,A,B}(\eta)=\frac1{1-\eta^{-2}}
\bigl[&\half\sigma^3_c\ab_B^*
\left(\sigma^3_aq^{-2D_B-1}\ab_B^*-\sigma^+_a(\eta+q^{-2D_B-1}\ab _B^*\ab _A)
\right)\nn\\
&-\sigma^+_c\sigma^+_a\left(\eta(q^{-2D_B}-\eta ^{-2})\ab_B^*\ab_A-q^{2D_B+1}
+\half (q+q^{-1})\right)\nn\\
&+\sigma^+_c\sigma ^3_a\eta(q^{-2D_B}-\eta^{-2})\ab_B^*\bigr]q^{\sigma ^3_a+2D_A}\,.\nn
\end{align}
The $q$-exact from in (\ref{exactaddition}) is singular (has pole at $\eta ^2=1$), but it is easy to see that the singularity is harmless when we substitute
$\kb (\xi ,\al)$ in the definition of either $\bar{\cb }(\z ,\al)$ or $\cb (\z ,\al)$,
so, (\ref{exactaddition}) does not contribute to the commutation relations
with $\bar{\cb }(\z ,\al)$ and $\cb (\z ,\al)$. 

Now it remains to change the order of $\mathbb{T}_{\{b,B\}}(\xi,\al)$ and
$\mathbb{T}_{\{a,A\}}(\z,\al)$ in order to compare it with (\ref{four}). Using
Yang-Baxter we come to necessity to calculate
\begin{align}
&
R_{\{a,A\},\{b,B\}}(\eta )^{-1}W^{(3)}_{a,b,c,A,B}(\eta)R_{\{a,A\},\{b,B\}}(\eta )\equiv-W^{(1)}_{a,b,c,A,B}
\,.\label{RR}
\end{align}
The latter identity is a result of straightforward, but really hard computation.
So,
$$\mathcal{X}_{c,[k,m]}(\z,\xi)\simeq_\xi 0\,.$$
\end{proof}

\subsection{Commutation relations for $\bb$,$\bar{\bb}$ and 
$\bb^{*}$}

We now move on to the commutation relations between operators with 
opposite spin, such as $\bb^{*}(\z)$ with $\bb(\z)$ or $\bar{\bb}(\z)$. 
The derivation of these relations will follow basically 
the same line as in the previous subsection. 
Hence we shall mainly focus on the points 
which need further elaboration. 

Recall that $\bb(\z,\al)$ and $\bar{\bb}(\z,\al)$ are defined 
from the residues of the operator
\be
\phi(\kb)(\xi,\al)(X_{[k,l]})
=q^{-1}N(\al-\bS-1)\Tr_{b,B}\left(
\sigma^-_b \bT^-_{\{b,B\}}(\xi,\al)\xi^{-\al+\mathbb{S}}
(q^{2S_{[k,l]}}X_{[k,l]})\right)\,.
\en
Here, monodromy matrices with superfix $-$ 
are defined in terms of the $L$ operators obtained by spin reversal,  
\begin{align}
L^-_{A,j}(\z)=\sigma^1_j\ L_{A,j}(\z)\ \sigma^1_j\,, 
\quad 
L^-_{\{a,A\},j}(\z)=\sigma^1_a\sigma^1_j\ L_{\{a,A\},j}(\z)\ \sigma^1_a\sigma^1_j
\,.\label{Lminus}
\end{align}
Within this subsection and in Appendix \ref{sec:appC}, 
\ref{sec:appD} the original $L$ operators will be denoted by 
$L^+_{x,j}(\z)=L_{x,j}(\z)$ ($x=A,\{a,A\}$)
and likewise for $\bT^+$. 
(Warning: this sign convention for 
$L^\pm$ is opposite to the one in the previous papers \cite{HGSI,FB}.
We apologize the reader for making this change.)

When dealing with $\bT^+$ together with $\bT^-$,
a technical obstacle is the absence of an R matrix
which ensures the Yang-Baxter relation to hold. 
This is due to the fact that the $q$-oscillator representations
$W^+\otimes W^-$ and $W^-\otimes W^+$ are not isomorphic to each other. 
Nevertheless they have the same composition factors, and 
in most cases this is   
sufficient for the computation of traces. 
Introduce the notation
\be
&&\UAB(\eta)=\eta\ \ao_A^*+\ao_B q^{2D_A},
\quad
\YAB(\eta)=(\eta q^2-\ao_A \ao_B)q^{2D_A}\,.
\en
Under the trace, 
the order of the monodromy matrices can be exchanged 
according to the following rule. There exists a
$4\times 4$  matrix $R^{\text{quasi}}_{\{aA\},\{bB\}}(\eta)$ such that, 
for any matrix 
$\mathcal{X}_{a,b,A,B}(\eta)$ in $V_a\otimes V_b$ which is a 
polynomial in $\ao_B, \UBA(\eta^{-1})$, $\YBA(\eta^{-1})^{\pm 1}$
and $q^{\pm (2(D_A-D_B)+\sigma^3_a+\sigma^3_b)}$, we have   
\bea
&&
\Tr_{a,b,A,B}\Bigl(q^{-\sigma^3_aD_B}
\mathcal{X}_{a,b,A,B}(\eta)q^{\sigma^3_aD_B}
 \bT_{\{b,B\}}^-(\xi,\al)\bT_{\{a,A\}}^+(\z,\al)\Bigr)
\label{RT+T-}\\
&&=
\Tr_{a,b,A,B}\Bigl(q^{\sigma^3_bD_A}    
R^{\text{quasi}}_{\{aA\},\{bB\}}(\eta)^{-1}\sigma\bigl(\mathcal{X}_{a,b,A,B}(\eta)\bigr)
\nn\\
&&\qquad\qquad\qquad\times 
R^{\text{quasi}}_{\{aA\},\{bB\}}(\eta)q^{-\sigma^3_bD_A}  
\bT_{\{a,A\}}^+(\z,\al)\bT_{\{b,B\}}^-(\xi,\al)\Bigr)\,.
\nn
\ena
Here $\eta=\z/\xi$, and 
$\sigma$ is a linear map satisfying $\sigma(PQ)=\sigma(P)\sigma(Q)$ and 
\be
&&\sigma\left((1-\eta\YBA(\eta^{-1})^{-1})\ao_B\right)=\UAB(\eta), 
\\
&&      
\sigma\left(\UBA(\eta^{-1})\right)=
(1-\eta^{-1}\YAB(\eta)^{-1})\ao_A,
\\
&&\sigma\left(\YBA(\eta^{-1})\right)=q^{2}\YAB(\eta)^{-1}\,.
\en
We shall refer to $R^{\text{quasi}}_{\{aA\},\{bB\}}(\eta)$ as `quasi R matrix'. 
The details about this formula 
will be presented in Appendix \ref{sec:appC}, 
Lemma \ref{lem:RabAB}, 
along with  
the explicit formula \eqref{quasiRmat}
for the qausi R matrix. From there we quote here another useful formula
(see Lemma \ref{lem:adjust2}):
\begin{align}
&\Tr_{a,b,A,B}\Bigl(
q^{-2(D_A-D_B)-\sigma^3_aD_B}
\mathcal{X}_{a,b,A,B}(\eta)q^{\sigma^3_aD_B}
\bT_{\{b,B\}}^-(\xi,\al)\bT_{\{a,A\}}^+(\z,\al)\Bigr)
\label{shift}\\
&=\eta
\frac{N(\al-\mathbb{S})}{N(\al-\mathbb{S}-1)}\Tr_{a,b,A,B}\Bigl(
q^{-\sigma^3_aD_B-1}
\YBA(\eta^{-1})
\mathcal{X}_{a,b,A,B}(\eta)q^{\sigma^3_aD_B}
\nn\\
&\qquad\qquad\qquad\quad\times
\bT_{\{b,B\}}^-(\xi,\al)\bT_{\{a,A\}}^+(\z,\al)\Bigr)
\,,
\nn
\end{align}
where $\eta$ and $\mathcal{X}_{a,b,A,B}(\eta)$ have the same meaning as above. 
These formulas (and their analogs wherein $a,A$ are interchanged with $b,B$) 
will be frequently used in this subsection. 

Let us start the calculation. 
Our first task is to find an `exact $2$-form' relation between 
$\kb(\z,\al)$ and  $\phi(\kb)(\xi,\al)$. 
\begin{lem}\label{lem:k+k-}
We have 
\bea
&&
\kb(\z,\al)_{[k,l]}\phi(\kb)_{[k,l]}(\xi,\al+1)
+\phi(\kb)_{[k,l]}(\xi,\al)\kb_{[k,l]}(\z,\al-1)
\label{k+k-}\\
&&\quad
=\Delta_\z \mathbf{m}^{(+-)}_{[k,l]}(\z,\xi,\al)+\Delta_\xi \mathbf{m}^{(-+)}_{[k,l]}(\xi,\z,\al)\,.
\nn
\ena
The operators on the right hand side are given by 
\begin{align}
& \mathbf{m}^{(+-)}(\z,\xi,\al)(X_{[k,l]})=N(\al-\mathbb{S})\Tr_{b,A,B}
\left(M'_{b,A,B}(\eta)
\bT_{A}^+(\z,\al)\bT^-_{\{b,B\}}(\xi,\al)(X_{[k,l]})
\right)\eta^{\al-\mathbb{S}},\nn
\\
& \mathbf{m}^{(-+)}_{[k,l]}(\z,\xi,\al)
=-\mathbb{J} \mathbf{m}^{(+-)}_{[k,l]}(\z,\xi,\al)\mathbb{J}\,,
\nn\\
&M'_{b,A,B}(\eta)=\frac{1}{\eta-\eta^{-1}}q^{\sigma^3_bD_A}
\left(\half(\eta +\eta ^{-1})\sigma^3_b+\eta^{-1}\UAB(\eta)\sigma^-_b\right)
q^{-\sigma^3_bD_A}, \nn
\end{align}
where we have set $\eta=\z/\xi$.
\end{lem}
\begin{proof}
Omitting common suffix $[k,l]$ rewrite (\ref{k+k-}) as
\bea
\kb(\z,\al)\phi(\kb)(\xi,\al+1)- \Delta_\z\mathbf{m}^{(+-)}(\z,\xi,\al)
=-\phi(\kb)(\xi,\al)\kb(\z,\al-1)+\Delta_\xi \mathbf{m}^{(-+)}(\xi,\z,\al)\,,
\nn
\ena
so, that in the left hand side the monodromy matrices 
under the trace are ordered
as $\bT_{\{a,A\}}^+(\z,\al)\bT_{\{b,B\}}^-(\xi,\al)$ while in the right hand
side the order is opposite. Applying (\ref{shift}) to $-\phi(\kb)(\xi,\al)\kb(\z,\al-1)$  and using (\ref{often2}) and suitable analogs of
(\ref{often1}) obtain
\begin{align}
(RHS)=&N(\al-\mathbb{S})
\Tr_{a,b,A,B}\Bigl(q^{-\sigma^3_aD_B}W_{a,b,A,B}(\eta ) q^{\sigma^3_aD_B} \bT_{\{b,B\}}^-(\xi,\al)\bT_{\{a,A\}}^+(\z,\al)\Bigr)
\eta^{\al-\mathbb{S}}\,,\nn
\end{align}
where 
$$W_{a,b,A,B}(\eta )=-q^{-1}\YBA(\eta^{-1})\sigma^+_a\sigma^-_b-
\frac {\eta ^2q^{2\sigma ^3_b} +1} {2(\eta ^2q^{2\sigma ^3_b} -1)}
\sigma ^3_a-
\frac {\eta}
{\eta ^2q^{2\sigma ^3_b} -1} U_{B,A}(\eta ^{-1})\sigma ^+_a\,.$$
Now apply (\ref{RT+T-}), and verify that
$$  
R^{\text{quasi}}_{\{aA\},\{bB\}}(\eta)^{-1}\sigma\(W_{a,b,A,B}(\eta )\) 
R^{\text{quasi}}_{\{aA\},\{bB\}}(\eta)\equiv
-\sigma ^1_a\sigma ^1_b
W_{b,a,B,A}(\eta ^{-1}) 
\sigma ^1_a\sigma ^1_b\,,$$
where $\equiv $ means identity up to terms proportional to
$\sigma ^-_a$ or $\sigma ^+_b$.







\end{proof}

Unlike the previous case treated in Lemma \ref{kk}, 
the `exact forms' appearing in Lemma \ref{lem:k+k-} have
a simple pole on the diagonal $\z^2=\xi^2$.  
Indeed, their residues are proportional to the identity:
\begin{lem}\label{lem:ResG}
As $\z\to\xi$, we have 
\bea
&& \mathbf{m}^{(+-)}_{[k,l]}(\z,\xi,\al)=\psi (\z/\xi,\al+\mathbb{S}_{[k,l]})+O(1)\,.
\label{ResG}
\ena
\end{lem}
The proof is given in Appendix C (see Lemma
\ref{lem:residueG}). 
 

As noted before, the singularity on the diagonal is irrelevant 
to the derivation of the anti-commutation relations 
for annihilation operators. Thus Lemma \ref{lem:k+k-} immediately 
implies
\begin{thm}
We have the anti-commutation relations for the annihilation operators
\be
[\cb(\z),\bb(\z')]_+=0,\quad 
[\bar{\cb}(\z),\bb(\z')]_+=0,\quad 
[\bar{\cb}(\z),\bar{\bb}(\z')]_+=0\,.
\en
\end{thm}

To deduce the anti-commutators between creation and annihilation operators, 
the pole of   $\mathbf{m}^{(\pm\mp)}$ on the diagonal plays an important role. 
For that matter, it is convenient to subtract  from them 
the singular part
\begin{align}
&\Delta_\z \mathbf{m}^{(+-)}_{[k,l]}(\z,\xi,\al)+\Delta_\xi \mathbf{m}^{(-+)}_{[k,l]}(\z,\xi,\al)(\xi,\z,\al)\nn
\\
&\quad=\Delta_\z ( \mathbf{m}^{(+-)}_{[k,l]}(\z,\xi,\al)-\psi(\z/\xi,\al+\mathbb{S}_{[k,l]}))
+\Delta_\xi (\mathbf{m}^{(-+)}_{[k,l]}(\xi,\z,\al)-\psi(\z/\xi,\al+\mathbb{S}_{[k,l]}))\,. \nn
\end{align}
The terms in the right hand side are $q$-exact forms in the strict sense
(i.e., they do not have singularities other than $\z^2=1$). 
Therefore by the same arguments as in Lemma \ref{lemmakf} we obtain 
\begin{align}
\fb_{[k,l]}(\z,\al)\phi(\kb)_{[k,l]}(\xi,\al+1)
+\phi(\kb)_{[k,l]}(\xi,\al)\fb_{[k,l]}(\z,\al-1)\label{fk-}\\
\simeq _{\xi} \mathbf{m}^{(+-)}_{[k,l]}(\z,\xi,\al)-\psi(\z/\xi,\al+\mathbb{S}_{[k,l]})\,.
\nn
\end{align}

\begin{thm}\label{thbb*}
The following anti-commutation relations hold. 
\begin{align}
&[\bb^*(\z),\bb(\z')]_+=-\psi(\z/\z^{'},\alb +\mathbb{S}), 
\label{bb*1}
\\
&[\bb^*(\z),\bar{\bb}(\z')]_+=
\bt^*(\z)\psi(\z/\z^{'},\alb +\mathbb{S})\,. 
\label{bb*2}
\end{align}
\end{thm}
\begin{proof}
Assume $l>m$. 
Reasoning as in the proof of Theorem \ref{thcb*} 
, we obtain the relations
\begin{align}
&
\left(
\phi(\kb_{[k,l]})(\xi,\al)\bb _{[k,l]}^*(\z ,\al-1)
+\bb_{[k,l]}^*(\z,\al)\phi(\kb_{[k,l]})(\xi,\al+1)
\right)
\bigl(X_{[k,m]}\bigr)
\nn
\\
&\quad\simeq _{\xi}
\Tr_c\mathbb{T}_{c,[m+1,l]}(\z)
\mathcal{X}'_{c,[k,m]}(\z,\xi)
\qquad\mod\,,\nn\\
&\mathcal{X}'_{c,[k,m]}(\z,\xi)=\{\phi(\kb)_{[k,m]\sqcup c}(\xi,\al)
\gb _{c}(\z ,\al -1)+\gb _{c}(\z ,\al )\phi(\kb)(\xi ,\al+1)\}(X_{[k,m]})\,.\nn
\end{align}
From the relation \eqref{fk-} we find that
$\mathcal{X}'_{c,[k,m]}(\z,\xi)=\mathcal{X}^{'\text{sing}}_{c,[k,m]}(\z,\xi)+\mathcal{X}^{'\text{reg}}_{c,[k,m]}(\z,\xi)$
\begin{align}
&\mathcal{X}^{'\text{reg}}_{c,[k,m]}(\z,\xi)
=
\half \mathbf{m}^{(+-)}(\z q,\xi, \al)+\half \mathbf{m}^{(+-)}(\z q^{-1},\xi, \al)\nn
\\&
 -\mathbb{T}_{c}(\z,\al)\mathbf{m}^{(+-)}(\z ,\xi, \al)
+\ub _c (\z ,\al )\phi(\kb)(\xi ,\al+1)
+\kb_{[k,m]\sqcup c}(\xi,\al)
\ub_{c}(\z,\al-1)
\,
\nn
\end{align}
and 
\be
\mathcal{X}^{'\text{sing}}_{c,[k,m]}(\z,\xi)
=
-\half \psi(\z q/\xi, \al+\mathbb{S})
-\half \psi(\z q^{-1}/\xi, \al+\mathbb{S})
+\mathbb{T}_{c}(\z,\al)\psi(\z/\xi, \al+\mathbb{S})
\,.
\en
From the residues of the last term, 
the right hand sides of the anti-commutation relations \eqref{bb*1},
\eqref{bb*2} arise: 
\begin{align}
&
{\rm res}_{\xi^2=1}
\mathcal{X}^{'\text{sing}}_{c,[k,m]}(\z,\xi)\psi(\z'/\xi,-\al-\mathbb{S})
\frac{d\xi^2}{\xi^2}
=\bT_c(\z,\al)
\psi(\z/\z',\al+\mathbb{S})\,,\nn
\\
&
\left({\rm res}_{\xi^2=q^2}+{\rm res}_{\xi^2=q^{-^2}}\right)
\mathcal{X}^{'\text{sing}}_{c,[k,m]}(\z,\xi)\nn
\\
&\quad\times\left(\psi(q\z'/\xi,-\al-\mathbb{S})
+\psi(q^{-1}\z'/\xi,-\al-\mathbb{S})\right)
\frac{d\xi^2}{\xi^2}
=-{\half}\psi(\z/\z',\al+\mathbb{S})\,.\nn
\end{align}

Hence the proof is reduced to showing that
$\mathcal{X}^{'\text{reg}}_{c,[k,m]}(\z,\xi)\simeq _{\xi} 0$. 
We now sketch this calculation. 

Define $\eta =\z/\xi$. We have:
\begin{align}
&\mathcal{X}^{'\text{reg}}_{c,[k,m]}(\z,\xi)=
\Tr_{a,b,A,B}\bigl(
{W'}^{(1)}_{a,b,c,A,B}(\eta) 
\bT^{+}_{\{a,A\}}(\z,\al)\bT^{-}_{\{b,B\}}(\xi,\al)
\eta^{\al-\mathbb{S}}N(\al-\mathbb{S})\nn\\
&+
{W'}^{(2)}_{a,b,c,A,B}(\eta) 
\bT^-_{\{b,B\}}(\xi,\al)\bT^+_{\{a,A\}}(\z,\al)
q^{-1}\eta^{\al-\mathbb{S}-1}N(\al-\mathbb{S}-1)
\bigr)(X_{[k,m]}),\nn
\end{align}
with
\begin{align}
&{W'}^{(1)}_{a,b,c,A,B}(\eta )
:=
\half\(q\beta(\eta)\tau^+_a
+q^{-1}\beta(\eta^{-1})\tau^-_a
\)M'_{b,A,B}(\eta)
-F_{a,A}^{-1}P_{ac}M'_{b,A,B}(\eta)
F_{a,A}
\nn\\&\quad
 + q^{-D_A}\YAB(\eta)q^{-D_A}Y_{a,c,A}
\sigma^-_b\,.\nn
\end{align}
where we have used \eqref{shift} to shift the argument of $N(\al-\mathbb{S}+1)$  
in the term with $\mathbf{u}_c\phi (\kb)$.  
Let us use $\equiv$ for calculations modulo terms proportional
to $\sigma ^-_a$, $\sigma ^+_b$. For ${W'}^{(2)}$ we have
\begin{align}
&{W'}^{(2)}_{a,b,c,A,B}(\eta) \equiv
(\sigma^-_b +\gamma(\eta^{-1})\tau^-_b \sigma^-_c)N'_{a,c,A,B}(\eta)
+\gamma(\eta)\tau^+_b N'_{a,c,A,B}(\eta)\sigma^-_c
\,,
\nn\\
&
N'_{a,c,A,B}(\eta):=-\theta_c\left(L^-_{B,c}(\eta^{-1})
q^{\sigma^3_c/2} Y_{a,c,A} q^{-\sigma^3_c/2}
L^-_{B,c}(\eta^{-1})^{-1}q^{-2D_A-\sigma^3_a}\right)\,.\nn
\end{align}
As was done in the previous section, it is simpler first to modify 
$W^{'(2)}_{a,b,c,A,B}(\eta) $ by an exact form, introducing
\be
{W'}^{(3)}_{a,b,c,A,B}(\eta) ={W'}^{(2)}_{a,b,c,A,B}(\eta) +{W'}^{(4)}_{a,b,c,A,B}(\eta) , 
\en
where
\begin{align}
&{W'}^{(4)}_{a,b,c,A,B}(\eta) 
(\eta)
:=\eta ^{-1}\(1-\half \Tr _c\) \(q^{-1}\beta (\eta)\tau ^+_b
 N'_{a,c,A,B}(\eta)\sigma ^-_c
 +
 q\beta (\eta ^{-1})\tau ^-_b\sigma ^-_c
 N'_{a,c,A,B}(\eta)
\right.\nn\\&\left. \qquad\qquad\qquad
 \qquad\qquad\qquad\qquad\qquad\qquad-\half 
 q^{-2D_A-\sigma ^3_a}\eta ^{-1}  \sigma ^3_b\sigma ^+_a\sigma ^-_c\)\,.\nn
 \end{align}
We have the exact form:
\begin{align}
&\Tr _{a,b,A,B} {W'}^{(4)}_{a,b,c,A,B}(\z/\xi)\mathbb{T}^-_{\{b,B\}}(\xi,\al)\mathbb{T}^+_{\{a,A\}}(\z,\al)\eta^{\al -\mathbb{S}-1}\label{exactaddition1}
\\
&=\Delta _{\xi}\Tr _{a,A,B} S'_{a,c,A,B}(\z/\xi)\mathbb{T}_{B,}(\xi,\al)\mathbb{T}_{\{a,A\}}(\z,\al)\eta^{\al -\mathbb{S}}\,,
\nn\end{align}
 where 
\begin{align}
& S'_{a,c,A,B}(\eta)=\frac 1{1-\eta^{2}}q^{-D_A-\sigma ^3_a}\bigl[\half \sigma ^3_c
\(-\sigma ^3_aq^{2D_B+1}+
\sigma ^+_aq^2U_{B,A}(\eta ^{-1})\)
\nn\\
& +\sigma ^-_c
\bigl(\sigma _a^3q^{2D_B+2}\ab _B+
\sigma ^+_a( q \eta ^{-1}Y_{B,A}(\eta ^{-1})
- \half (1+q^2))\bigr)
\bigr]q^{-D_A}\,.\nn
\end{align}
Noting that 
$$ 
q^{D_B\sigma ^3_a}{W'}^{(3)}_{a,b,c,A,B} q^{-D_B\sigma ^3_a}
$$
consists of right admissible quantities
(for the definition, 
see the end of Subsection \ref{appC2} and 
the paragraph after Corollary \ref{corC3}), 
we can change the order 
of monodromy matrices by applying the quasi $R$-matrix.
After a straightforward, albeit extremely lengthy, calculation
we get:
\begin{align}
&R^{\text{quasi}}_{\{a,A\}, \{b,B\}}(\eta )^{-1}\sigma\bigl(
q^{D_B\sigma ^3_a}{W'}^{(3)}_{a,b,c,A,B}(\eta)  q^{-D_B\sigma ^3_a}
\bigr)
R^{\text{quasi}}_{\{a,A\}, \{b,B\}}(\eta )\nn\\&\equiv-
q^{-2D_A+2D_B}
Y_{A,B}(\eta)q^{-\sigma^3_b D_A}{W'}^{(1)}_{a,b,c,A,B}(\eta) q^{\sigma^3_b D_A}\,,\nn
\end{align}
Now, shifting as usual the argument  $N(\al-\mathbb{S}-1)$  we finish
the proof of the theorem.
\end{proof}

 \subsection{Commutation relations of $\tb^*$ with $\bb^*$, $\cb^*$.}
 
In this paper we shall not prove all the commutation relations 
between the creation operators, a weak variant of these
relations is sufficient for our goals as it will be discussed later.
However, we give the proof of commutativity of $\tb^*$ and
$\bb ^*$ because it is important from general point of view. For the lack of space
we consider the homogeneous case only.
\begin{thm}\label{tht*b*}
The following commutation relation holds in the homogeneous case:
\begin{align}
\bb ^*(\xi ) \tb ^*(\z)=\tb ^*(\z)\bb ^* (\xi )
\label{comt*b*}
\end{align}
\end{thm}

\begin{proof}
Consider the formula (\ref{auxhom}). 
We used it for the commutation 
with annihilation operators, for that reason 
we dropped $q$-exact forms in $\xi$. 
Written in full in the case 
$Y_{c,[k,m]}=\mathbb{T}_{c,[k,m]}(\z ,\al +1)(X_{[k,m]})$
it looks as follows:
\begin{align}
&\kb _{[k,l]}(\xi ,\al) \tb ^*(\z,\al+1)\bigl(X_{[k,m]}\bigr)\label{auxhom1}\\ &=
\tb _{[k,l]}^*(\z,\al)\(\kb _{[k,m]}(\xi ,\al) +\Delta _{\xi}\mathbf{v}_{[k,l]}(\xi ,\al)\)
(X_{[k,m]})\bigr)
\qquad
\mod\,.\nn
\end{align}
Using the definition of $\bb ^*(\z,\al)$ we obtain:
\begin{align}
&\bb ^*_{[k,l]}(\xi ,\al)\tb_{[k,l]}^*(\z,\al +1)(X_{[k,m]})\nn\\
&=\(\fb _{[k,l]}(\xi q ,\al)+\fb _{[k,l]}(\xi q^{-1} ,\al)-
\tb _{[k,l]}^*(\xi,\al)\fb _{[k,l]}(\xi  ,\al)\)\tb_{[k,l]}^*(\z,\al +1)(X_{[k,m]})\nn\\
&=\tb^*_{[k,l]}(\z,\al)\bigl(\fb _{[k,m]}(\xi q,\al)+\fb _{[k,m]}(\xi q^{-1},\al)-
\tb _{[k,l]}^*(\xi,\al)\fb _{[k,m]}(\xi ,\al)\nn\\
&+\mathbf{v}_{[k,l]}(\xi q,\al)+\mathbf{v}_{[k,l]}(\xi q^{-1},\al)-
\tb ^*_{[k,l]}(\xi,\al)\mathbf{v}_{[k,l]}(\xi ,\al)\bigr)(X_{[k,m]})\,.\nn
\end{align}
From the proof of Proposition \ref{APPB} (see Appendix 
\ref{sec:appB}, \eqref{Truc} and \eqref{ucIII})
one extracts\begin{align}
\bigl(\mathbf{v}_{[k,l]}(\xi q,\al)&+\mathbf{v}_{[k,l]}(\xi q^{-1},\al)-
\tb ^*_{[k,l]}(\xi,\al)\mathbf{v}_{[k,l]}(\xi ,\al)\bigr)(X_{[k,m]})
\nn\\ & =
\Tr _c \ \mathbb{T}_{c,[m+1,l]}(\xi)\mathbf{u}_{c,[k,m]}(\xi,\al)(X_{[k,m]})\,.\nn
\end{align}
Now the statement of the Theorem  follows from the reduction
relation (\ref{BBTU}).
\end{proof}
\subsection {Commutation relations for inhomogeneous case.}

Now let us consider the inhomogeneous case.
Analyzing the proofs given in this section we realize 
that they consist of two parts. First the 
interval $[k,l]$ is reduced to $[k,m]$ 
by using Lemma \ref{auxlemma},  
then the proofs consist of 
algebraic manipulations with operators on this, small, interval. 
So, if we find a direct analog of Lemma \ref{auxlemma} 
in the inhomogeneous case, the rest is simple. 
This analog is
\begin{lem}
In the inhomogeneous case we have for $l<j\le m$:
\begin{align}
&\kb _{[k,l]}(\xi ,\al) \Tr _c\mathbb{T}_{c,[m+1,l]}
(\xi _j)\bigl(Y_{c,[k,m]}\bigr)\simeq _{\xi}
\Tr _c \mathbb{T}_{c,[m+1,l]}(\xi _j)\mathbf{k}_{[k,m]\sqcup c}(\xi ,\al)(Y_{[k,m],c})\,,
\nn
\end{align}
where the inhomogeneity parameter associated with $c$ is $\xi _j$.
\end{lem}
Since the proof is simple, we leave it to the reader.
Using the above Lemma we easily repeat the proof of 
Theorem \ref{thcb*}, Theorem \ref{thbb*}, 
Theorem \ref{tht*b*} in the inhomogeneous case,  
deducing that 
\begin{thm}\label{thm:comm-inhom}
In the inhomogeneous case the following commutation relations  
hold on $(\mathcal{W})_{(-\infty,n-1]}$:
\begin{align}
&[\cb (\z), \bt ^*(\xi _n)]
=0,\quad\quad [\bar{\cb} (\z), \bt ^*(\xi _n)]
=0\,,\nn\\
&[\cb (\z), \bb ^*(\xi _n)]_+
=0,\quad\ [\bar{\cb} (\z), \bb ^*(\xi _n)]_+
=0\,,\nn\\
&[\bb (\z), \bb ^*(\xi _n)]_+
=-\psi (\xi _n/\z,\alb+\mathbb{S})
\,,\nn\\
&[\bar{\bb} (\z), \bb ^*(\xi _n)]_+
=\bt^*(\xi_n)\psi (\xi _n/\z ,\alb+\mathbb{S})
\,.
\label{cominhom}
\end{align}
In addition we have
\begin{align}
&[\tb ^*(\xi_{p}), \bb ^*(\xi _{q})]
=0,\quad\ [\tb ^*(\xi_{p}), \cb ^*(\xi _{q})]
=0\,,\label{t*inhom}
\end{align}
for  $ p\ge n,\ q\ge n,\  p\ne q$.
 \end{thm}


\section{Vacuum expectation values}\label{sec:5}

We are now in a position to discuss the construction of a 
fermionic basis of quasi-local operators, 
and calculate the vacuum expectation values (VEV). 
First we construct the basis in the inhomogeneous case, 
and prove its completeness. 
In Subsection \ref{sec:VEV-homogeneous} we give 
the construction in the case of the infinite homogeneous chain. 
While the completeness is still conjectural for the homogeneous case, 
the VEV's of the base vectors are given by a determinant 
as in the inhomogeneous case. 

\subsection{Fermionic basis}\label{sec:fbasis}

Let us consider the inhomogeneous chain. 
We want to construct a basis of the subspace 
$(\mathcal{W}^{(\al)})_{[1,\infty)}$
using the
operators $\bb ^*(\xi _k)$, $\cb ^*(\xi _k)$, $\tb ^*(\xi _k)$. 
Starting from the primary field $q^{2\al S(0)}$,  
define inductively the quasi-local operators 
$X^{\la_1,\cdots,\la _n}(\xi _1,\cdots ,\xi _n;\al)$ 
labeled by $\la _j\in\{+,-,0,\emptyset\}$: 
\begin{eqnarray*}
X^{\lambda_{1}, \ldots , \lambda_{n}}(\xi_{1}, \ldots , \xi_{n}; \alpha):=
\left\{ 
\begin{array}{ll}
{\bf b}^{*}(\xi_{n})X^{\lambda_{1}, \ldots , \lambda_{n-1}}
(\xi_{1}, \ldots , \xi_{n-1}; \alpha) & (\lambda_{n}=+), \\
{\bf c}^{*}(\xi_{n})(-1)^{\mathbb{S}}X^{\lambda_{1}, \ldots , \lambda_{n-1}}
(\xi_{1}, \ldots , \xi_{n-1}; \alpha) & (\lambda_{n}=-), \\
\half{\bf t}^{*}(\xi_{n})X^{\lambda_{1}, \ldots , \lambda_{n-1}}
(\xi_{1}, \ldots , \xi_{n-1}; \alpha) & (\lambda_{n}=0),\\
X^{\lambda_{1}, \ldots , \lambda_{n-1}}
(\xi_{1}, \ldots , \xi_{n-1}; \alpha) & (\lambda_{n}=\emptyset). 
\end{array}
\right.
\end{eqnarray*} 
This operator has spin determined by the rule (\ref{spins}).
We have
\begin{lem}
For generic values of $\xi_1,\xi _2\cdots $, the set 
\be
\{X^{\la _1,\cdots ,\la _n}(\xi _1,\cdots ,\xi _n;\al),\ n=0,1,2\cdots\}
\en
span $(\mathcal{W}^{(\al)})_{[1,\infty)}$. 
\end{lem}
\begin{proof}
 Since for any $n$ there are as many
$X^{\la _1,\cdots ,\la _n}(\xi _1,\cdots ,\xi _n;\al)$ as
$\dim\bigl((\mathcal{W}^{(\al)})_{[1,n]}\bigr)$, 
it suffices to prove their linear independence. 
Let 
$Y_\pm,Y_{\emptyset},Y_{0}\in (\mathcal{W}^{(\al)})_{[1,n-1]}$,  
and suppose we have a linear relation 
\bea
Y_{\emptyset}+\tb^*(\xi_{n})(Y_{0})+
\bb^{*}(\xi_{n})(Y_+)+\cb^{*}(\xi_{n})(Y_{-})=0\,. 
\nonumber 
\ena
Apply $\bb(\z)$ or $\cb(\z)$ to both sides and take the residue at $\z=\xi _n$. Then from the commutation relations
(\ref{cominhom}) we find that 
$Y_\pm=0$. 
Furthermore, in the limit $\xi_n\to\infty$ we have
\be
\tb^{*}(\xi_{n})(Y_{0})=
q^{\al\sigma^3_n+\sigma^3_n\bS_{[1,n-1]}}(Y_{0})+O(\xi_n^{-1}).
\en
Comparing the $n$-th tensor component
we find $Y_{\emptyset}=Y_{0}=0$. 
The assertion follows from these by induction.
\end{proof}


\subsection{$\kappa$-trace}\label{subsec:4.3}

In this subsection, we prepare a Lemma which will be used to 
calculate the weighted traces of fermionic basis elements. 

Introducing a new parameter $\kappa$, we set 
\begin{eqnarray}
\mathbf{tr}_{[k,l]}^{\kappa}(X_{[k,l]})=
\frac{\tr_{[k,l]}(q^{-\kappa S_{[k,l]}}X_{[k,l]})}{\tr_{[k,l]}(q^{-\kappa S_{[k,l]}})}. 
\label{al-trace} 
\end{eqnarray}
Note that if $i,j\in[k,l]$, 
\bea
\mathbf{tr}^\kappa_{[k,l]}\check{\mathbb{R}}_{i,j}(\z)(X_{[k,l]})
=\mathbf{tr}^\kappa_{[k,l]}(X_{[k,l]}). 
\label{eq:R-symmetry-of-kappa-trace}
\ena
We shall use this property in the form 
\bea
\mathbf{tr}^\kappa_{[k,l]}\bT_{c,[k,l]}(\xi_l)(Y_{[k,m],c})
=\mathbf{tr}^\kappa_{[k,l]}(Y_{[k,m],l})\,,
\label{bT-trace}
\ena
where $k\le m<l$ and $Y_{[k,m],c}\in M_{[k,m]}\otimes M_c$.


Now set 
\bea
\omega_{0}(\z,\al)=-
\left(\frac{1-q^{\alpha}}{1+q^{\alpha}}\right)^{2}
\Delta_{\z}\psi(\zeta, \alpha)
\label{omega1}
\ena
The following formulas will be used in the next subsection.
\begin{lem}\label{kappatr} 
Assume that $k\le m<l$. Then we have
\begin{eqnarray}
&&
\mathbf{tr}^\kappa_{[k,l]}\tb^{*}_{[k,l]}(\xi_l,\al)(X_{[k,m]})
=2\frac{q^\al+q^{\kappa-\al}}{1+q^\kappa}
\mathbf{tr}^\kappa_{[k,m]}(X_{[k,m]})\,,
\label{eq:ktrt*}
\\
&&
\mathbf{tr}^\kappa_{[k,l]}{\bf b}^{*}_{[k,l]}(\xi_l,\al)(X_{[k,m]})
=\left(
\frac{1-q^\kappa}{1+q^\kappa} \cdot \frac{1+q^{\alpha}}{1-q^{\alpha}} \right)^2
\label{eq:ktrcc*}\\
&& \qquad {}\times 
\frac{1}{2\pi i}\oint_{\Gamma}
\omega_0(\xi_l/\xi,\al)
\mathbf{tr}^\kappa_{[k,m]}\cb_{[k,m]}(\xi,\al)(X_{[k,m]})
\frac{d\xi^2}{\xi^2}\,. 
\nn
\end{eqnarray}
Here the contour $\Gamma$ encircles $\xi_j^2$ while 
keeping $q^{\pm 2}\xi_{j}^{2} \, (j\in [k,m])$ and $q^{\pm2}\xi_l^2$ outside. 
\end{lem}
\begin{proof}
Formula \eqref{eq:ktrt*} follows from (\ref{bT-trace}).
Let us consider \eqref{eq:ktrcc*}. 
In the rest of the proof we set $J=[k,m]$ and $K=[m+1, l]$. 
We may restrict to the case when the spin of $X_J$ is $s=-1$, 
since  otherwise the trace is zero.

Substituting $\z=\xi_l$ in \eqref{BBTU} and using \eqref{bT-trace}, 
we obtain
\be
\mathbf{tr}^\kappa_{J \sqcup K}{\bf b}^{*}_{J \sqcup K}(\xi_l,\al)(X_{J})
=
2\mathbf{tr}^\kappa_{J \sqcup \{l\}}{\bf g}_{l,J}(\xi_l,\al)(X_{J})
=\mathbf{tr}^\kappa_{J}H_J(\xi_l),
\en
where 
\be
H_J(\z)=\left(
\fb(q\z,\al)+\fb(q^{-1}\z,\al)
-2\frac{q^\al+q^{\kappa-\al}}{1+q^\kappa}\,
\fb(\z,\al)
-\frac{1-q^\kappa}{1+q^\kappa} \, 
\kb(\z,\al)\right)(X_J)\,.
\en
The second equality follows after taking the trace over the space $l$. 
Hence, with the notation
\be
H'_J(\z)=\left(
\frac{1-q^\kappa}{1+q^\kappa}\right)^2
\frac{1}{2\pi i}\oint_{\Gamma}
\Delta_\z\psi(\z/\xi,\al)\ 
\mathbf{tr}^\kappa_{[k,m]}\cb_{[k,m]}(\xi,\al)(X_{[k,m]})
\frac{d\xi^2}{\xi^2}\,,  
\en
the proof is reduced to showing the identity 
\bea
\mathbf{tr}^\kappa_{J}\left(H_J(\z)+H'_J(\z)\right)=0.
\label{H=H'}
\ena

The left hand side of \eqref{H=H'} has the form $\z^{\al}F(\z^2)$
with some rational function $F(\z^2)$ 
(we recall that $X_J$ has spin $-1$). 
By Lemma \ref{regularity}, $H_J(\z)$ is regular at $\z^2=\xi_j^2$,  
and from the definition $H'_J(\z)$ is also regular there. 

Let us calculate the residues of 
\eqref{H=H'} at $\xi_{l}=q^{\pm 1}\xi_{j}$. 
{}From the $R$-matrix symmetry, Lemma \ref{FQ} and the relations
$
q^{\kappa/2}{\rm tr}^\kappa(\sigma^+x)=
q^{-\kappa/2}{\rm tr}^\kappa(x\sigma^+)={\rm tr}(\sigma^{+}x), 
$
the two residues ${\bf tr}_{J}^{\kappa}{\bf k}_{J}(\zeta, \alpha)(X_{J})$ at 
$\zeta=q^{-1}\xi_{j}$ and $q\xi_{j} \, (j \in J)$ are proportional to each other. 
{}From this fact and the definition of 
${\bf f}_{J}(\zeta, \alpha)$ and ${\bf c}_{J}(\zeta, \alpha)$, 
we obtain 
\begin{eqnarray}
\res_{\xi=\xi_j}\xi^{-\al}
\mathbf{tr}^\kappa_{J}\fb_{J}(\xi,\al)(X_{J})\frac{d\xi^2}{\xi^2}
&=&
\frac{1-q^\kappa}{1+q^\kappa}C_{j}, \nonumber \\
\res_{\xi=q^{\pm 1}\xi_j}\xi^{-\al}
\mathbf{tr}^\kappa_{J}\kb_{J}(\xi,\al)(X_{J})\frac{d\xi^2}{\xi^2}
&=&
\frac{2q^{\mp \al}}{1+q^{\mp\kappa}}C_j, 
\label{eq:kappa-trace-of-k-to-c}
\end{eqnarray}
where 
\begin{eqnarray*}
C_{j}=\mathop{\rm res}_{\xi=\xi_j}\xi^{-\al}
\mathbf{tr}^\kappa_{J}{\bf c}_{J}(\xi,\al)(X_{J})
\frac{d\xi^2}{\xi^2}\,. 
\end{eqnarray*}
Combining these we conclude that $F(\z^2)$ is regular at 
$\z^2=q^{\pm2}\xi_j^2$. 
Clearly it is also regular at $\z^2=0$ and $\infty$. 
Hence $F$ must be a constant. 
The value at $\infty$ can be calculated using 
\be
&&\lim_{\z^2 \to \infty}\z^{-\alpha}\mathbf{tr}^\kappa_{J}
\kb_J(\z,\al)(X_J)=0, 
\\
&&
\lim_{\z^2 \to \infty}\z^{-\alpha}\mathbf{tr}^\kappa_{J}
\fb_J(\z,\al)(X_J)
=\frac{1-q^\kappa}{2(1+q^\kappa)}\sum_{j\in J}C_j,
\\
&&
\lim_{\z^2 \to \infty}\z^{-\alpha}\mathbf{tr}^\kappa_{J}H'_J(\z)
={\half}(q^\al-q^{-\al})
\left(\frac{1-q^\kappa}{1+q^\kappa}\right)^2\sum_{j\in J}C_j.
\en
It follows that $F(\infty)=0$ and hence $F(\z^2)\equiv 0$. 
This completes the proof.

\end{proof}

\begin{rem}
For the purpose of calculating the VEV of 
quasi-local operators $q^{2\al S(0)}\mathcal{O}$,  
we will need only the case $\kappa=\al$
(see the next subsection). 
However Lemma \ref{kappatr}  holds for all 
$\kappa$, and in particular, for the ordinary trace we have
\be
\tr_{[k,l]}{\bf b}^{*}_{[k,l]}(\xi_l,\al)(X_{[k,m]})=0. 
\en
\end{rem}

\subsection{Determinant formula for expectation values}\label{sec:VEV-determinant-formula}

The weighted trace $\mathbf{tr}^{\al}$ is a well defined 
linear map on $\mathcal{W}_\al$. 
According to the main formula of \cite{HGSI},  
the VEV of a quasi-local operator 
$q^{2\al S(0)}\mathcal{O}$ is expressed as follows. 
\bea
\frac {\langle \text{vac}|q^{2\al S(0)}\mathcal{O} |\text{vac}\rangle}
{\langle \text{vac}|q^{2\al S(0)} |\text{vac}\rangle}
=
\mathbf{tr}^{\al}\(e^{\mbox{\scriptsize\boldmath{$\Omega$}}} 
\bigl(q^{2\al S(0)}\mathcal{O} \bigr) \)\,.
\label{mainofmains}
\ena
Here $\Ob$ is an operator on $\mathcal{W}$ given by
\be
&&\Ob=\frac{1}{(2\pi i)^2}
\oint\limits_{\Gamma}\!\!\!\oint\limits_{\Gamma} 
\omega(\zeta_{1}/\zeta_{2}, \alb+\mathbb{S})
{\bf b}(\z_1){\bf c}(\z _2)
\frac {d\z^2_1}{\z^2_1}\frac {d\z^2_2}{\z^2_2}\,. 
\en
The scalar function $\omega(\zeta, \alpha)$ consists of 
two pieces, 
\bea
&&\omega(\z,\alpha)=\omega_{\rm trans}(\z,\alpha)-\frac{4q^{\alpha}}{\(1-q^{\alpha}\)^2}\,\omega_0(\z,\alpha)\,.  
\label{omega-fun}
\ena 
The elementary piece $\omega_{0}(\zeta, \alpha)$ 
is defined by \eqref{omega1}, and the transcendental piece 
$\omega_{\rm trans}(\z,\alpha)$ is given by 
\be
&&\omega_{\rm trans}(\z,\alpha)
=P\!\!\int\limits_{-i\infty}^{i\infty}\z ^{u+\alpha}
\frac {\sin \frac {\pi} 2\(u-\nu(u+\alpha)\)}{\sin \frac {\pi} 2 u
\cos \frac {\pi\nu} 2\(u+\alpha\)}du\,, 
\label{omega-0}
\en
where $P\!\!\int\limits_{-i\infty}^{i\infty}$ 
means the principal value 
$(1/2)(\int\limits_{-i\infty-0}^{i\infty-0}
+\int\limits_{-i\infty+0}^{i\infty+0})$.
%

Consider the operator 
\begin{eqnarray}
\Ob_{0}=\frac{1}{(2\pi i)^2}
\oint\limits_{\Gamma}\!\!\!\oint\limits_{\Gamma} 
\omega_{0}(\zeta_{1}/\zeta_{2}, \alb+\mathbb{S})
{\bf b}(\z_{1})\,{\bf c}(\z_{2})
\frac {d\z^2_1}{\z^2_1}\frac {d\z^2_2}{\z^2_2}\,. 
\label{eq:def-omega-zero}
\end{eqnarray}
and define the linear functional $\mathrm{v}^{\alpha}$ by
\begin{eqnarray*}
\mathrm{v}^{\alpha}(\cdot)=
{\bf tr}^{\alpha}\bigl( e^{\mbox{\scriptsize\boldmath{$\Omega$}}_{0}}(\cdot) \bigr). 
\end{eqnarray*}
{}From the commutation relations (\ref{cominhom}) and Lemma \ref{kappatr}, 
for $X \in (\mathcal{W}^{(\alpha)})_{[1,n-1]}$ we find
\begin{eqnarray*}
\mathrm{v}^{\alpha}({\bf t}^{*}(\xi_{n})(X))=
\mathrm{v}^{\alpha}(X), \quad \mathrm{v}^{\alpha}({\bf b}^{*}(\xi_{n})(X))=
\mathrm{v}^{\alpha}({\bf c}^{*}(\xi_{n})(X))=0. \label{vacuum}
\end{eqnarray*}
Thus the functional $\mathrm{v}^{\alpha}$
plays a role of the dual vacuum. 

Now let us calculate the expectation value of 
an element of the fermionic basis 
$X^{\lambda_1,\cdots,\la _n}(\xi _1,\cdots,\xi _n;\al)$. 
Since $\Ob$ commutes with $\Ob_{0}$, 
we have 
\begin{eqnarray*}
{\bf tr}^{\alpha}\left(
e^{\mbox{\scriptsize\boldmath{$\Omega$}}}(X^{\lambda_1,\cdots,\la _n}(\xi _1,\cdots,\xi _n;\al))
\right)=
\mathrm{v}^{\alpha}\left(
e^{\mbox{\scriptsize\boldmath{$\Omega$}}-\mbox{\scriptsize\boldmath{$\Omega$}}_{0}}
(X^{\lambda_1,\cdots,\la _n}(\xi _1,\cdots,\xi _n;\al))\right). 
\end{eqnarray*}
Together with (\ref{vacuum}) it gives immediately:
\begin{thm}\label{determinant}
The vacuum expectation value of $X^{\lambda_1,\cdots,\la _n}(\xi _1,\cdots,\xi _n;\al)$ 
is $0$ unless it has spin $0$.
In the latter case it is given by the determinant 
\begin{align}
\frac {\langle \mathrm{vac}| 
X^{\lambda_1,\cdots,\la _n}(\xi _1,\cdots,\xi _n;\al)|\mathrm{vac}\rangle}
{\langle \mathrm{vac}| q^{2\al S(0)}|\mathrm{vac}\rangle}
=
\det \left(
(\omega-\omega_{0})(\xi_{i^+_p}/\xi_{i^-_q},\al)\right)
_{1\le p,q\le m}\,.
\label{det}
\end{align}
Here the indices $i^\pm_p$ are defined by 
\be
&&\{i\mid \lambda_i=\pm\}=\{i^\pm_1,\cdots,i^\pm_m\}
\quad (i^\pm_1<\cdots<i^\pm_m). 
\en
\end{thm}

\begin{rem}
The VEV in the massive regime $0<q<1$ is also given by the formula \eqref{det},   
where the transcendental part $\omega_{\rm trans}$ in the definition of $\omega$ is replaced by 
\be
\omega_{\rm trans}(\z,\alpha)=2\z^\alpha
\left(1-(\z+\z^{-1})\sum_{n\ge1}(-1)^n
\Bigl(\frac{q^{(\alpha+2)n}\z}{1-q^{2n}\z^2}
+\frac{q^{(-\alpha+2)n}\z^{-1}}{1-q^{2n}\z^{-2}}\Bigr)\right)\,.
\en 
The other parts are the same as in the massless regime. 
\end{rem}

\subsection{The homogeneous case}\label{sec:VEV-homogeneous}


The calculation of VEV carries over to the homogeneous chain
as well,  on the basis of 
the following analog of Lemma \ref{kappatr}.
We shall restrict to the case of $\kappa=\alpha$. 

\begin{lem}\label{lem:alpha-trace}
We have 
\begin{eqnarray}
&& 
\mathbf{tr}^{\alpha}\tb^{*}(\zeta)(X)=2\,\mathbf{tr}^{\alpha}(X), 
\label{eq:homogeneous-alpha-trace1} \\
&&
\mathbf{tr}^{\alpha}{\bf b}^{*}(\zeta)(X)=\frac{1}{2\pi i}\oint_{\Gamma}
\omega_0(\zeta/\xi,\al)\,\mathbf{tr}^{\alpha}\cb(\xi)(X)\frac{d\xi^2}{\xi^2}\,, 
\label{eq:homogeneous-alpha-trace2} 
\end{eqnarray}
where $X \in \mathcal{W}_{\alpha}$ in 
\eqref{eq:homogeneous-alpha-trace1}, and 
$X \in \mathcal{W}_{\alpha+1}$ in 
\eqref{eq:homogeneous-alpha-trace2}. 
\end{lem}
\begin{proof}
Let $Y_{[k,m],c}\in M_{[k,m]}\otimes M_c$ and $k\le m<l$. 
Using Lemma \ref{EXPANSION} and noting that 
\be
\mathbf{tr}^\al_{[k,l]}({\bf r}_{i,j}(\, \cdot \,))=0   
\quad (i,j\in[k,l]),  
\en
we obtain modulo $(\z^2-1)^{l-m}$ 
\be
\mathbf{tr}^\al_{[k,l]}\bT_{c,[k,l]}(\z)(Y_{[k,m],c})
&\equiv&\mathbf{tr}^\al_{[k,l]}
\z^{\bbS_{m+1}}\bT_{m+1,[k,m]}(\z)(Y_{[k,m],m+1})
\\
&=&\mathbf{tr}^\al_{[k,m+1]}(Y_{[k,m],m+1})\,.
\en
Hence if $X=q^{2\alpha S(k-1)}X_{[k,m]}$ then we have
\be
{\half}\mathbf{tr}^\al\tb^*(\z,\al)(X)
=\lim_{l\to\infty}\mathbf{tr}^\al_{[k,l]}{\half}\Tr_c
\bT_{c,[k,l]}(\z)(q^{\al\sigma^3_c}X_{[k,m]})
=\mathbf{tr}^\al_{[k,m]}(X_{[k,m]}),
\en
proving \eqref{eq:homogeneous-alpha-trace1}. 
Similarly, by the reduction relation \eqref{BBTU}, 
\eqref{eq:homogeneous-alpha-trace2}
is reduced to the identity  \eqref{H=H'}, 
which has been proved in Lemma \ref{kappatr}. 
\end{proof}

Now let us introduce 
generating functions of quasi-local operators.  
Let 
$\epsilon=(\epsilon_{1}, \ldots , \epsilon_{n})$ be a sequence
in $\{0, +, -\}^{n}$.  
(Notice that $\emptyset$ is not allowed.)
We define 
$X^{\epsilon}(\zeta_{1}, \ldots , \zeta_{n}; \alpha)$ from
the primary field $q^{2\alpha S(0)}$ inductively by 
\begin{eqnarray*}
X^{\epsilon_{1}, \ldots , \epsilon_{n}}(\zeta_{1}, \ldots , \zeta_{n}; \alpha):=
\left\{ 
\begin{array}{ll}
{\bf b}^{*}(\zeta_{n})X^{\epsilon_{1}, \ldots , \epsilon_{n-1}}
(\zeta_{1}, \ldots , \zeta_{n-1}; \alpha) & (\epsilon_{n}=+), \\
{\bf c}^{*}(\zeta_{n})(-1)^{\mathbb{S}}X^{\epsilon_{1}, \ldots , \epsilon_{n-1}}
(\zeta_{1}, \ldots , \zeta_{n-1}; \alpha) & (\epsilon_{n}=-), \\
\half{\bf t}^{*}(\zeta_{n})X^{\epsilon_{1}, \ldots , \epsilon_{n-1}}
(\zeta_{1}, \ldots , \zeta_{n-1}; \alpha) & (\epsilon_{n}=0).
\end{array}
\right.
\end{eqnarray*} 
Even though the notations are similar, this object is different
from the fermionic basis $X^{\lambda_{1}, \ldots,\lambda_{n}}
(\xi_{1}, \ldots , \xi_{n}; \alpha)$ considered 
in the inhomogeneous case. 
The former is a power series in the variables 
$(\z_j^2-1)$, each coefficient being 
a quasi-local operator in $\W^{(\al)}$. 


Now define the operator $\Ob_{0}$ by \eqref{eq:def-omega-zero}. 
{}From the canonical commutation relations given by Theorem \ref{thcb*}, Theorem \ref{thbb*}
and Lemma \ref{lem:alpha-trace}, 
the functional 
$\mathrm{v}^{\alpha}(\cdot):={\bf tr}^{\alpha}\left(
 e^{\mbox{\scriptsize\boldmath{$\Omega$}}_{0}}(\cdot)\right)$ 
plays a role of the dual vacuum as in the inhomogeneous case. 
Then  a calculation similar to the one in Subsection \ref{sec:VEV-determinant-formula}
leads us to the following determinant formula 
for the vacuum expectation values: 

\begin{thm}\label{det-homog}
The vacuum expectation value of $X^{\epsilon_1,\cdots,\epsilon _n}(\z _1,\cdots,\z _n;\al)$ 
is $0$ unless it has spin $0$.
In the latter case it is given by the determinant 

\begin{align}
\frac {\langle \mathrm{vac}| 
X^{\epsilon}(\zeta_{1}, \ldots , \zeta_{n}; \alpha)|\mathrm{vac}\rangle}
{\langle \mathrm{vac}| q^{2\al S(0)}|\mathrm{vac}\rangle}
=
\det \left(
(\omega-\omega_{0})(\zeta_{i^+_p}/\zeta_{i^-_q},\al)\right)
_{1\le p,q\le m}\,.
\label{dethom}
\end{align}
Here the indices $i^{\pm}_{p}$ 
are defined as in Theorem \ref{determinant}. 
\end{thm}
 In spite of different meaning of the 
 operators in the left hand sides, the formulae (\ref{det}) and (\ref{dethom}) look identical. Why is it so? The formulae
 (\ref{dethom}) must be understood as generating
 function of VEV for quasi-local operators created by 
 $\bb ^*_p$, $\cb ^*_p$, $\tb ^*_p$ (the latter operators do not change
 VEV). The formula (\ref{dethom}) gives for such quasi-local
 operators determinants composed of Taylor coefficients
 of function $\omega-\omega_{0}$. On the other hand
 consider (\ref{det}). In order to take the homogeneous
 limit one has to construct suitable linear combinations of
 $X(\xi;\al)$'s, often with singular coefficients, and
 then send $\xi _j\to 1$. We could give examples, but
 for the lack of space we leave it for a future publication.
 The result will be again 
 determinants composed of Taylor coefficients
 of  $\omega-\omega_{0}$. Establishing exact correspondence between operators in homogeneous and 
 inhomogenious cases is related to the problem of completeness
 for the former case, again, it is left for a future publication.

\appendix

\section{Representations of $U_q\mathfrak{b}^+$ and $L$ operators}
\label{sec:appA}

In this appendix, 
we collect several facts about quantum affine algebras and 
$L$ operators used in the text. 

\subsection{Quantum algebras}

Consider the quantum affine algebra $U_q(\slth)$ 
with Chevalley generators $e_i,f_i,t_i=q^{h_i}$ 
($i=0,1$) and $q^{d}$, equipped with the coproduct $\Delta$ 
\be
&&\Delta(e_i)=e_i\otimes 1+t_i\otimes e_i, 
\ 
\Delta(f_i)=f_i\otimes t_i^{-1}+1\otimes f_i, 
\ 
\Delta(q^h)=q^h\otimes q^h \quad (h=h_i,d).
\en
We shall follow closely the notational 
convention in \cite{JMbk}.   
However, in this paper we denote the antipode by $S$: 
\be
&&S(e_i)=-t_i^{-1}e_i,\quad S(f_i)=-f_it_i,\quad 
S(q^h)=q^{-h}\quad (h=h_i,d).
\en
(We use $S$ only in this appendix; it is not to be confused 
with the total spin.)
We denote by $U'_q(\slth)$ the subalgebra generated by 
$e_{i}, f_{i},t_{i} \, (i=0, 1)$, and  
by $U_q\mathfrak{b}^+$ (resp. $U_q\mathfrak{b}^-$) 
the Borel subalgebra generated by $e_i,t_i$ (resp. $f_i,t_i$) 
($i=0,1$).

Let further $E,F,q^{H}$ be the standard generators of
$U_q(\slt)$. 
For $\z\in\C^\times$, the evaluation homomorphism 
$ev_\z:U'_q(\slth)\to U_q(\slt)$
is defined by 
\be
&&ev_\z(e_0)=\z F,\quad ev_\z(f_0)=\z^{-1} E,\quad
ev_\z(t_0)=q^{-H}, 
\\
&&ev_\z(e_1)=\z E,\quad ev_\z(f_1)=\z^{-1} F,\quad
ev_\z(t_1)=q^{H}\,. 
\en
A representation $\varpi:U_q(\slt)\to\End(W)$ 
gives rise to the evaluation representation 
$\varpi_\z=\varpi\circ ev_\z:U'_q(\slth)\to \End(W)$.  
We write the latter also as $W_\z$. 
Of frequent use is the case of two-dimensional representation 
$(\varpi,W)=(\pi^{(1)},V)$, $V=\C^2$, with  
\be
\pi^{(1)}(E)=\sigma^+,\quad \pi^{(1)}(F)=\sigma^-,
\quad \pi^{(1)}(q^H)=q^{\sigma^3}\,.
\en

\subsection{$q$-oscillator representations}

The $q$-oscillator algebra $Osc$ is an associative
$\C(q^\al)$-algebra with generators
$\ao,\ao^*,q^{D}$ and defining relations
\be
&&q^D \ao\, q^{-D}=q^{-1} \ao,
\quad q^D\, \ao^* q^{-D}=q\ \ao^*,
\quad
\\
&&\ao\ \ao^*=1-q^{2D+2},
\quad \ao^*\ao=1-q^{2D}\,.
\en
Representations of $Osc$ relevant to us are 
$\rho^\pm:Osc\to \End(W^\pm)$ defined by 
\bea
&&W^+=\oplus_{k\ge0}\C\ket{k},\quad 
W^-=\oplus_{k<0}\C\ket{k},
\label{eq:WApm}
\\
&&q^D\ket{k}=q^{k}\ket{k},\ 
\ao\ket{k}=(1-q^{2k})\ket{k-1},\
\ao^*\ket{k}=(1-\delta_{k,-1})\ket{k+1}\,.
\nn
\ena
We shall use the trace functional $\Tr(q^{2\al D}\cdot):Osc\to \C(q^\al)$
given as follows.   
For each $x\in Osc$ and $y\in\C$, the ordinary trace 
$\pm \Tr_{W^\pm}(y^{D} x)$ on $W^\pm$ 
is well-defined for sufficiently small $|y|^{\pm1}$,  
and gives the same rational function $g_x(y)$ in $y$. 
By definition, $\Tr(q^{2\al D}x)$ means $g_x(q^{2\al})\in \C(q^\al)$. 
Notice that $\Tr(q^{2\al D}\cdot)$ 
is a purely algebraic operation characterized as 
the unique linear map with the properties 
\be
&&\Tr(q^{2\al D}XY)=\Tr(q^{2\al D}q^{2\al d(X)}YX) 
\quad(X,Y\in Osc, q^D X q^{-D}=q^{d(X)}X),
\\  
&&\Tr(q^{2\al D}q^{mD})=\frac{1}{1-q^{2\al+m}}
\quad (m\in \Z). 
\en

There is a homomorphism of algebras 
$o_\z:U_q\mathfrak{b}^+\to Osc$ given by
\be
&&o_\z(e_0)=\frac{\z}{q-q^{-1}}\ao,
\quad
o_\z(e_1)=\frac{\z}{q-q^{-1}}\ao^*,
\quad
o_\z(t_0)=q^{-2D},\quad o_\z(t_1)=q^{2D}\,.
\en
We define representations 
$o^\pm_\z:U_q\mathfrak{b}^+\to \End(W^\pm)$ by 
\be
o_\z^+=\rho^+\circ o_\z, \quad 
o_\z^-=\rho^-\circ o_\z\circ\iota\,,  
\en
where $\iota$ denotes the involution $e_i\to e_{1-i}$, 
$t_i\to t_{1-i}$ of $U_q\mathfrak{b}^+$. 


\subsection{$L$ operators}

In the main text, we make use of 
$L$ operators associated with 
the auxiliary space $V_a$ or $W^+_A$,  
and the quantum space $V_j$. 
They are given as images
of the universal $R$ matrix of $U_q(\slth)$, 
\be
&&\cR=\overline{\cR}
\cdot q^{-(h_1\otimes h_1/2+c\otimes d+d\otimes c)},
\\
&&\overline{\cR}=1-(q-q^{-1})\sum_{i=0,1}e_i\otimes f_i+\cdots
\in U_q\mathfrak{b}^+\otimes U_q\mathfrak{b}^-\,. 
\en
Set $\mathcal{R}':=\mathcal{R} \cdot q^{c\otimes d+d\otimes c}$. 
From the standard product formula for the universal
$R$ matrix \cite{TKh} we obtain 
\begin{eqnarray*}
&& 
(\pi_\z\otimes\pi_\xi)\mathcal{R}'=
\rho(\z/\xi)\cdot L^\circ_{aj}(\z/\xi), 
\\
&& 
(o^\pm_\z \otimes\pi_\xi)\mathcal{R}'=
\sigma(\z/\xi)\cdot{L}^{\circ}_{Aj}{}^\pm(\z/\xi), 
\end{eqnarray*}
Here 
$L_{aj}^\circ(\z)$, 
${L}^{\circ}_{Aj}{}^+(\z)={L}^{\circ}_{Aj}(\z)$ 
are given by 
\eqref{eq:Rmat}, 
\eqref{eq:Lplus} 
in the text and  
\be
{L}^{\circ}_{Aj}{}^-(\z)=\sigma^1_j{L}^{\circ}_{Aj}(\z)\sigma^1_j.  
\en
The scalar factors $\rho(\z)$, $\sigma(\z)$ 
are formal power series in $\z^2$ subject to the relations
\bea
\rho(\z)=q^{-1/2}\frac{\sigma(q^{-1}\z)}{\sigma(\z)}\,,
\quad
\sigma(\z)\sigma(q^{-1}\z)=\frac{1}{1-\z^2}\,.
\label{SIGMA}
\ena
Their explicit formulas will not be used in this paper. 

\subsection{$R$ matrix for $W^+_{\z_1}\otimes W^+_{\z_2}$}
\label{subsec:RWW}

For generic $\z_1,\z_2$, the tensor product 
$W^+_{\z_1}\otimes W^+_{\z_2}$ is irreducible, and 
is isomorphic to the tensor product in the opposite order.  
As in the main text, let us use $A$ (resp. $B$) to label  
the first (resp. second) tensor component. 
An $R$ matrix intertwining the $L$ operators
\be
R_{A,B}(\z_1/\z_2)L_{A,j}(\z_1)L_{B,j}(\z_2) 
=L_{B,j}(\z_2)L_{A,j}(\z_1)R_{A,B}(\z_1/\z_2)
\en
can be written as
\bea
&&R_{A,B}(\z)=P_{A,B}h(\z,u_{A,B})\z^{D_A+D_B},  
\label{RAB}
\ena
where we have set 
$u_{A,B}=\ao_A^*q^{-2D_A}\ao_B$,  
and $h(\z,u)$ is the unique formal power series in $u$ 
satisfying 
\bea
&&
(1+\z u)h(\z,u)=(1+\z^{-1} u)h(\z,q^2u),
\label{hfcn1}\\
&&h(\z,u)=(1+\z^{-1}u)(1+q^{-2}\z u)h(q^{-2}\z,u)
\label{hfcn2}
\ena
and $h(\z,0)=1$.  

The $R$ matrix intertwining the fused $L$ operators 
\be
R_{\{a,A\},\{b,B\}}(\z_1/\z_2)
L_{\{A,a\},j}(\z_1)L_{\{B,b\},j}(\z_2)
=
L_{\{B,b\},j}(\z_2)L_{\{A,a\},j}(\z_1)
R_{\{a,A\},\{b,B\}}(\z_1/\z_2)
\en
is composed of \eqref{RAB} and the $L$ operators as 
\footnote{This matrix is not the same as 
a similar matrix given in (A.2), \cite{FB}.}
\be
R_{\{a,A\},\{b,B\}}(\z)=
(F_{a,A}F_{b,B})^{-1}L^-_{B,a}(\z^{-1})^{-1}
R_{a,b}(\z)R_{A,B}(\z)L^-_{A,b}(\z)F_{a,A}F_{b,B}\,.
\en
We have the Yang-Baxter relation with the monodromy matrices
with twist, 
\bea
&&R_{\{a,A\},\{b,B\}}(\z_1/\z_2)
\bT_{\{A,a\}}(\z_1,\al)\bT_{\{B,b\}}(\z_2,\al)
\label{YBabAB}
\\
&&\quad =
\bT_{\{B,b\}}(\z_2,\al)\bT_{\{A,a\}}(\z_1,\al)
R_{\{a,A\},\{b,B\}}(\z_1/\z_2)\,.\nn
\ena

Taking the standard basis $v_+,v_-$ of $V=\C^2$, 
we regard it as a $4\times 4$ matrix with respect to the basis  
$v_+\otimes v_+,v_+\otimes v_-,v_-\otimes v_+,v_-\otimes v_-$ 
of $V_a\otimes V_b$, and set 
\bea
R_{\{a,A\},\{b,B\}}(\z)=
\begin{pmatrix}
R_{11} & 0 & 0 & 0 \\
R_{21}&R_{22} &0 &0\\
R_{31}&0 &R_{33} &0\\
R_{41}&R_{42} &R_{43} &R_{44}\\
\end{pmatrix}\,.
\label{RWW}
\ena
The entries are given as follows. 
\begin{align}
&R_{11}= -\z^2 q^{D_A}R_{A,B}(\z)q^{-D_B}, 
\qquad\qquad\qquad\ \ \  R_{21}=\frac{q\z^3}{1-q^2\z^2}
R_{A,B}(q\z)\ao_A^*q^{-D_A-D_B},\nn\\
&R_{22}=-\frac{q\z^2}{1-q^2\z^2}
q^{-D_A}R_{A,B}(q^2\z)q^{-D_B},\nn\\\nn\\
& R_{33}=-\z(q^{-1}\z-q\z^{-1})
q^{D_A}R_{A,B}(q^{-2}\z)q^{D_B},
\quad R_{43}=-\z R_{A,B}(q^{-1}\z)\ao_A^*q^{-D_A+D_B},\nn\\
&R_{44}=q^{-D_A}R_{A,B}(\z)q^{D_B},
\nn\\\nn\\
&R_{31} =-q \z\,q^{-2D_B}\ao^*_BR_{A,B}(q^{-1}\z)q^{D_A+D_B},
\qquad\ \  R_{41}=\frac{q\z^2}{1-q^2\z^2}
q^{-D_B}\ao^*_BR_{A,B}(\z)\ao_A^*q^{-D_A},
\nn\\
&R_{42}=-\frac{\z}{1-q^2\z^2}
\ao^*_B R_{A,B}(q\z)q^{-D_A-D_B}.\nn
\end{align}
\section{Proof of Lemma \ref{APPB}}\label{sec:appB}

In this section we denote $J=[k,m]$ and $K=[m+1,l]$.
We decompose the left hand side of \eqref{BBTU} as
\be
&&\bb^*_{J\sqcup K}(\z,\al)({X}_J)=\bb^*(\z,\al)({X}_J)+I+II+III+IV,
\en
where
\begin{eqnarray*}
I&=&
\left(\fb_{J\sqcup K}(q\z,\al)-\fb(q\z,\al)\right)(X_J)\,, \\
II&=&
\left(\fb_{J\sqcup K}(q^{-1}\z,\al)-\fb(q^{-1}\z,\al)\right)(X_J)\,, \\ 
III&=&-\tb^*_{J\sqcup K}(\z,\al)\left(\fb_{J\sqcup K}(\z,\al)
-\fb(\z,\al)\right)(X_J)\,,\\
IV&=&\left(\tb^*_{J\sqcup K}(\z,\al)-\tb^*(\z,\al)\right)
\fb(\z,\al)(X_J).
\end{eqnarray*}
Note that
\be
\Tr_c\left\{\bT_{c,K}(\z)\left(\gb_c(\z,\al)-\ub_c(\z,\al)\right)(X_J)\right\}
=\bb^*(\z,\al)(X_J)+IV.
\en
We want to show that
\bea
\Tr_c\left\{\bT_{c,K}(\z)\ub_c(\z,\al)(X_J)\right\}=I+II+III.
\label{Truc}
\ena
The main part of the proof is computing $III$.
Set
\be
\tilde {X}_{\{a,A\},J}=\bT_{\{a,A\}}(\z,\al)\z^{\al-\bS}(q^{-2S_J}X_J).
\en
Separating the $K$ part from the $J$ part, we have
\be
&&III=\Tr_{A,a}\left\{\theta_K(Z_{a,A,K})
\tilde {X}_{\{a,A\},J}\right\},\\
&&Z_{a,A,K}=F_{a,A}^{-1}T_{a,K}(q^{-1}\z)C_{A,K}(\z)q^{S_K}F_{a,A}
T_{\{a,A\},K}(q\z)^{-1}\\
&&=\begin{pmatrix}1&\ab_A\\&1\end{pmatrix}
\begin{pmatrix}a_K(q^{-1}\z)&b_K(q^{-1}\z)\\c_K(q^{-1}\z)&d_K(q^{-1}\z)
\end{pmatrix}
C_{A,K}(\z)q^{S_K}
\begin{pmatrix}1&-\ab_A\\&1\end{pmatrix}\\
&&\quad\times
\begin{pmatrix}
1&0\\
-q^{-S_K}T_{A,K}(\z)^{-1}C_{A,K}(q\z)&1
\end{pmatrix}
\begin{pmatrix}
q^{S_K}T_{A,K}(q^2\z)^{-1}&0\\
0&q^{-S_K}T_{A,K}(\z)^{-1}
\end{pmatrix}
\en
We use several identities.

Fusion relation:
\bea
&&\begin{pmatrix}1&\ab_A\\&1\end{pmatrix}
\begin{pmatrix}a_K(\z)&b_K(\z)\\c_K(\z)&d_K(\z)\end{pmatrix}
T_{A,K}(\z)
\begin{pmatrix}1&-\ab_A\\&1\end{pmatrix}\label{FUSION}\\
&&\quad=
\begin{pmatrix}
T_{A,K}(q\z)q^{-S_K}&0\\
C_{A,K}(\z)&T_{A,K}(q^{-1}\z)q^{S_K}
\end{pmatrix},\nn
\ena
or equivalently,
\bea
&&T_{A,K}(\z)=
\begin{pmatrix}d_K(q^{-1}\z)&-b_K(q^{-1}\z)\\-c_K(q^{-1}\z)&a_K(q^{-1}\z)
\end{pmatrix}
\begin{pmatrix}1&-\ab_A\\&1\end{pmatrix}\label{FUSION2}\\
&&\times\begin{pmatrix}
T_{A,K}(q\z)q^{-S_K}&0\\
C_{A,K}(\z)&T_{A,K}(q^{-1}\z)q^{S_K}
\end{pmatrix}
\begin{pmatrix}1&\ab_A\\&1\end{pmatrix}.\nn
\ena

Crossing symmetry:
\bea
\begin{pmatrix}d_K(q^{-1}\z)&-b_K(q^{-1}\z)\\-c_K(q^{-1}\z)&a_K(q^{-1}\z)
\end{pmatrix}
\begin{pmatrix}a_K(\z)&b_K(\z)\\c_K(\z)&d_K(\z)\end{pmatrix}
=\begin{pmatrix}1&0\\0&1\end{pmatrix}.\label{CROSSING}
\ena

We define $x^3_K(\z),x^\pm_K(\z)$ by
\bea
&&\begin{pmatrix}a_K(q^{-1}\z)&b_K(q^{-1}\z)\\c_K(q^{-1}\z)&d_K(q^{-1}\z)
\end{pmatrix}
\begin{pmatrix}d_K(\z)&-b_K(\z)\\-c_K(\z)&a_K(\z)\end{pmatrix}\label{DEFX}\\
&&\quad=1+x^3_K(\z)\s^3+x^+_K(\z)\s^++x^-_K(\z)\s^-.\nn
\ena

\medskip\noindent
$(Z_{a,A,K})_{1,2}$:
From the $(2,1)$ element in \eqref{FUSION} we obtain
$C_{A,K}(\z)=c_K(\z)T_{A,K}(\z)$. Using this we obtain
\be
&&(Z_{a,A,K})_{1,2}=
-(a_K(q^{-1}\z)+\ab_Ac_K(q^{-1}\z))c_K(\z)T_{A,K}(\z)\ab_AT_{A,K}(\z)^{-1}\\
&&\quad+(b_K(q^{-1}\z)+\ab_Ad_K(q^{-1}\z))c_K(\z).
\en
From the $(2,1)$ element in \eqref{CROSSING} we obtain
$a_K(q^{-1}\z)c_K(\z)=c_K(q^{-1}\z)a_K(\z)$.
From the $(1,2)$ element in \eqref{FUSION} we obtain
\be
\left(a_K(\z)+\ab_Ac_K(\z)\right)T_{A,K}(\z)\ab_AT_{A,K}(\z)^{-1}=
b_K(\z)+\ab_Ad_K(\z).
\en
Using these equalities we have
\be
(Z_{a,A,K})_{1,2}&=&b_K(q^{-1}\z)c_K(\z)-c_K(q^{-1}\z)b_K(\z)
+\ab_A\left(d_K(q^{-1}\z)c_K(\z)-c_K(q^{-1}\z)d_K(\z)\right)\\
&=&-x^3_K(\z)-\ab_Ax^-_K(\z).
\en

\medskip\noindent
$(Z_{a,A,K})_{2,2}$:
The calculation is similar, but we use the (2,2) element in \eqref{FUSION}.
We obtain
\be
(Z_{a,A,K})_{2,2}=-x^-_K(\z)+C_{A,K}(q^{-1}\z)q^{S_K}T_{A,K}(\z)^{-1}.
\en

\medskip\noindent
$(Z_{a,A,K})_{1,1}$:
Reducing a part of the computation to that for $(Z_{a,A,K})_{1,2}$, we obtain
\be
&&(Z_{a,A,K})_{1,1}=
\left(a_K(q^{-1}\z)+\ab_Ac_K(q^{-1}\z)\right)
C_{A,K}(\z)q^{2S_K}T_{A,K}(q^2\z)^{-1}\\
&&\quad+\left(x^3_K(\z)+\ab_Ax^-_K(\z)\right)
C_{A,K}(q\z)q^{S_K}T_{A,K}(q^2\z)^{-1}.
\en
The $(2,1)$ element of \eqref{FUSION2} can be rewritten as
\be
\left(a_K(q^{-1}\z)+\ab_Ac_K(q^{-1}\z)\right)C_{A,K}(\z)
=c_K(q^{-1}\z)T_{A,K}(q\z)q^{-S_K}.
\en
Multiply $\left(c_K(q^{-1}\z),d_K(q^{-1}\z)\right)$ from the left
to \eqref{FUSION2} with $\z$ replaced by $q\z$. Use \eqref{DEFX} and
take the first component. The result is
\be
c_K(q^{-1}\z)T_{A,K}(q\z)=x^-_K(\z)\left(T_{A,K}(q^2\z)q^{-S_K}
-\ab_AC_{A,K}(q\z)\right)+(1-x^3_K(\z))C_{A,K}(q\z).
\en
Combining all these we obtain
\be
(Z_{a,A,K})_{1,1}=x^-_K(\z)+C_{A,K}(q\z)q^{S_K}T_{A,K}(q^2\z)^{-1}.
\en
Now the proof is easy. Recall that $\tilde {X}_{\{a,A\},J}$ is lower triangular.
If we substitute
\bea
&&Z_{a,A,K}=
\begin{pmatrix}
x^-_K(\z)&-x^3_K(\z)-\ab_Ax^-_K(\z)\\
*&x^-_K(\z)
\end{pmatrix}
+D_{a,A,K},
\label{ucIII}
\\
&&D_{a,A,K}=
\begin{pmatrix}
C_{A,K}(q\z)q^{S_K}T_{A,K}(q^2\z)^{-1}&0\\
0&C_{A,K}(q^{-1}\z)q^{S_K}T_{A,K}(\z)^{-1}
\end{pmatrix},
\nn
\ena
we obtain $III$.
The second diagonal term cancels with $I+II$.
The last trick is introducing another auxiliary space in order to
write the formula without using $\theta_K$. In fact, we have
\be
&&\Tr_c\left\{\bT_{c,K}(\z)\ub_c(\z,\al)({X}_J)\right\}
=\Tr_{A,a,c}\theta_K(1+x^3_K(\z)\s^3+x^+_K(\z)\s^++x^-_K(\z)\s^-)\\
&&\quad\times\left(-\frac12\s^3_c\s^+_a+\s^+_cs^3_a-\ab_a\s^+_a)\right)
\tilde {X}_{\{a,A\},J}\\
&&\quad=\Tr_{A,a}\theta_K\left(x^-_K(\z)\s^3_a-(x^3_K(\z)+\ab_Ax^-_K(\z))\s^+_a
\right)\tilde {X}_{\{a,A\},J}.
\en
The proof is over.
\section{Bazhanov-Lukyanov-Zamolodchikov construction}\label{sec:appC}

As has been mentioned in the main text, 
there is no $R$-matrix which intertwines the tensor products 
$W_{\z_1}^+\otimes W_{\z_2}^-$ 
and $W_{\z_2}^-\otimes W_{\z_1}^+$. 
In order to calculate the commutation 
relations between operators such as $\bb$ and $\bb ^*$,  
it is necessary to find some substitute for the $R$-matrix. 
In this appendix we explain that the 
Bazhanov-Lukyanov-Zamolodchikov (BLZ) 
construction \cite{BLZ} offers a way to do that. 
We shall use the notation (\ref{Lminus})
for the $L$ operators $L^{\pm}_{x,j}$.

\subsection{Shifted Verma modules}

The key point of the BLZ construction is to 
relate the tensor products $W^\pm\otimes W^\mp$ with 
Verma modules of $U_q(\slt)$.  
Let $\varpi^{(\Lambda)}:U_q(\slt)\to \End(V(\Lambda))$ 
be the lowest weight Verma representation
with lowest weight $\Lambda$, 
and let 
$\varpi^{(\Lambda)}_\zeta=\varpi^{(\Lambda)}\circ ev_\z$ 
be the evaluation representation.  
Up to a scalar multiple, the corresponding $L$-operator 
$(\varpi^{(\Lambda)}_{\z}\otimes\pi^{(1)}_\xi)\mathcal{R}'$ 
is given by 
\bea
L(\eta)=\begin{pmatrix}
1-\eta^2q^{H+1} & -(q-q^{-1})\eta F\\
-(q-q^{-1})\eta E & 1-\eta ^2 q^{-H+1}
\end{pmatrix}
q^{-H \sigma^3/ 2 }\,, 
\label{Lverma}
\ena
where $\eta=\z/\xi$. 
Where necessary, 
we use the letter $v$ as a label for the Verma module.
As usual the monodromy matrix is defined by 
\be
&&\bT_{v,[k,l]}(\z,\al)(X_{[k,l]})
=T_{v,[k,l]}(\z)\ q^{\al H}X_{[k,l]}\ T_{v,[k,l]}(\z)^{-1},
\\
&&T_{v,[k,l]}(\z)=L_{v,l}(\z/\xi_l)\cdots L_{v,k}(\z/\xi_k). 
\en

For our purposes it is convenient to twist 
$\varpi^{(\Lambda)}_\eta$ 
by the automorphism of $U_q\mathfrak{b}^+$,  
\be
\gamma_m(e_0)=e_0,\ \gamma_m(e_1)=e_1,\ 
\gamma_m(t_0)=q^{-m}t_0,\ \gamma_m(t_1)=q^mt_1\,, 
\en 
where $m\in\C$ is a parameter.  
We call 
$\varpi^{(\Lambda)}_{\eta,m}=
\varpi^{(\Lambda)}_\eta\circ\gamma_m
:U_q\mathfrak{b}^+\to \End(V(\Lambda))$ 
shifted Verma representation,  
and denote it by $V_{\eta,m}(\Lambda)$.  
In an appropriate basis $\{v_j\}_{j\ge 0}$,  
the generators act on $V_{\eta,m}(\Lambda)$ by   
\bea
&&(q-q^{-1})^2 e_0 v_j
=\eta^2q^{-\Lambda+1}
(q^{\Lambda-H-2}-1)(q^{\Lambda+H}-1)v_{j-1},
\quad 
e_1 v_j=v_{j+1},
\label{shiftVerma1}\\
&&t_0 v_j=q^{-H-m}v_j,
\quad
t_1 v_j=q^{H+m}v_j\,,
\label{shiftVerma2}
\ena
where $q^Hv_j=q^{\Lambda+2j}v_j$ and $v_{-1}=0$. 
The shift parameter $m$ enters the corresponding 
$L$ operator simply via 
\be
(\varpi^{(\Lambda)}_{\eta,m}\otimes\pi^{(1)}_\xi)\mathcal{R}'
=\mbox{ (scalar) }\cdot L_{v,j}(\eta)q^{-m \sigma^3/2}.
\en

\subsection{Filtrations of the tensor product 
$W^\pm\otimes W^\mp$}\label{appC2}

Let us introduce the following elements of $Osc^{\otimes 2}$ 
which will play a role in the sequel. 
\bea
&&U_{A,B}(\z)=\z \ab_A^*+\ab _Bq^{2D_A}\,,
\label{UAB}
\\
&&V_{A,B}(\z)= \z \ab_B^*+\ab _Aq^{2D_B}
\,,
\label{VAB}
\\
&&Y_{A,B}(\z)=(\z q^{2}-\ab _A\ab_B)q^{2D_A}\,,
\label{YAB}
\\
&&Z_{A,B}(\z)=\z^{-1}q^{2D_B+2}-\ab _A^*\ab _B^*q^{-2D_A}\,.
\label{ZAB}
\ena
These operators 
appear as matrix elements of products of $L$-operators,  
\begin{align}
&L^{\circ}_A{}^+(\z _1)L^{\circ}_B{}^-(\z _2)
\label{LL^-}
\\&
=\begin{pmatrix}
1-\z_1\z_2Y_{A,B}(\z)
&-\z _1(1-\z ^{-1}q^{-2}Z_{A,B}(\z))\ab_A-
c(\z_1,\z_2)V_{A,B}(\z)\\
-\z_2 U_{A,B}(\z)
&1-\z_1\z_2Z_{A,B}(\z)
\end{pmatrix}
q^{-(D_A-D_B)\sigma ^3}
\,,\nn
\\
&L_B^{\circ}{}^-(\z _2)L_A^{\circ}{}^+
(\z _1)\label{LL^+}
\\&
=\begin{pmatrix}
1-\z_1\z_2Z_{B,A}(\z^{-1})
&-\z_1U_{B,A}(\z^{-1})
\ \\
-\z _2(1-\z q^{-2}Z_{B,A}(\z^{-1}))\ab_B-
c(\z_2,\z_1)V_{B,A}(\z^{-1})
&1-\z_1\z_2Y_{B,A}(\z^{-1})
\end{pmatrix}
q^{(D_B-D_A)\sigma ^3}
\,.\nn
\end{align}
Here we have set $\z=\z_1/\z_2$, 
and $c(\z_1,\z_2)=\z_1^{-1}\z_2^2(1-\z _1^2q^2)$. 

Let us list the 
commutation relations that are relevant to us.

\noindent
$\bullet$ $U_{A,B}$, $Y_{A,B}$, $Z_{A,B}$, $\ao_A$
among themselves:
\begin{align}
&Y_{A,B}(\z)U_{A,B}(\z)=q^2U_{A,B}(\z)Y_{A,B}(\z)\,,\nn\\
&Z_{A,B}(\z)U_{A,B}(\z)=q^{-2}U_{A,B}(\z)Z_{A,B}(\z)\,,\nn\\
& \ab _AU_{A,B}(\z)-q^2 U_{A,B}(\z)\ab _A= \z (1-q^2),
\label{commut1}\\
&Y_{A,B}(\z)\ao_A=q^{-2}\ao_A Y_{A,B}(\z)\nn\\
&Z_{A,B}(\z)\ao_A=
q^2\ao_A Z_{A,B}(\z)+q^2(1-q^2)\z^{-1}V_{A,B}(\z)\,,\nn\\
&Y_{A,B}(\z)Z_{A,B}(\z)=q^2 -\z ^{-1}q^4U_{A,B}(\z)V_{A,B}(\z)\,,\nn\\
&Z_{A,B}(\z)Y_{A,B}(\z)=q^2  -\z ^{-1}q^2U_{A,B}(\z)V_{A,B}(\z)\,.\nn
\end{align}
$\bullet$ $U_{A,B}$, $Y_{A,B}$, $Z_{A,B}$, $\ab_A$, $\ab_B$
 with $V_{A,B}$:
\bea
&&[V_{A,B}(\z),X]=0\quad
\mbox{for $X=U_{A,B}(\z), Y_{A,B}(\z), Z_{A,B}(\z), \ao_A$},
\label{commut2}\\
&&V_{A,B}(\z)\ab _B-q^{-2}\ab_BV_{A,B}(\z)=\z(1-q^{-2})
\,.
\nn
\ena
$\bullet$ 
$U_{A,B}$, $Y_{A,B}$, $Z_{A,B}$, $\ab_A$ with $\ab_B$:
\bea
&&[U_{A,B}(\z),\ab_B]=[Y_{A,B}(\z),\ab_B]=[\ao_A,\ao_B]=0,
\label{commut3}\\
&&Z_{A,B}(\z)\ab _B=\ab _BZ_{A,B}(\z)+\z^{-1}(1-q^2)
U_{A,B}(\z )q^{2(D_B-D_A)}\,.
\nn
\ena

The following result can be extracted from \cite{BLZ}. 
\begin{lem}\label{BLZfil}
Set 
\bea
\z=\frac{\z_1}{\z_2},\quad q^{\Lambda}=q\z\,.
\ena
The tensor product $W_{\z_1}^+\otimes W_{\z_2}^-$
has an increasing filtration by $U_q\mathfrak{b}^+$-submodules  
\bea
&&\{0\}=W^{(-1)}_L\subset W^{(0)}_L\subset W_L^{(1)}
\subset \cdots \subset W_L^{(m)}\subset\cdots
\subset W_{\z_1}^+\otimes W_{\z_2}^-\,,
\label{increase}\\ 
&&\bigcup\limits _{m=-1}^{\infty} W_L^{(m)}
=W_{\z_1}^+\otimes W_{\z_2}^-, \nn
\ena
such that 
each subquotient is isomorphic to a shifted Verma module
\bea
\iota_L: 
W_L^{(m)}/ W_L^{(m-1)}\overset{\sim}{\to}
V_{\sqrt{\z _1\z _2},2m}(\Lambda)\,. 
\label{WLV}
\ena

The tensor product $W_{\z_2}^-\otimes W_{\z_1}^+$
in the opposite order has a decreasing filtration by  
$U_q\mathfrak{b}^+$-submodules  
\bea
&&W_{\z_2}^-\otimes W_{\z_1}^+
=W_R^{{(-1)}}\supset W_R^{(0)}
\supset \cdots \supset W_R^{(m)}\supset\cdots
\,, 
\label{decrease}\\
&&
\bigcap\limits_{l=-1}^{\infty} W_R^{(m)}=0 , \nn
\ena
such that 
each subquotient is isomorphic to a shifted Verma module
\bea
\iota_R:W_R^{(m-1)}/W_R^{(m)}\overset{\sim}{\to}
V_{\sqrt{\z _1\z _2},2m}(\Lambda)\,.
\label{WRV}
\ena
\end{lem}
\begin{proof}
The vector space $W^+\otimes W^-$ has the following basis
$$
e_{j,p}=U_{A,B}(\z)^j\ab _B^p \ |0\rangle \otimes |-1\rangle
\quad (j,p\in \Z_{\ge 0})\,.
$$
Let $W_L^{(m)}$ denote the linear span of 
$e_{j,p}$ with $j\ge 0$ and $p\le m$. 
Introduce the operator $H$ by 
$q^He_{j,m}=\z  q^{2j+1}e_{j,m}$. 
A direct calculation using \eqref{commut1}--\eqref{commut3} 
shows that (with $\star$ denoting an irrelevant constant)  
\begin{align}
&U_{A,B}(\z)e_{j,m}=e_{j+1,m}\,,\nn\\
&Y_{A,B}(\z)e_{j,m}=q^{H+1}e_{j,m}\,,\nn\\
&Z_{A,B}(\z)e_{j,m}=q^{-H+1}e_{j,m}+\star\ e_{j+1,m-1}\,,\nn\\
&
(1-\z ^{-1}q^{-2}Z_{A,B}(\z))
\ab _A e_{j,m}=\z^{-1}(\z q^{-H-1}-1)(\z q^{H+1}-1)
e _{j-1,m}+\star\ e_{j,m-1}\,,\nn\\
&V_{A,B}(\z )e_{j,m}=\star \ e_{j,m-1}\nn\\
&q^{2(D_A-D_B)}e_{j,m}=\z ^{-1}q^{H+2m+1}e_{j,m}
\nn\,.
\end{align}
In view of the relations
\be
&&(q-q^{-1})\Delta(e_0)=\z_1(1-q^{-2}\z^{-1}Z_{A,B}(\z))\ao_A
+\z_1^{-1}\z_2^2 V_{A,B}(\z),
\\
&&(q-q^{-1})\Delta(e_1)=\z_2 U_{A,B}(\z),
\\
&&\Delta(t_0)^{-1}=\Delta(t_1)=q^{2(D_A-D_B)},
\en
we see that $W^{(m)}_L$ are $U_q\mathfrak{b}^+$-submodules.
Comparing these with \eqref{shiftVerma1}, \eqref{shiftVerma2}
we obtain the first statement of Lemma. 

Similarly, for $W_{\z_2}^-\otimes W_{\z_1}^+$ 
we introduce a basis 
$$
f_{j,p}=\((1-\z q^{-2}
Z_{B,A}(\z ^{-1}))\ab_B\)^jV_{B,A}(\z ^{-1})^p
|-1\rangle\otimes|0\rangle\,.
$$
Let $W^{(m)}_R$ be the linear span of $f_{j,p}$ with 
$j\ge 0$ and $p>m$.
Setting $q^{H}f_{j,m}=\z q^{2j+1}f_{j,m}$, we have 
\begin{align}
&(1-\z q^{-2}Z_{B,A}(\z ^{-1}))\ \ab _Bf_{j,m}
=f_{j+1,m}\,,\\
& 
U_{B,A}(\z ^{-1})f_{j,m}=
\z^{-1}(\z q^{-H-1}-1)(\z q^{H+1}-1)f _{j-1,m}\,,
\nn\\
&Z_{B,A}(\z ^{-1})f_{j,m}=q^{H+1}f_{j,m}
+\star f_{j-1,m+1}\,,\nn\\
&Y_{B,A}(\z ^{-1})f_{j,m}=q^{-H+1}f_{j,m}\,,\nn\\
&V_{B,A}(\z ^{-1})f_{j,m}=f_{j,m+1}\,,\nn\\
&q^{2(D_A-D_B)}f_{j,m}=\z ^{-1}q^{H+2m+1}f_{j,m}\,.\nn
\end{align}
The second statement follows from these. 
\end{proof}

Let us say that an operator 
$\mathcal{X}^L(\z)\in \End(W^+\otimes W^-)$
(resp. $\mathcal{X}^R(\z)\in \End(W^-\otimes W^+)$) 
is left (resp. right) admissible if it preserves the filtration 
\eqref{increase} (resp. \eqref{decrease}). 
The operators 
\be
U_{A,B}(\z), V_{A,B}(\z), Y_{A,B}(\z), Z_{A,B}(\z),  \ab_A, 
q^{2(D_A-D_B)} 
\en
are left admissible, and 
\be
U_{B,A}(\z^{-1}), V_{B,A}(\z^{-1}),  
Y_{B,A}(\z^{-1}), Z_{B,A}(\z^{-1}),  \ab_B, q^{2(D_A-D_B)} 
\en
are right admissible. 
By the isomorphisms \eqref{WLV},\eqref{WRV}, 
we have the correspondence of operators on each subquotient, 
\be
\iota_L\circ\mathcal{X}^L(\z)\circ\iota_L^{-1}=\mathcal{X}(\z)
=\iota_R\circ\mathcal{X}^R(\z)\circ\iota_R^{-1}, 
\en
where $\mathcal{X}^L(\z)$, 
$\mathcal{X}^R(\z)$ and $\mathcal{X}(\z)$ 
are related to each other via the following table \ref{TABLE}.  

\begin{table}[h]
\caption{Correspondence of operators:
$W^+_{\z_1}\otimes W^{-}_{\z_2}$ (left), 
$V_{\z}(\Lambda)$ (middle), 
$W^{-}_{\z_2}\otimes W^{+}_{\z_1}$ (right)}
\label{TABLE}
\begin{center}
\begin{tabular}{c|c|c}
\hline
$\mathcal{X}^L$ & $\mathcal{X}$ & $\mathcal{X}^R$ \\
\hline
$V_{A,B}(\z)$ &$0$ & $\star$ \\
 $\star$            &$0$ & $V_{B,A}(\z^{-1})$ \\
$Y_{A,B}(\z)$ &$q^{H+1}$ & $Z_{B,A}(\z^{-1})$ \\
$Z_{A,B}(\z)$ &$q^{-H+1}$ & $Y_{B,A}(\z^{-1})$ \\
$\sqrt{\z}(1-\z^{-1}q^{-2}Z_{A,B}(\z))\ab_A$ &
$(q-q^{-1})F$&  $\sqrt{\z}\,U_{B,A}(\z ^{-1})$\\
$\sqrt{\z}^{-1}U_{A,B}(\z)$ & $(q-q^{-1})E$ & 
$\sqrt{\z}^{-1}(1-\z q^{-2}Z_{B,A}(\z ^{-1}))\ab_B$ \\
\hline 
\end{tabular}
\end{center}
\end{table}

\subsection{Exchange relations under the trace}

Lemma \ref{BLZfil} has two corollaries which are important to us. 
We shall omit writing the intervals $[k,l]$. 
\begin{cor}
If $\mathcal{X}^L(\z)$ is left admissible, then 
\bea
&&N(\al-\mathbb{S})
\Tr_{A,B}\left\{\mathcal{X}^L(\z)\ 
\mathbb{T}^+_{A}(\z_1,\al)\mathbb{T}^-_{B}(\z_2,\al)\right\}
\z ^{\al -\mathbb{S}}
\nn\\ 
&&\quad =-\Tr_{V(\Lambda)} 
\left\{\mathcal{X}(\z)\ 
\mathbb{T}_{v}(\sqrt{\z_1\z _2},\al)\right\}
\label{1quasicom}\\ 
&&\quad = N(\al -\mathbb{S})\Tr_{A,B}\left\{\mathcal{X}^R(\z)\ 
\mathbb{T}^-_{B}(\z_2,\al)\mathbb{T}^+_{A}(\z_1,\al)\right\}
\z ^{\al-\mathbb{S}}\,.
\nn
\ena
The operators $\mathcal{X}(\z)$, $\mathcal{X}^R(\z)$ are obtained from 
$\mathcal{X}^L(\z)$ via the table \eqref{TABLE}. 
\end{cor}\label{corC3}
\begin{proof}
Comparing \eqref{LL^-}, \eqref{LL^+} with \eqref{Lverma}, 
we have the relations on each subquotient 
$V_{\sqrt{\z_1\z_2},2m}(\Lambda)$
\be
&& \iota_L\circ\bT^+_{A}(\z_1,\al)\bT^-_{B}(\z_2,\al)
\circ\iota_L^{-1}
=\bT_{v}(\sqrt{\z_1\z_2},\al)
\z^{-\al+\bbS}q^{(\al-\bbS)(2m+1)}\,,
\\
&& \iota_R\circ\bT^-_{B}(\z_2,\al)
\bT^+_{A}(\z_1,\al)\circ\iota_R^{-1}
=\bT_{v}(\sqrt{\z_1\z_2},\al)
\z^{-\al+\bbS}q^{(\al-\bbS)(2m+1)}\,.
\en
Taking traces and summing geometric series over $m$, 
we obtain the desired relations. 
The minus sign enters because $\Tr_B=-\Tr_{W^-_B}$
(see the definition of the trace functional $\Tr$ 
after \eqref{eq:WApm} in Appendix \ref{sec:appA}).
\end{proof}

Often it becomes necessary to compare the traces which have 
multipliers $N(\al-\bbS)$ with shifted arguments. 
The following Lemma tells how to do that. 
\begin{cor}
For left admissible $\mathcal{X}_{A,B}(\z)$, we have 
\begin{align}
&N(\al-\mathbb{S}+1)\Tr \(q^{2D_A-2D_B}\mathcal{X}_{A,B}(\z)
\mathbb{T}^+_{A}(\z_1,\al)\mathbb{T}^-_{B}(\z _2,\al)\)
\label{adjust}
\\ &=\z ^{-1}
N (\al-\mathbb{S})\Tr \(q^{-1}Y_{A,B}(\z)\mathcal{X}_{A,B}(\z)
\mathbb{T}^+_{A}(\z_1,\al)\mathbb{T}^-_{B}(\z _2,\al)\)
\,. 
\nn
\end{align}
\end{cor}


Let us proceed to traces of products of 
fused monodromy matrices 
$\mathbb{T}^+_{\{a,A\}}(\z _1,\al)$ 
and $\mathbb{T}^-_{\{b,B\}}(\z_2,\al)$. 
As it turns out, proper analogues of 
the left and right admissible operators in this case 
are elements of $M_a\otimes M_b$ of the form
\begin{align}
&q^{\sigma ^3_bD_A}\mathcal{X}^L_{A,B,a,b}(\z)q^{-\sigma ^3_bD_A}
\,,
\qquad
q^{-\sigma ^3_aD_B}\mathcal{X}^R_{a,b,A,B}(\z)q^{\sigma ^3_aD_B}
\,,\nn
\end{align}
where each entry of 
$\mathcal{X}^L_{A,B,a,b}(\z)$ (resp. $\mathcal{X}^R_{a,b,A,B}(\z)$) 
is a left (resp. right) admissible operator 
in the sense defined already. 
Let us explain the origin of this definition 
taking as an example the case of right admissible operators, 
$$
\Tr \(q^{m(2D_A-2D_B+\sigma ^3_a+\sigma ^3_b)}
q^{-\sigma ^3_aD_B}\mathcal{X}^R_{A,B,a,b}(\z)q^{\sigma ^3_aD_B}
\mathbb{T}_{\{b,B\}}^-(\z _2,\al)
\mathbb{T}_{\{a,A\}}^+(\z _1,\al)\)\,.
$$
First undo the fusion, 
\begin{align}
&\mathbb{T}_{\{b,B\}}^-(\z _2,\al)
\mathbb{T}_{\{a,A\}}^+(\z _1,\al)
\nn\\
&=(F^+_{a,A}F^-_{b,B})^{-1}
\mathbb{T}_{b}(\z _2,\al)\mathbb{T}_{B}^-(\z _2,\al)
\mathbb{T}_{a}(\z _1,\al)\mathbb{T}_{A}^+(\z _1,\al)
F^+_{a,A}F^-_{b,B}\,.\nn
\end{align}
Now move $\mathbb{T}_{B}^-(\z _2,\al)$ 
and $\mathbb{T}_{A}^+(\z _1,\al)$ next 
to each other using the Yang-Baxter equation:
\begin{align}
\mathbb{T}_{B}^-(\z _2,\al)\mathbb{T}_{a}(\z _1,\al)
=
L^-_{B,a}(\z _2/\z_1)^{-1}\mathbb{T}_{a}(\z _1,\al)
\mathbb{T}_{B}^-(\z _2,\al)L^-_{B,a}(\z _2/\z_1)\,.
\nn
\end{align}
Using the cyclicity of the trace, we obtain
$$
\Tr_{a,b, A,B} \ 
\mathcal{V}_{a,b, A,B}
\mathcal{X}^R_{a,b,A,B}(\z)\mathcal{V}_{a,b, A,B}^{-1}(\z)
\mathbb{T}_{b}(\z _2,\al)\mathbb{T}_{a}(\z _1,\al)
\mathbb{T}_{B}^-(\z _2,\al)\mathbb{T}_{A}^+(\z _1,\al)
$$
where we have set 
$$
\mathcal{V}_{a,b, A,B}(\z)=
L^-_{B,a}(\z)F^+_{a,A}F^-_{b,B}
q^{-D_B\sigma ^3_a}\,.
$$
Using the explicit formula for  $L^-_{B,a}(\z)$ and $F^\pm$, 
it can be shown that 
each entry of the matrix 
$\mathcal{V}_{a,b, A,B}(\z)\in M_a\otimes M_b$ is a
right admissible operator. 
Hence we can change their order according to table 
(\ref{TABLE}). 
After some calculations we obtain the following result. 

\begin{lem}\label{lem:RabAB}
Let $\mathcal{X}^R_{a,b,A,B}(\z)$ be an element of $M_a\otimes M_b$
with right admissible matrix elements. 
Then
\begin{align}
&\Tr \(q^{m(2D_A-2D_B+\sigma ^3_a+\sigma ^3_b)}
q^{-\sigma ^3_aD_B}\mathcal{X}_{a,b,A,B}^R(\z) q^{\sigma ^3_aD_B}
\mathbb{T}_{\{b,B\}}^-(\z _2,\al)
\mathbb{T}_{\{a,A\}}^+(\z _1,\al)\)
\label{quasicom}\\
&=
\Tr \(q^{m(2D_A-2D_B+\sigma ^3_a+\sigma ^3_b)}
q^{\sigma ^3_bD_A}\mathcal{Y}_{A,B,a,b}(\z)q^{-\sigma ^3_bD_A}
\mathbb{T}_{\{a,A\}}^+(\z _1,\al)
\mathbb{T}_{\{b,B\}}^-(\z _2,\al)\)
\nn
\end{align}
and 
\begin{align}
&
\mathcal{Y}_{A,B,a,b}(\z)=R^{\quasi}_{\{a,A\},\{b,B\}}(\z) ^{-1}
\mathcal{X}^L_{a,b,A,B}(\z)
R^{\quasi}_{\{a,A\},\{b,B\}}(\z)
\nn\,,
\end{align}
where $\mathcal{X}^L_{a,b,A,B}(\z)$ is obtained from 
$\mathcal{X}^R_{a,b,A,B}(\z)$ according to the table (\ref{TABLE}).
The quasi-$R$-matrix $R^{\quasi}_{\{a,A\},\{b,B\}}(\z)$ is given by
\bea
R^{\quasi}_{\{a,A\},\{b,B\}}(\z) =
\begin{pmatrix}
R^{\quasi}_{11} & R^{\quasi}_{12}& 0 &0\\
0 &R^{\quasi}_{22} &0&0\\
R^{\quasi}_{3,1}&R^{\quasi}_{3,2}&R^{\quasi}_{3,3} 
&R^{\quasi}_{3,4}\\
0 & R^{\quasi}_{4,2}&0&R^{\quasi}_{4,4}
\end{pmatrix}_{a,b}
\label{quasiRmat}
\ena
Here, setting $Y=Y_{A,B}(\z)$ and $U=U_{A,B}(\z)$, we have
\begin{align}
&R^{\quasi}_{11}=Y^{-1}(1-\z Y)(1-\z q^2Y)\,,
\quad R^{\quasi}_{12}=-\z q^2Y^{-1}(1-\z Y)\ab _A\,,\nn\\
&R^{\quasi}_{22}=q(1-\z ^2 q^2)Y^{-1}\,,\nn\\\nn\\
&R^{\quasi}_{33}=-(1-\z^2q^2)q^{-1}Y\,,\qquad\quad \quad 
R^{\quasi}_{34}=(1-\z^2q^2)q^{-3}\z\frac Y {1-\z q^{-2}Y}\ab _A\,,\nn\\
&R^{\quasi}_{44}=-q^{-2}(1-\z^2q^2)(1-\z^2q^{-2})\frac {Y}{(1-\z q^{-2}Y)
(1-\z q^{-4}Y)},\nn\\ 
\nn\\ &R^{\quasi}_{31} =  -U,\qquad\qquad \qquad \qquad \qquad  R^{\quasi}_{32} = 
-\z q \frac {q^{-3}Y-\z q}{1-\z q^{-2}Y}\,,\nn\\
&
 R^{\quasi}_{42}= -q\frac {1-\z^2q^2}{(1-\z q^{-2}Y)(1-\z q^{-4}Y)}U\,,\nn
 \\\,.\nn
\end{align}
\end{lem}

The following analogs of (\ref{adjust}) are also useful. 
\begin{lem}\label{lem:adjust2}
 We have 
\begin{align}
&
\Tr \(q^{-2D_A+2D_B}
q^{-\sigma ^3_aD_B}\mathcal{X}^R_{a,b,A,B}(\z)q^{\sigma ^3_aD_B}
\mathbb{T}_{\{b,B\}}^-(\z _2,\al)
\mathbb{T}_{\{a,A\}}^+(\z _1,\al)\)
\nn\\
& =\zeta
\frac{N (\al -\mathbb{S})}{N (\al -\mathbb{S}-1)}
\Tr \(
q^{-\sigma ^3_aD_B-1}Y_{B,A}(\z ^{-1})\mathcal{X}^R_{a,b,A,B}(\z)
q^{\sigma ^3_aD_B}
\mathbb{T}_{\{b,B\}}^-(\z _2,\al)
\mathbb{T}_{\{a,A\}}^+(\z _1,\al)\)
,
\nn
\end{align}
similar formula for opposite order of multipliers is obtained
by spin reversal and $\al\to -\al$.
\end{lem}


\subsection{Proof of Lemma \ref{lem:ResG}}

Here we prove the following Lemma. 
\begin{lem}\label{lem:residueG} 
Let $\mathbf{m}^{(+-)}(\z,\xi,\al)$ be as in \eqref{k+k-}, and  
set $\eta=\z/\xi$. Then as $\eta\to1$ we have 
\be
&&\mathbf{m}^{(+-)}(\z,\xi,\al)=\frac{1}{\eta-\eta^{-1}}+O(1)\,.
\en
\end{lem}
\begin{proof}
The proof is based on the following identity:
\begin{align}
&\ \ \ 
-N(\al-\mathbb{S})\Tr_{a,b,A,B}\left\{
q^{-2D_A}\sigma^+_a\sigma^-_b \mathbb{T}^+_{\{a,A\}}(\z ,\al)
\mathbb{T}^-_{\{b,B\}}(\xi,\al)\right\}\eta^{\al-\mathbb{S}}
\label{B0} \\
&+
N(\al-\mathbb{S})\Delta_{\z }
\Tr_{a,b,A,B}
\left\{\widetilde{M}_{b,A,B}(\eta)\mathbb{T}^+_{A}(\z ,\al)
\mathbb{T}^-_{\{b,B\}}(\xi,\al)\right\}\eta^{\al-\mathbb{S}}
\nn\\
&=\Tr_{a,b}\left\{B^0_{a,b}(\eta)\mathbb{T}_a(\z,\al)\mathbb{T}_b(\xi,\al)\right\}
\Tr_{V(\Lambda)}\mathbb{T}_v(\sqrt{\z \xi},\al)\nn \\
&-\frac {1}{\eta q-\eta ^{-1}q^{-1}}
\Tr_{V(\Lambda+2)}
\mathbb{T}_v(\sqrt{\z \xi},\al)-
\frac {1}{\eta q^{-1}-\eta ^{-1} q}
\Tr_{V(\Lambda-2)}\mathbb{T}_v(\sqrt{\z \xi},\al)\,,
\nn
\end{align}
where  
$q^{\Lambda}=q\eta$,
$$
B^0(\eta)
=\frac{(\eta -\eta ^{-1})}{(\eta q -\eta ^{-1}q^{-1})
(\eta q^{-1}-\eta ^{-1} q)}
\(q\tau^+_a\tau^-_b+q^{-1}\tau^-_a\tau^+_b
-\eta^{-1}\sigma ^+_a\sigma ^-_b-\eta \sigma ^-_a\sigma ^+_b\),
$$
and
$$
\widetilde{M}_{b,A,B}(\eta)=
\frac{1}{\eta-\eta^{-1}}q^{\sigma^3_bD_A}
\left(\sigma^3_b+\UAB(\eta)\sigma^-_b\right)
q^{-\sigma^3_bD_A}\,.
$$

To prove the identity (\ref{B0}), start with 
\begin{align}
&\Tr_{a,b}\left\{B^0_{a,b}(\eta)\mathbb{T}_a(\z,\al)\mathbb{T}_b(\xi,\al)\right\}
\Tr_{V_{\Lambda}}\mathbb{T}_v(\sqrt{\z \xi},\al)\nn\\ 
&=-N(\al -\mathbb{S})
\Tr_{a,b,A,B}\ B^0_{a,b}(\eta)\mathbb{T}_a(\z,\al)
\mathbb{T}_b(\xi,\al)
\mathbb{T}^+_A(\z,\al)\mathbb{T}^-_B(\xi,\al)
\eta^{\al-\mathbb{S}}\nn
\\
&=-N(\al -\mathbb{S})\Tr_{a,b,A,B}H_{a,b,A,B}(\eta)
\mathbb{T}^+_{\{a,A\}}(\z,\al)\mathbb{T}^-_{\{b,B\}}(\xi,\al)
\eta^{\al-\mathbb{S}}\,,\nn
\end{align}
where 
$$
H_{a,b,A,B}(\eta)=\(F_{a,A}^+F_{b,B}^-\)^{-1}
L_{A,b}(\eta)^{-1}B^0_{a,b}(\eta)L_{A,b}(\eta)F_{a,A}^+F_{b,B}^-\,.
$$
The rest of the proof is a direct computation of 
$H_{a,b,A,B}(\eta)$.

Consider the residue of both sides of (\ref{B0}) 
at $\eta ^2=q^{-2}$.
Since $\widetilde{M}_{b,A,B}(\eta)$ and $M'_{b,A,B}(\eta)$
have the same residue at $\eta=1$, the residue of the left hand side 
gives $\text{res}_{\eta ^2=1}\mathbf{m}^{(+-)}(\z,\xi,\al)$. 
On the other hand, the quantum determinant relation gives 
$\text{res}_{\eta^2=1}\Tr _{a,b}\ 
B^0_{a,b}(\eta)\mathbb{T}_a(\z)\mathbb{T}_b(\xi)=1$. 
Hence in the right hand side we have
$$ 
\(\Tr_{V(0)}-\Tr_{V(2)}\)\mathbb{T}_v(\z q,\al)=1\,.
$$
This completes the proof. 
\end{proof} 
\section{Equivalence to the previous definition}\label{sec:appD}

In our previous papers \cite{HGSI},\cite{FB}, 
annihilation operators were introduced 
in a way different from the one in the present paper. 
The old definition in \cite{HGSI} reads
\be
&&\cb^{OLD}_{[k,l]}(\z,\al+1)={\rm sing}\, 
(1-q^{2(\al-\bS+1)})\kb^{OLD}_{[k,l]}(\z,\al+1), 
\\
&&\bb^{OLD}_{[k,l]}(\z,\al)
=(q^{-\al+\bS}-q^{\al-\bS})^{-1}\J\ \cb^{OLD}(\z,-\al)\ \J\,,
\en
where 
\be
&&\kb^{OLD}(\z,\al+1)(X_{[k,l]})
=\Tr_{a,A}\left[q^{2\al D_A}
\sigma^+_a\bT_{a}(\z)^{-1}\bT^-_{A}(\z)^{-1}
(q\z)^{\al-\bS}(X_{[k,l]})\right]\,,
\en
and
\be
{\rm sing}\, f(\z)=\int_{\Gamma}\frac{d\xi}{2\pi i}
\frac{f(\xi)}{\z-\xi}\,.
\en
They are related to the present definition via 
\bea
&&\cb^{OLD}_{[k,l]}(\z,\al+1)=-2q^{\al}(1-q^{2(\al-\bS+1)})
{\rm sing}\, \cb_{[k,l]}(\z,\al)\,,
\label{old-new}
\\
&&\bb^{OLD}_{[k,l]}(\z,\al-1)=2q^{-\al+1}\frac{1}{1-q^{2(\al-\bS-1)}}
{\rm sing}\, \bb_{[k,l]}(\z,\al)\,.
\label{old-new2}
\ena
In particular we have 
\be
\bb^{OLD}_{[k,l]}(\z_1,\al-1)\cb^{OLD}_{[k,l]}(\z_2,\al)\frac{d\z_1}{\z_1}
\frac{d\z_2}{\z_2}
=
-{\rm sing}\, 
\bb_{[k,l]}(\z_1,\al)\cb_{[k,l]}(\z_2,\al-1)
\frac{d\z^2_1}{\z^2_1}\frac{d\z^2_2}{\z^2_2}\,.
\en
In the following we shall show \eqref{old-new}. 

Introduce an anti-involution $\tau$ 
of the $q$-oscillator algebra $Osc$ by 
\be
&&\tau(\ao)=-\ao^*q^{-2D-1},
\quad
\tau(\ao^*)=-\ao\ q^{2D-1},
\quad 
\tau(q^D)=q^D\,.
\en
We have 
\be
&&\tau(L^\pm_{A,j}(\zeta)^{-1})=L^\mp_{A,j}(q^{-1}\zeta)\,,
\en
and
\be
&&\Tr_A\left(q^{2\al D}\tau(x)\right)=\Tr_A(q^{2\al D}x)
\quad (x\in Osc). 
\en
Applying $\tau\circ \theta_a$ inside the trace, 
we obtain 
\bea
&&\kb^{OLD}(\z,\al+1)(X_{[k,l]})
=-\Tr_{a,A}\left[T_{\{a,A\}}(q^{-1}\z)q^{2\al D_A}
\sigma^+_a (q\z)^{\al-\bbS}(X_{[k,l]})
T_{\{a,A\}}(q\z)^{-1}\right]
\nn\\
&&\quad\quad =-q^\al\Bigl\{
\Tr_A\left[T_A(\z)q^{-2S_{[k,l]}+2\al D_A}
\z^{\al-\bbS}(X_{[k,l]})\tilde{C}_A(q\z)\right]
\label{res-k0}\\
&&\quad\qquad +
\Tr_A\left[C_A(q^{-1}\z)q^{-S_{[k,l]}+2\al D_A}
\z^{\al-\bbS}(X_{[k,l]})
T_A(\z)^{-1}\right]\Bigr\}\,.\nn
\ena
Here $\tilde{C}_A(\z)$ denotes the $(2,1)$ block of 
$T_{\{a,A\}}(\z)^{-1}$. 

On the other hand, the operator $\kb_{[k,l]}(\z,\al)$ 
is written as 
\be
&&\kb(\z,\al)(X_{[k,l]})=\Tr_A
\Bigl[T_A(q^{-1}\z)q^{-2S_{[k,l]}+2\al D_A}
(q^{-1}\z)^{\al-\bS}(X_{[k,l]})\tilde{C}_A(\z)\Bigr]
\\
&&\qquad\qquad\qquad\quad +
\Tr_{A}\Bigl[C_A(\z)q^{-S_{[k,l]}+2\al D_A}
(q\z)^{\al-\bS}(X_{[k,l]})
T_A(q\z)^{-1}\Bigr]\,. 
\en
Using the explicit formula for $L_{\{a,A\},j}(\z)$  
it can be shown that in the last line  
only one term is singular
at each of the poles $\z^2=q^{\pm 2}\xi_j^2$,  i.e., 
\bea
&&{\rm sing}\, \kb(q\z,\al)(X_{[k,l]}) 
={\rm sing}\, 
\Tr_A\left[T_A(\z)q^{-2S_{[k,l]}+2\al D_A}\z^{\al-\bbS}(X_{[k,l]})
\tilde{C}_A(q\z)\right]\,,
\label{res-k1}\\
&&
{\rm sing}\, \kb(q^{-1}\z,\al)(X_{[k,l]}) 
={\rm sing}\, 
\Tr_A\left[C_A(q^{-1}\z)q^{-S_{[k,l]}+2\al D_A}
\z^{\al-\bbS}(X_{[k,l]}) T_A(\z)^{-1}\right]\,.
\label{res-k2}
\ena

It follows from \eqref{res-k1},\eqref{res-k2}, \eqref{res-k0}
that 
\bea
{\rm sing}\,
\left(\kb_{[k,l]}(q\z,\al)+\kb_{[k,l]}(q^{-1}\z,\al)\right)
={\rm sing}\,\kb^{OLD}_{[k,l]}(\z,\al+1)\,, 
\ena
which implies the desired relation \eqref{old-new}.

\bigskip

\noindent
{\it Acknowledgments.}\quad
HB is grateful to the Volkswagen Foundation and to the
'Gradui- ertenkolleg' DFG project:
"Representation theory  and its application in
mathematics and physics" for the financial support.
Research of MJ is supported by 
the Grant-in-Aid for Scientific Research B--18340035 and A--18204012. 
Research of TM is supported by 
the Grant-in-Aid for Scientific Research B--17340038.
Research of FS is supported by  EC networks   "ENIGMA",
contract number MRTN-CT-2004-5652
and GIMP program (ANR), contract number ANR-05-BLAN-0029-01. 
Research of YT is supported by the Grant-in-Aid for Young Scientists B--17740089. 

The authors are grateful to O. Babelon, F. G{\"o}hmann, A. Kl{\"u}mper,
J.-M. Maillet and J. Suzuki for their interest and friendly support.

FS is grateful for hospitality to Theory group at DESY, Hamburg (visit was supported by EU-grant MEXT-CT-2006-042695)  where an important part of this work was done, special thanks a due to J. Teschner. 

HB and FS are grateful to Tokyo University for hospitality.

\end{document}